\pdfoutput=1 
\documentclass[showpacs,prb,aps,superscriptaddress,twocolumn,floatfix]{revtex4}
\usepackage{graphicx,float,amsmath,amssymb}
\usepackage[francais,english]{babel}
\newfont{\ensmathquatorze}{msbm10 scaled 1400}
\newfont{\ensmathonze}{msbm10 scaled 1100}
\newfont{\ensmathdix}{msbm10}
\newfont{\ensmathneuf}{msbm10 scaled 833}
\newfont{\ensmathhuit}{msbm10 scaled 694}
\newfam\ensmathfam
\textfont\ensmathfam=\ensmathonze
\scriptfont\ensmathfam=\ensmathdix
\scriptscriptfont\ensmathfam=\ensmathhuit
\def\ensmf{\fam\ensmathfam\ensmathonze}         

\renewcommand\em{\it}

\renewcommand{\leq}{\leqslant}
\renewcommand{\geq}{\geqslant}

\def\eqdef{\stackrel{\mbox{\tiny def}}{=}}     
\def\eqlaw{\stackrel{\mbox{\tiny (law)}}{=}}     
\newcommand{\ket}[1]{|\kern.3ex#1\kern.3ex\rangle}
\newcommand{\bra}[1]{\langle\kern.3ex #1 \kern.3ex|}
\newcommand{\mean}[1]{\left\langle #1 \right\rangle} 
\newcommand{\smean}[1]{\langle #1 \rangle} 

\newcommand{\EXP}[1]{{\mbox{\large e}}^{#1}}         
\newcommand{\argcosh}{\mathop{\mathrm{argch}}\nolimits}
\newcommand{\argsinh}{\mathop{\mathrm{argsh}}\nolimits}

\newcommand{\re}{\mathop{\mathrm{Re}}\nolimits}      
\newcommand{\tr}[1]{\mathop{\mathrm{Tr}}\nolimits\left\{ #1 \right\}}  
\newcommand{\cotg}{\mathop{\mathrm{cotg}}\nolimits}  
\renewcommand{\min}[2]{\mathop{\mathrm{min}}\nolimits\left( #1 , #2\right)}

\def\NN{{\ensmf N}}                 
\def\ZZ{{\ensmf Z}}                 
\def\RR{{\ensmf R}}                 

\def\I{\mathrm{i}}                  
\def\D{\mathrm{d}}                  

\newcommand{\derivp}[2]{\frac{\partial #1}{\partial #2}}

\newcommand\ab{{\alpha\beta}}

\newcommand{\diagram}[3]{\raisebox{#3}{\includegraphics[scale=#2]{#1}}}
\newcommand\sw{s}

\begin{document}

\title{ Quantum oscillations and decoherence due to electron-electron 
  \\interaction in metallic networks and hollow cylinders }

\author{Christophe Texier}
\affiliation{Laboratoire de Physique Th\'eorique et Mod\`eles Statistiques,
              UMR 8626 du CNRS, Universit\'e Paris-Sud, 91405 Orsay, 
              France.}
\affiliation{Laboratoire de Physique des Solides, UMR 8502 du CNRS,
              Universit\'e Paris-Sud, 91405 Orsay, France.}

\author{Pierre Delplace}
\affiliation{Laboratoire de Physique des Solides, UMR 8502 du CNRS,
              Universit\'e Paris-Sud, 91405 Orsay, France.}

\author{Gilles Montambaux}
\affiliation{Laboratoire de Physique des Solides, UMR 8502 du CNRS,
              Universit\'e Paris-Sud, 91405 Orsay, France.}

\date{July 17, 2009}

\begin{abstract}
We have studied the quantum oscillations of the conductance for arrays
of connected mesoscopic metallic rings, in the presence of an
external magnetic field. Several geometries have been considered: a
linear array of rings connected with short or long wires compared to
the phase coherence length, square networks and hollow
cylinders. Compared to the well-known case of the isolated ring, we
show that for connected rings, the winding of the Brownian
trajectories around the rings is modified, leading to a different harmonics
content of the quantum oscillations. We relate this harmonics content
to the distribution of winding numbers. We consider the
limits where coherence length $L_\varphi$ is small or large compared to the
perimeter $L$ of each ring constituting the network. In the latter case,
the coherent diffusive trajectories explore a region larger than $L$,
whence a network dependent harmonics content. Our analysis is based on
the calculation of the spectral determinant of the diffusion equation
for which we have a simple expression on any network. It is also based
on the hypothesis that the time dependence of the dephasing between
diffusive trajectories can be described by an exponential decay with a
single characteristic time $\tau_\varphi$ (model A) . 

At low temperature, decoherence is limited by electron-electron
interaction, and can be modelled in a one-electron picture by the
fluctuating electric field created by other electrons (model B).  It
is described by a functional of the trajectories and thus the
dependence on geometry is crucial.  Expressions for the
magnetoconductance oscillations are derived within this model and
compared to the results of model A. It is shown that they involve
several temperature-dependent length scales. 
\end{abstract}


\maketitle





\section{Introduction}


Understanding which processes limit phase coherence in electronic
transport is an important issue
in mesoscopic physics. Such phenomena like weak localization or
universal conductance fluctuations are well understood to result from
phase coherence effects limited at a given time (or length) scale
called phase coherence time $\tau_\varphi$ (or phase coherent length
$L_\varphi$).   
This explains the interest in studying quantum corrections to
the classical conductivity~: to provide a powerful experimental probe of phase
coherence in weakly disordered metals and furnish some
informations on the microscopic mechanisms responsible for the
limitation of quantum coherence.  
This limitation originates from  the coupling of electrons
to external degrees of freedom like magnetic impurities or phonons
\cite{ChaSch86,AkkMon07}. It also results from the 
interaction among electrons themselves. The physical origin of this
decoherence in weakly disordered metals has been understood 
in the pioneering paper of Altshuler, Aronov \& Khmelnitskii (AAK)
\cite{AltAroKhm82}. 
In a one electron picture, it is due  to the fluctuations of the
electric field created by the other electrons.
In a quasi-1d wire, these authors have shown that this mechanism leads
to the following temperature dependence of 
the dephasing time $\tau_\varphi(T)\propto{}T^{-2/3}$.
This power-law can be understood qualitatively as follows:
 the typical dephasing is proportional to the fluctuations of the
 electric potential, which themselves are known from Nyquist theorem 
to be proportional to the temperature $T$ and to the resistance of the
sample. For an infinite wire, 
the relevant fluctuations are limited to the scale of the coherence
length itself. Consequently, the dephasing time 
has the structure~: ${\hbar}/{\tau_\varphi}={k_BT}/{g(L_\varphi)}$,
where $g(L_\varphi)$ is the dimensionless conductance at the length
scale $L_\varphi$. For a quasi-1d conductor, the
conductance is linear in length and the length scales as the
square-root of time. Therefore the 
function $g(L_\varphi)$ scales as $\sqrt{\tau_\varphi}$, whence the
above power law. 

More recently, Ludwig \& Mirlin \cite{LudMir04} and 
two of the authors 
\cite{TexMon05b} 
have considered 
the geometry of a ring, and they have shown that the damping of
magnetoresistance oscillations 
could be described with a different temperature dependence of the
dephasing time $\tau_\varphi(T) \propto{}T^{-1}$. 
This new behaviour can be qualitatively understood by considering that
the diffusive trajectories encircle the 
ring and have all a length equal to the perimeter $L$ of the ring, so
that the relevant resistance is the 
resistance of the ring itself. As a result~: 
${\hbar}/{\tau_\varphi}={k_BT}/{g(L)}$. 

In Ref.~\cite{TexMon05b} we have shown how the dephasing on a
ring depends on the nature of the diffusive  trajectories~:
the fluctuations of the electric potential  affect differently
trajectories which 
encircle the ring and trajectories which do not encircle it. Within
this framework,
we have analyzed magnetoresistance experiments performed
on a square network of quasi-1d wires, and we have found that indeed
two characteristic lengths 
with two different temperature dependence could be extracted from the
data \cite{FerRowGueBouTexMon08}.
These recent considerations have led us to the general conclusion that
the dephasing depends on 
the geometry
of the system considered.

The purpose of this paper  is to analyze the dephasing process 
and to calculate the weak localization correction in different 
geometries, where the decoherence induced by electron-electron
interaction may have a more  complex structure. 
In order to address this question, it is important to understand that
the weak localization correction depends on two ingredients, one is
the probability to have pairs of reversed trajectories, which is
related to the return probability $\mathcal{P}(t)$ for a diffusive
particle after a time $t$, the other is the nature of the dephasing
process itself.  
Schematically, the weak localization correction to the conductivity
can be written as
\begin{equation}
  \label{eq:1}
  \Delta\sigma\sim -\int_0^\infty \D t\, \mathcal{P}(t)\, 
  \langle \EXP{\I\Phi(t)} \rangle
  \:,
\end{equation}
where $\langle\EXP{\I\Phi(t)}\rangle$ is the average dephasing
accumulated along a diffusive trajectory for a time $t$.
The return probability  has been analyzed in Ref.~\cite{TexMon05} for
various types of networks.  Its Laplace transform, 
the spectral determinant, can be simply calculated from the parameters
of the network. 
More complex is the analysis of the dephasing process itself.
A simple and natural ansatz would be to assume an exponential decay
$\langle\EXP{\I\Phi(t)}\rangle=\EXP{-t/\tau_\varphi}$. This assumption
is correct 
when the dephasing is due to random magnetic impurities or
electron-phonon 
scattering. For electron-electron interaction
the analysis of the AAK result for a wire shows that time dependence
is not exponential \cite{MonAkk05}~:
$\langle\EXP{\I\Phi(t)}\rangle_\mathrm{ee}\neq\EXP{-t/\tau_\varphi}$. The
qualitative reason stands again on the 
fact that dephasing can be described as due to the fluctuations of
the electric potential due to other 
electrons. Then, one may understand that this dephasing depends on
the nature of the trajectories and is not 
exponential.
The main goal of this paper is to describe this dephasing for complex
networks and to generalize 
the known results of the infinite wire and the ring.

\vspace{0.25cm}

The paper is organized as follows :
In section~\ref{sec:background}, we recall the physical basis at the
origin of this work 
and in section~\ref{sec:gf} we  present 
the general formalism appropriate for our study. In the next
sections, we consider successively more and more complex geometries. 
In section~\ref{sec:wr}, we recall known results for the infinite wire and the
ring. In section~\ref{sec:conring}, we consider 
the case of a ring attached to arms and show how the harmonics of the
magnetoresistance oscillations are reduced 
by the existence of the arms. The situation is the same for a chain
of rings connected through  arms longer than the coherence
length. When rings become close to each other the dephasing in one
ring 
is strongly modified by the winding trajectories in the neighboring
rings. This is discussed in section~\ref{sec:chain}. 
The case of an infinite regular network is much more
difficult to address since the hierarchy of diffusive trajectories
is difficult to analyze, and we 
have used the limit
of the infinite plane as a guideline (section~\ref{sec:sn}). Finally the case
of a hollow cylinder (section~\ref{sec:cyl}) 
is quite interesting since it combines trajectories winding around
the axis of the cylinder and two-dimensional 
trajectories.  Throughout the paper, we shall consider two situations,
respectively denoted by {\it model A} and {\it model B}~: the case where the
dephasing has a simple exponential time dependence, 
and the case where the dephasing is induced by electron-electron
interaction. We shall systematically discuss the analogies 
and the differences between these two situations. 


\section{Background\label{sec:background}}

In a weakly disordered metal, due to elastic scattering by impurities,
the classical conductivity reaches a finite value at low temperature,
given by the Drude conductivity $\sigma_0=\frac{ne^2\tau_e}{m}$, where
$n$ is the electronic density and $\tau_e$ the elastic scattering
time. Quantum interferences are responsible for small quantum
corrections to the Drude result. One important contribution, that
survives averaging over the
disorder~\cite{UmbHaeLaiWasWeb86,WasWeb86,SchMalMaiTexMonSamBau07},  
comes from interferences of
reversed closed electronic trajectories, and therefore diminishes the
conductivity. This quantum contribution to the {\it average}
conductivity is called the {\it weak localization} (WL) correction.
It has been expressed as (\ref{eq:1}) where the function
$\smean{\EXP{\I\Phi(t)}}$ describes dephasing and cut off the large time
contributions. A simple exponential decay
$\smean{\EXP{\I\Phi(t)}}=\EXP{-t/\tau_\varphi}$ is usually assumed
(denoted {\it model A} in the present paper). $\tau_\varphi$ is the
{\it phase coherence time}, related to the {\it phase coherence
  length} $L_\varphi=\sqrt{D\tau_\varphi}$, where $D$ is the diffusion
constant of electrons in the disordered metal. 
From eq.~\eqref{eq:1}, we obtain the WL correction in a wire  
$\Delta\sigma^\mathrm{(1d)}\sim-\int_{\tau_e}^{\tau_\varphi}\frac{\D{}t}{\sqrt{t}}\sim-\sqrt{\tau_\varphi}\sim-L_\varphi$
and in a plane
$\Delta\sigma^\mathrm{(2d)}\sim-\int_{\tau_e}^{\tau_\varphi}\frac{\D{}t}{t}\sim-\ln(\tau_\varphi/\tau_e)$
(diffusion sets in after a time $\tau_e$, whence the lower cutoff in
the integrals).
In practice, the WL is a small correction to the Drude conductivity
and it can be extracted thanks to its sensitivity to a magnetic
field. In the presence of a magnetic field, the contribution of a
closed diffusive trajectory $\mathcal{C}$ is multiplied by 
$\EXP{2\I{}e\phi_\mathcal{C}/\hbar}$, where
$\phi_\mathcal{C}$ is the magnetic flux through the loop. 
This phase factor comes from the interference of the {\it two} reversed
electronic trajectories, whence the factor $2$.  
After summation over
all loops, the additional magnetic phase is responsible for the 
vanishing of  the contributions of loops such that
$\phi_\mathcal{C}\gtrsim\phi_0$, where $\phi_0=h/e$ is the 
flux quantum. Therefore the magnetic field provides an additional
cutoff at time $\tau_\mathcal{B}$ corresponding to diffusive
trajectories encircling one flux quantum. In a narrow wire of width
$w$ we have
$\tau^\mathrm{(1d)}_\mathcal{B}\sim\phi_0^2/(Dw^2\mathcal{B}^2)$ and
in a thin film (plane)
$\tau^\mathrm{(2d)}_\mathcal{B}\sim\phi_0/(D\mathcal{B})$ (see 
Refs.~\cite{ChaSch86,AkkMon07}).  
The two cutoffs are added ``\`a la Matthiessen'' \cite{AltAro81,Ber84} as
$1/\tau_\varphi\to1/\tau_\varphi+1/\tau_\mathcal{B}$~; this leads to a
smooth dependence of the WL correction as a function of the magnetic field.

The above discussion concerns homogeneous devices (like a wire or a plane).
Another experimental setup appropriate to study quantum interferences 
and extract the phase coherence length is a metallic ring or an array of 
rings. 
In this case the
topology constrains the magnetic flux intercepted by the rings to be
an integer multiple of the flux per ring $\phi$ (we neglect for the moment the
penetration of the magnetic field in the wires)~:
$\phi_\mathcal{C}=n\,\phi$ with $n\in\ZZ$. 
This gives rise to Aharonov-Bohm (AB) oscillations of the conductance
as a function of the flux with period $\phi_0$. 
Disorder averaging is responsible for the vanishing of these
$\phi_0$-periodic oscillations~: only survive
the contributions of the reversed electronic trajectories leading to
WL correction oscillations, known as Al'tshuler-Aronov-Spivak (AAS)
oscillations~\cite{AltAroSpi81,AroSha87}, with a period {\it half} of the
flux quantum. 
It will be convenient to introduce the harmonics $\Delta\sigma_n$ of the
periodic WL correction.  
An important motivation for considering the harmonic content $\Delta\sigma_n$
in networks, is that it allows to
decouple the two effects of the magnetic
field~\cite{FerAngRowGueBouTexMonMai04}~:
the rapid AAS oscillations ($\Delta\sigma_{n\neq0}$) and the
penetration of the magnetic field in the wires, responsible for a
smooth decrease of the MC at 
large field ($\Delta\sigma_{0}$).
Since the $n$-th harmonic is given by contributions of
loops encircling $n$ fluxes we can write
\begin{equation}
  \label{harmonics}
  \Delta\sigma_n =
  -\frac{2e^2D}{\pi\sw}\int_0^\infty\D{t}\,\mathcal{P}_n(t)\,\EXP{-t/\tau_\varphi}
  \:,
\end{equation}
where $\mathcal{P}_n(t)$ is the return probability after a time $t$
having encircled $n$ fluxes. 
$\sw$ is the cross section of the wires.
 In an isolated ring of perimeter $L$, this probability reads
$\mathcal{P}_n^\mathrm{ring}(t)=\frac1{\sqrt{4\pi Dt}}\exp-\frac{(nL)^2}{4Dt}$.
Integral (\ref{harmonics}) gives~\cite{AltAroSpi81}
\begin{equation}
  \label{AAS}
  \Delta\sigma_n^\mathrm{AAS} = -\frac{2e^2}{h\,\sw}\,L_\varphi\,
  \EXP{-|n|L/L_\varphi}
  \:,
\end{equation}
where 
$L_\varphi=\sqrt{D\tau_\varphi}$ is the phase coherence length. 
Note that $\Delta\sigma_{-n}=\Delta\sigma_{n}$ follows from the
symmetry under reversing the magnetic field~; in the following we will
simply consider $n\geq0$.
The exponential decay
of the harmonics directly originates from the diffusive nature of
the winding around the ring~: for a time $t\sim\tau_\varphi$, the
typical winding scales as $n_t\sim\sqrt{Dt}/L\sim{}L_\varphi/L$. 
The AAS oscillations were first observed in narrow metallic hollow
cylinders~\cite{ShaSha81,AltAroSpiShaSha82} and in large metallic
networks \cite{PanChaRamGan84,PanChaRamGan85,AroSha87}.

Although the simple behaviour (\ref{AAS}) has been used to analyze
AAS or AB oscillations~\cite{footnote1}
in many experiments until recently
(see for example Refs.~\cite{WasWeb86,PieBir02}), a realistic
description of a network made of connected rings leads to harmonics
with a $L/L_\varphi$ dependence {\it a priori} quite different from the simple
exponential prediction (\ref{AAS})  for two
reasons related to the nontrivial {\it topology} of the networks.

\vspace{0.15cm}

\begin{figure}[htbp]
\begin{center}
\includegraphics[scale=0.75]{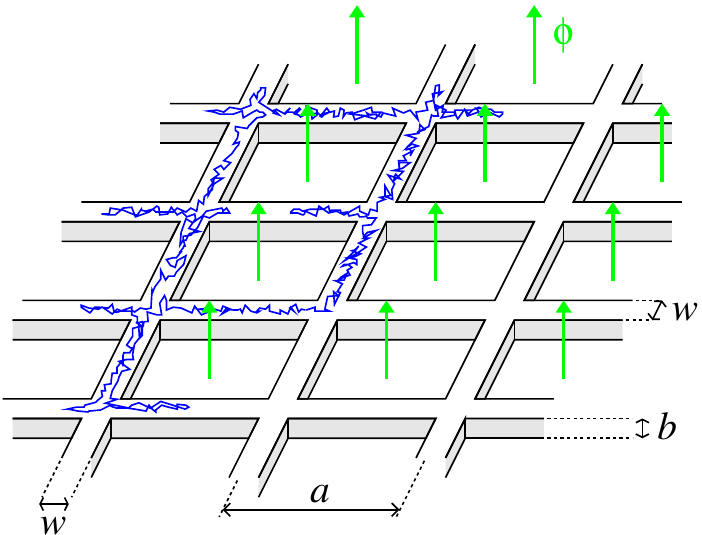}
\caption{\it A square metallic network submitted to a magnetic field.
  Schematic picture of a
  closed diffusive trajectory winding a flux $2\phi$ is represented
  ($\phi$ is the flux per elementary plaquette).}
\label{fig:sn}
\end{center}
\end{figure}

\noindent
({\it i}) {\it Winding properties of diffusive loops in networks.--} 
Consider for example the square network of figure~\ref{fig:sn} made of
rings of perimeter $L=4a$.
For $L_\varphi\ll{}L$, an electron unlikely keeps its phase coherence
around a ring, therefore AAS oscillations are dominated by trajectories
enlacing one ring only and all rings can be considered as independent.
In the opposite regime~\cite{footnote2}
$L_\varphi\gtrsim{}L$, the interfering
electronic trajectories 
explore regions much larger than the ring perimeter $L$. In this
case,  winding properties are more
complicated (figure~\ref{fig:sn}) and the probability
$\mathcal{P}_n(t)$ may strongly differ from the one obtained for a
single ring $\mathcal{P}_n^\mathrm{ring}(t)$.
A theory must be developed to account for these
topological effects, which leads to an harmonic content quite different from
(\ref{AAS}).
This was done in
Refs.~\cite{DouRam85,DouRam86,Pas98,PasMon99,AkkComDesMonTex00} 
for large regular networks. This theory was later extended in 
Ref.~\cite{TexMon04} in order to deal with arbitrary networks, properly 
accounting for their connections to contacts~\cite{footnote19}.

\vspace{0.15cm}

\vspace{0.15cm}

\noindent
({\it ii}) {\it e-e interaction leads to geometry dependent decoherence.--}
Not only the winding probability $\mathcal{P}_n(t)$
involved in eq.~(\ref{harmonics}) is affected by the nontrivial
topology of networks, but also 
$\smean{\EXP{\I\Phi(t)}}=\EXP{-t/\tau_\varphi}$
describing  the nature of phase coherence relaxation is replaced by a
more complex function.
When decoherence is due to e-e interaction, the dominant
phase-breaking mechanism at low temperature, this relaxation is not
described by a simple exponential anymore. 
This situation will be refered as {\it model B} throughout this paper.
Such decoherence can be modeled in a one-electron picture by including
dephasing due to the fluctuating electromagnetic field created by the
other electrons \cite{AltAroKhm82}. 
Therefore the pair of reversed interfering trajectories picks up an
additional phase $\Phi[\mathcal{C}]$ that depends on the electric
potential $V$. Averaging over the fluctuations of the potential leads
to the relaxation of phase coherence. The harmonics present the
structure 
\begin{equation}
  \label{harmFV}
  \Delta\sigma_n\sim-\int_0^\infty\D{t}\,\mathcal{P}_n(t)\:
  \smean{ \EXP{\I\Phi[\mathcal{C}_t^n] 
               } }_{V,\mathcal{C}_t^n}
  \:,
\end{equation}
where averaging is taken over the potential fluctuations
$\smean{\cdots}_V$ and over the loops with winding $n$ for time $t$,
$\smean{\cdots}_{\mathcal{C}_t^n}$.
In a quasi 1d-wire, the relaxation of phase coherence
involves an important length scale, the {\it Nyquist length} $L_N$,
characterizing the efficiency of the electron-electron interaction to destroy
the phase coherence in the wire.
We will see that, in networks also, the Nyquist length is the
intrinsic length characterizing decoherence due to e-e interaction. 
It is given by
\cite{AltAroKhm82,AltAro85,footnote3,footnote4}
\begin{equation}
  \label{nyquist}
  \boxed{
  L_N = \left( \frac{\hbar^2\sigma_0D\sw}{e^2k_BT} \right)^{1/3} 
  = \left( \frac{\alpha_d}{\pi} N_c\ell_e L_T^2 \right)^{1/3} 
  }
\end{equation}
where $\sw$ is the cross-section of the wire.  We have rewritten
the Nyquist length in terms of the thermal length
$L_T=\sqrt{\hbar{}D/k_BT}$, the elastic mean free path $\ell_e$ 
and the number of 
conducting channels  $N_c$ (not including spin degeneracy)~; $\alpha_d$ is a 
dimensionless constant depending on the dimension
($\alpha_d=V_d/V_{d-1}$ where $V_d$ is the volume of the unit sphere in dimension
$d$)~\cite{footnote5}.
In the following we will set $\hbar=k_B=1$.
In the infinite wire, the decaying function
$\smean{\EXP{\I\Phi}}_{V,\mathcal{C}_t}$ can only involve the unique length
$L_N$. 
A precise analysis of the
magnetoconductance (MC) of the wire shows that, in this case, the calculation
of (\ref{harmonics}) and (\ref{harmFV}) for the infinite wire, 
for which $\mathcal{P}(t)\propto1/\sqrt{t}$,
leads to almost indistinguishable results provided~\cite{Pie00,AkkMon07}
$L_\varphi\to\sqrt{2}\,L_N$. 
Therefore the analysis of the MC of the wire suggests
that the sophisticated calculation of (\ref{harmFV}) can be replaced by the
simpler one (\ref{harmonics}) with
$\EXP{-t/\tau_\varphi}\to\EXP{-t/2\tau_N}$ where
$\tau_N=L_N^2/D\propto{}T^{-2/3}$ is the Nyquist time.
However this is {\it a priori} not true anymore as soon as we consider
networks with a nontrivial topology {\it because
  electric potential fluctuations
  depend on the geometry, and therefore the decoherence is
  geometry-dependent}.

Let us formulate this idea more precisely.
Being related to the potential as $\Phi\sim\int^tV$, fluctuations of
the phase $\Phi$ picked by the two reversed electronic trajectories
can be related to the power spectrum of the potential, given by the
fluctuation-dissipation theorem (FDT)~:
$\frac{\D}{\D{}t}\smean{\Phi^2}_{V,\,\mathcal{C}_t}\sim{}e^2T\mathcal{R}_t\sim\frac{e^2T}{\sigma_0\sw}x(t)$
where $\mathcal{R}_t=x(t)/(\sigma_0\sw)$ is the resistance of a wire of length
$x(t)$. The average $\smean{\cdots}_{V,\,\mathcal{C}_t}$ is taken over
potential fluctuations and closed diffusive trajectories
$\mathcal{C}_t$ for a time scale $t$. The length $x(t)$ is the typical
length probed by electronic trajectories. For an infinite wire it
scales like $x(t)\sim\sqrt{Dt}$ therefore
$\smean{\Phi^2}_{V,\,\mathcal{C}_t}\sim(t/\tau_N)^{3/2}$ where
$\tau_N=L_N^2/D\sim{T}^{-2/3}$ is the Nyquist time. On the other hand,
in a ring, diffusion is constrained by the geometry~: harmonics
of the MC of a ring involve winding trajectories for which the length
scale probed is therefore the perimeter $x(t)\sim{L}$, leading to
$\smean{\Phi^2}_{V,\,\mathcal{C}_t}\sim(t/\tau_c)$
where~$\tau_c\sim{T}^{-1}$. 
Therefore, in a ring, depending on
their winding, trajectories probe different length scales~:
$L_N\propto{T}^{-1/3}$ or~$L_c\propto{T}^{-1/2}$.

Let us summarize. At the level of
eqs.~(\ref{harmonics},\ref{AAS}), $L_\varphi$ is a
phenomenological parameter put by hand. The modelization of
decoherence due to e-e interaction of AAK shows that, 
in an infinite wire, the WL correction probes the Nyquist length
$L_N\propto{T}^{-1/3}$ (the only length scale of the problem).
This shows that, in the MC of the infinite wire, the phenomenological
parameter $L_\varphi$ must be replaced by $L_\varphi\to{}L_N\propto{T}^{-1/3}$.
On the other hand the MC of a ring involves a new length scale
$L_c=L_N^{3/2}/L^{1/2}$ 
combination  of the Nyquist length and the perimeter. In this case, assuming
the simple AAS behaviour $\Delta\sigma_n\propto\EXP{-nL/L_\varphi}$, the
phenomenological parameter should be substituted
by~$L_\varphi\to{}L_c\propto{T}^{-1/2}$.  

\vspace{0.25cm}

\noindent{\it Geometry dependent decoherence in ballistic rings.--}
It is worth pointing that such a geometry dependent decoherence can
also be observed in ballistic systems~: potential fluctuations
responsible for decoherence depend on the precise distribution of
currents inside the device, that are affected by the way the current
is injected through different contacts 
\cite{PedLanBut98}.  Depending whether the measurement is local or
nonlocal, different phase coherence lengths have been extracted from
the damping of AB oscillations~\cite{KobAikKatIye02}.  
The different $\tau_\varphi$ are probed by changing the contact
configuration (current/voltage probes)~\cite{SeePilJorBut03}, whereas in the
diffusive ring, the different length scales are probed by considering
different harmonics.


\section{General formalism\label{sec:gf}}

We first recall the basic formalism and apply precisely the ideas given in the
introduction. We will consider the reduced conductivity $\tilde\sigma$, 
defined by
\begin{equation}
  \sigma=\frac{2e^2}{h\,\sw}\,\tilde\sigma
\end{equation}
where $\sw$ is the cross-section of the wire. The reduced WL
correction has the dimension of a length.  As
mentioned above, it is a sum of contributions of interfering closed
reversed electronic trajectories, which can be conveniently written as
a path integral~:
\begin{align}
   \label{PathIntegral}
   &\Delta\tilde\sigma(x) \equiv -2\,P_c(x,x) 
   \\ \nonumber
   &= -2 \int_0^\infty \D{t}\,
   \int_{x(0)=x}^{x(t)=x}\hspace{-0.5cm}\mathcal{D}x(\tau)\,
   \EXP{ -\int_0^t\D\tau\, \big(\frac14\dot{x}^2 + 2\I{e}\dot{x}A(x)\big) }\,
   \EXP{\I\Phi[x(\tau)]}
  \:.
\end{align}
$P_c(x,x)$ is the so-called {\it Cooperon}.
Summation over diffusive paths for time $t$
involves the Wiener measure
$\mathcal{D}x(\tau)\exp{-\int_0^t\D\tau\,\frac14\dot{x}^2}$ 
(we have performed a change of
variable $t\to{}t/D$ so that ``time'' has now the dimension of a
squared length). Each loop
receives a phase proportional to the magnetic flux
$2\int_0^t\D\tau\,\dot{x}A(x)$ intercepted by the reversed interfering
trajectories, where $A(x)$ is the vector potential. The factor $2$
originates from the fact that the Cooperon 
measures interference between two closed electronic trajectories undergoing the
same sequence of scattering events in a reversed order. 
Finally we have introduced an additional phase $\Phi[x(\tau)]$
to account for dephasing~: dephasing due to penetration of the magnetic
field in the wires~\cite{AltAro81} or decoherence due to electron-electron
interaction. In this latter case the phase $\Phi$ depends also on the
environment dynamics over which one should average.

\vspace{0.25cm}

\noindent
{\it The $x$ dependence of $\Delta\tilde\sigma(x)$ in eq.~(\ref{PathIntegral}).--}
In a general network, in the absence of translational invariance, 
the WL correction to the conductance was shown to be given by an
integration of $\Delta\tilde\sigma(x)$ over the network, with some nontrivial 
weights attributed to the wires. 
Let us write the classical dimensionless 
conductance as 
$g^\mathrm{class}=\alpha_dN_c\ell_e/\mathcal{L}$
where the effective length $\mathcal{L}$ is obtained from addition
(Kirchhoff) laws  
of classical resistances (dimensionless parameter $\alpha_d$ was defined above~:
$\alpha_3=4/3$, $\alpha_2=\pi/2$ and $\alpha_1=2$). Then the WL
correction to the conductance is~\cite{TexMon04}
\begin{equation}
  \label{Res2004}
  \Delta{}g = \frac1{\mathcal{L}^2}  \sum_i 
   \derivp{\mathcal{L}}{l_i}\,
   \int_{\mathrm{wire}\:i}\D x\, \Delta\tilde\sigma(x)
   \:,
\end{equation}
where the summation runs over all wires of the networks and $l_i$ is
the length of the wire $i$. 
Eq.~\eqref{Res2004} was demonstrated in Ref.~\cite{TexMon04} for the
conductance matrix elements of multiterminal networks with arbitrary
topology. This result relies on a careful discussion of current
conservation (derivation of current conserving quantum corrections can
be found in Refs.~\cite{KanSerLee88,HerAmb88,Her89,HasStoBar94}).
This point will play a relatively minor
role in the present paper. Eq.~(\ref{Res2004}) may be used in order
to calculate geometry dependent prefactors. 

\vspace{0.25cm}

\noindent{\it Magnetoconductance oscillations and winding properties.--}
In an array of metallic rings of same perimeter, the magnetic flux is an
integer multiple of the flux $\phi$ per ring
$\int_0^t\D\tau\,\dot{x}A(x)=\phi\times\mathcal{N}[x(\tau)]$ where
$\mathcal{N}[x(\tau)]\in\ZZ$ is the winding number of the closed trajectory
(the number of fluxes encircled). This makes the WL correction a
periodic function of 
the flux $\phi=\theta\frac{\phi_0}{4\pi}$, where $\phi_0=h/e$.
The $n$-th harmonic of the MC
\begin{equation}
  \Delta\tilde\sigma_n =
  \int_0^{2\pi}\frac{\D\theta}{2\pi}
  \Delta\tilde\sigma(\theta)\:\EXP{-\I{n}\theta}
\end{equation}
involves trajectories with winding number $n$. We can write the
harmonics as 
\begin{align}
  &\Delta\tilde\sigma_n(x) \equiv -2\,P_c^{(n)}(x,x) = -2\,
  \int_0^\infty \D{t}\,
 \\ \nonumber
 &\times \int_{x(0)=x}^{x(t)=x}\mathcal{D}x(\tau)\,
   \delta_{n,\mathcal{N}[x(\tau)]}\,
   \EXP{ -\int_0^t\D\tau\, \frac14\dot{x}^2}\,
   \EXP{\I\Phi[x(\tau)]}
  \:,
\end{align}
where the Kronecker symbol selects only trajectories for a given
winding number $n$. Let 
us introduce the probability for a diffusive particle to return to its
starting point after a time $t$, with the condition of winding $n$ fluxes
\begin{equation}
  \label{eq:PI1}
  \mathcal{P}_n(x,x;t) = \int_{x(0)=x}^{x(t)=x}\mathcal{D}x(\tau)\,
   \delta_{n,\mathcal{N}[x(\tau)]}\,
   \EXP{ -\int_0^t\D\tau\, \frac14\dot{x}^2}
  \:.
\end{equation}
For example, in an isolated ring of perimeter $L$, this probability is simply
given by
\begin{equation}
  \label{PnRing}
  \mathcal{P}_n^\mathrm{ring}(x,x;t)
  = \frac1{\sqrt{4\pi{t}}}\,\EXP{-\frac{(nL)^2}{4t}}
  \:.
\end{equation}
Then, we can rewrite the harmonics as
\begin{equation}
  \label{StartingPoint}
  \boxed{
  \Delta\tilde\sigma_n(x) = -2\,
  \int_0^\infty \D{t}\,\mathcal{P}_n(x,x;t)\,
  \smean{ \EXP{\I\Phi[\mathcal{C}^n_t]} }_{\mathcal{C}^n_t}
  }
  \:,
\end{equation}
which is the structure given in eq.~\eqref{harmFV}.
In eq.~(\ref{StartingPoint}) we have introduced the notation
$\mathcal{C}^n_t\equiv(x(\tau),\,\tau\in[0,t]\,|\,x(0)=x(t)=x\,;\,\mathcal{N}[x(\tau)]=n)$
for a closed diffusive path winding $n$ times.
$\mean{\cdots}_{\mathcal{C}^n_t}$ designates averaging over all such
paths, with the measure of the path integral~\eqref{eq:PI1}.
The phase $\Phi[\mathcal{C}]$ accounts for dephasing and eliminates
the contributions of diffusing trajectories at large time.
We now discuss two possible modelizations for this function,
denoted by ``A'' and ``B''.

\subsection{Model A~: Exponential relaxation}
The simplest
choice is an exponential relaxation, with a dephasing rate
$\gamma=1/\tau_\varphi=1/L_\varphi^2$~:
\begin{equation}
  \label{relaxexp}
  \smean{ \EXP{\I\Phi[\mathcal{C}_t]} }_{\mathcal{C}_t} = \EXP{-\gamma t}
  \:.
\end{equation}
This simple prescription correctly describes dephasing due to
spin-orbit coupling, magnetic impurities~\cite{HikLarNag80,Ber84}, effect of
penetration of the magnetic field in the wires~\cite{AltAro81},
or decoherence due to electron-phonon scattering~\cite{ChaSch86,footnote6}.
Using (\ref{PnRing},\ref{StartingPoint}) with this exponential decay
yields the familiar result~(\ref{AAS}) for the isolated ring.

\subsection{Model B : geometry dependent decoherence from electron-electron interaction}
It turns out that the simple exponential
relaxation does not describe correctly the decoherence due to
electron-electron interaction, the physical reason being that this
decoherence is due to electromagnetic field fluctuations that depend
on the geometry of the system. AAK have proposed a
microscopic description~\cite{AltAroKhm82,AltAro85} that we can
rephrase as follows. In eq.~(\ref{PathIntegral}), the phase $\Phi$
picked up by the reversed trajectories depends on the environment
(the potential $V$ created by the other electrons due to electron-electron
interaction)~:
$\Phi_V[\mathcal{C}_t]=\int_0^t\D\tau\,[V(x(\tau),\tau)-V(x(\tau),t-\tau)]$.
Averaging over the Gaussian fluctuations of $V$ leads to
$\smean{\EXP{\I\Phi_V[\mathcal{C}_t]}}_{V}=\EXP{-\frac12\smean{\Phi_V[\mathcal{C}_t]^2}_V}$
where the fluctuation-dissipation theorem (written for $\omega\ll{}T$
describing classical fluctuations)  
$\smean{V(r,t)V(r',t')}_V\simeq\frac{e^2T}{\sigma_0}\delta(t-t')\,P_d(r,r')$
gives
$\frac12\smean{\Phi_V[\mathcal{C}_t]^2}_V=\Gamma[\mathcal{C}_t]\,t$
with~\cite{footnote7}
\begin{align}
  \label{localFDT}
  &\boxed{
  \Gamma[\mathcal{C}_t]\, t
  =\frac{2}{L_N^3}\int_0^t\D\tau\,  W(x(\tau),x(t-\tau))
 }
 \\
  \label{localFDT2}
  &=\frac{k_BT}{\hbar}\frac{2\pi}{R_K}
  \int_0^t\D\tau\,  \mathcal{R}(x(\tau),x(t-\tau))
  \:,
\end{align}
where $R_K=h/e^2$ is the quantum of resistance.
The function $W(x,x')$ is related to the diffuson, solution of
$-\Delta{P_d}(x,x')=\delta(x-x')$, by
\begin{equation}
  \label{eq:defW}
  W(x,x') = \frac{ P_d(x,x) + P_d(x',x') }{2}  - P_d(x,x')
  \:.
\end{equation}
This function has a physical interpretation discussed in the
appendix~\ref{app:classresist}~: it is proportional to the 
equivalent resistance
$\mathcal{R}(x,x')$ between the points $x$ and $x'$
(figure~\ref{fig:rixix}).  
With this remark, we see that eq.~(\ref{localFDT2}) can be
understood as a local version of the Johnson-Nyquist theorem relating
the potential fluctuations to the resistance.

In eqs.~(\ref{localFDT},\ref{localFDT2}) we have introduced
a decoherence 
rate $\Gamma[\mathcal{C}_t]$ which depends not only on the time but on the 
trajectory itself. Therefore the decay of phase coherence is now described by
\begin{equation}
  \label{Decoherence}
  \boxed{
  \smean{ \EXP{\I\Phi_V[\mathcal{C}_t]} }_{V,\mathcal{C}_t}
  = \smean{ \EXP{-\Gamma[\mathcal{C}_t]\, t} }_{\mathcal{C}_t}
  }
  \:.
\end{equation}
Within this framework, relaxation of phase coherence is not described
by a simple exponential decay like in eq.~(\ref{relaxexp}) but is
controlled by a {\it functional of the trajectories}~\cite{footnote8}
$x(\tau)$. Therefore
the nature of decoherence depends on the network, through the resistance
$\mathcal{R}(x,x')$ between $x$ and $x'$, and on the winding properties of the
trajectories. 

The central problem of the present paper is to compute the path integral
\begin{align}
  \label{eq:CQ}
   &\Delta\tilde\sigma_n(x)  =-2\,
  \int_0^\infty \D{t}\, \EXP{-\gamma t}
\int_{x(0)=x}^{x(t)=x}\hspace{-0.5cm}\mathcal{D}x(\tau)\,
   \delta_{n,\mathcal{N}[x(\tau)]}\,
  \nonumber\\
  &\hspace{2cm}
   \times   \EXP{ -\int_0^t\D\tau\, \big[
            \frac14\dot{x}^2+\frac{2}{L_N^3} W(x(\tau),x(t-\tau))
        \big]}
\end{align}
for the different networks.
Such a calculation has been already performed in two cases~: the infinite
wire~\cite{AltAroKhm82} and the isolated ring~\cite{LudMir04,TexMon05b}.

The logic of the following sections is the following~:
first we study the winding properties in the network.
For that purpose we first compute the WL correction
$\Delta\tilde\sigma_n^{(A)}$ within  {\it model A},
eq.~\eqref{StartingPoint} with \eqref{relaxexp}. The 
probability $\mathcal{P}_n(x,x;t)$ can be extracted from an
inverse Laplace transform with respect to the parameter $\gamma$.
Having fully characterized the winding properties, we use this
information  in order to study the
harmonics $\Delta\tilde\sigma_n^{(B)}$ within the {\it model B}
describing decoherence due to electron-electron 
interaction, eq.~\eqref{StartingPoint} with~\eqref{Decoherence}.


\section{The wire and the ring \label{sec:wr}}

We first recall known results within the framework of {\it model B}
concerning the simplest geometries that will 
be useful for the following.

\subsection{Phase coherence relaxation in an infinite wire}
\begin{figure}[!ht]
  \centering
  \includegraphics[scale=0.4]{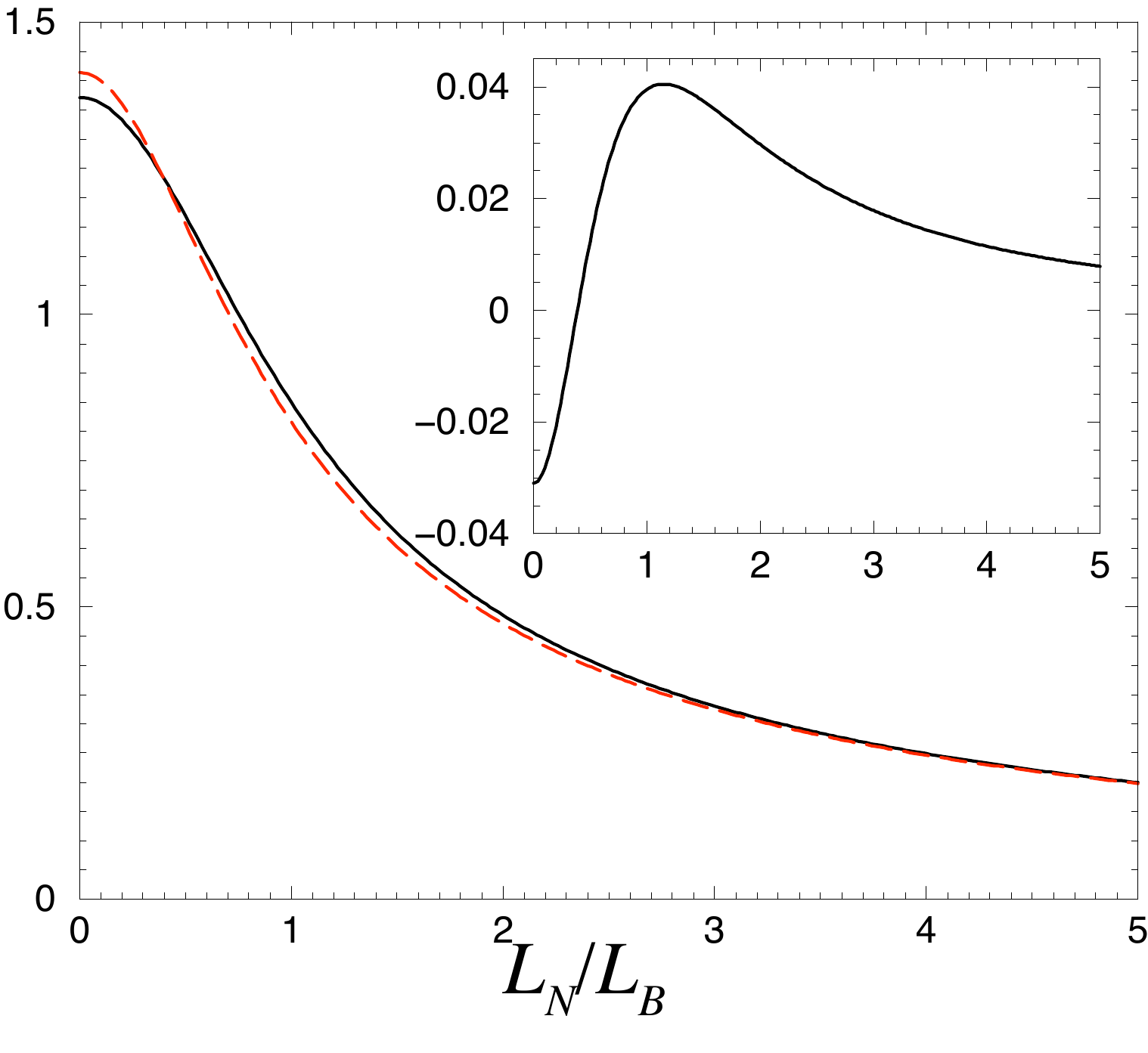}
  \caption{
      {\it Comparison between
      $-\mathrm{Ai}(x^2)/\mathrm{Ai}'(x^2)$ (black continuous line) and
      $1/\sqrt{1/2+x^2}$ (red dashed line). Relative
      difference (inset) does not exceed 4\%}.
      }
  \label{fig:penetBfield}
\end{figure}

The case of an infinite wire was originally solved in
Ref.~\cite{AltAroKhm82}. In this case we have
$W_\mathrm{wire}(x,x')=\frac12|x-x'|$ and the path integral
\begin{align}
  \label{PathIntInfWire}
   \Delta\tilde\sigma =& -2 \int_0^\infty \D{t}\,\EXP{-\gamma t}
   \\\nonumber
  &\times\int_{x(0)=x}^{x(t)=x}\hspace{-0.5cm}\mathcal{D}x(\tau)\,
   \EXP{ -\int_0^t\D\tau\,
       \big[
          \frac14\dot{x}^2 +  \frac1{L_N^3}|x(\tau)-x(t-\tau)|
       \big]}
\end{align}
can be computed thanks to translational invariance (as pointed in
Ref.~\cite{ComDesTex05}, using the symmetry of the path integral 
we can perform the substitution $x(\tau)-x(t-\tau)\to{}x(\tau)$, provided that
the starting point of the path integral is set to $x\to0$, see
appendix~\ref{app:winding}). 
Combining exponential relaxation ({\it model A}) and decoherence due to e-e
interaction ({\it model B})  allows to extract the
function (\ref{Decoherence}) with an inverse Laplace transform of the
AAK result~\cite{AltAroKhm82,AltAro85,AleAltGer99,AkkMon07}
\begin{align}
  \label{eq:3}
  \Delta\tilde\sigma&=-\int_0^\infty\frac{\D{t}}{\sqrt{\pi{}t}}\,
  \EXP{-\gamma t}\,\smean{
    \EXP{\I\Phi_V[\mathcal{C}_t]}}_{V,\mathcal{C}_t}
  \\
  &=L_N\frac{\mathrm{Ai}(\gamma\,L_N^2)}{\mathrm{Ai}'(\gamma\,L_N^2)}
  \:,
\end{align}
where $\mathrm{Ai}(z)$ is the Airy function~\cite{AbrSte64}. 
As mentioned above, this expression is very close to~\cite{Pie00,AkkMon07}
$\Delta\tilde\sigma\simeq-\big(\frac1{L_\varphi^2}+\frac1{2L_N^2}\big)^{-1/2}$,
the result obtained by performing the substitution
$\smean{\EXP{\I\Phi_V[\mathcal{C}_t]}}_{V,\mathcal{C}_t}\to\EXP{-t/2L_N^2}$
(see figure~\ref{fig:penetBfield}).

The inverse Laplace transform of \eqref{eq:3} was computed in
Ref.~\cite{MonAkk05} with residue's theorem~: 
\begin{align}
  f_\mathrm{wire}\left({t}/{\tau_N}\right)
  &=\smean{\EXP{\I\Phi_V[\mathcal{C}_t]}}_{V,\mathcal{C}_t}
  \\
  \label{dephasingwire}
  &=\sqrt{\frac{\pi t}{\tau_N}} 
  \sum_{m=1}^\infty\frac1{|u_m|}\EXP{-|u_m|t/\tau_N}
  \:,
\end{align}
where $u_m$ are zeros of $\mathrm{Ai}'(z)$.
In particular $u_1\simeq-1.019$ and 
$u_m\simeq-[\frac{3\pi}{2}(m-\frac34)]^{2/3}$ for~$m\to\infty$.
The limiting behaviours are
\begin{align}
  \label{relaxwire1}    
 f_\mathrm{wire}\left({t}/{\tau_N}\right)  
  &\simeq 1 -\frac{\sqrt\pi}4\,\left(\frac{t}{\tau_N}\right)^{3/2} 
      \hspace{0.5cm} \mbox{for } t \ll \tau_N \\
  \label{relaxwire2}
  &\simeq \frac1{|u_1|}\sqrt{\frac{\pi t}{\tau_N}} \,
           \EXP{-|u_1|t/\tau_N}
      \hspace{0.35cm} \mbox{for } \tau_N \ll t
  \:.
\end{align}
Note that the short time behaviour can be obtained by expanding 
$
f\left({t}/{\tau_N}\right)=
\smean{ \EXP{\I\Phi} }_{V,\mathcal{C}_t}
=
\smean{ 
  \EXP{ -\frac12\smean{ \Phi^2 }_V } 
      }_{\mathcal{C}_t}
\simeq1-\frac12\smean{\Phi_V[\mathcal{C}_t]^2}_{V,\mathcal{C}_t}
$.
This limit can be simply obtained by noticing that in the wire
$W(x(\tau),x(t-\tau))\sim{}x(\tau)\sim\sqrt{t}$, therefore
$\Gamma[\mathcal{C}_t]\,t\sim{}T\,t^{3/2}\sim({t}/{\tau_N})^{3/2}$ where
we recover that the Nyquist time scales as~$\tau_N\propto{}T^{-2/3}$. 

\subsection{Phase coherence relaxation in the isolated ring}
For the isolated ring of perimeter $L$, we have
$W_\mathrm{ring}(x,x')=\frac12|x-x'|(1-\frac{|x-x'|}{L})$. 
The path
integral \eqref{eq:CQ} can be computed exactly~\cite{TexMon05b}
(see appendix~\ref{app:winding}). Up to a dimensionless prefactor, we
obtain 
\begin{align}
  \label{LMTM}
  \Delta\tilde\sigma_n
  &\sim -L_N\,\EXP{-\frac\pi8n(L/L_N)^{3/2}}
      \hspace{0.5cm} \mbox{for } L_N\ll L
  \\
  &\sim -T^{-1/3}\,\EXP{-n L^{3/2}T^{1/2}}
  \:.
\end{align}
This result can be simply understood as follows~: in the ring,
trajectories with finite  winding  necessarily explore the whole ring. This
``ergodicity'' implies that $W(x(\tau),x(t-\tau))\sim{}x(\tau)\sim{}L$ and
therefore the decoherence rate
$\Gamma[\mathcal{C}_t]\,t\sim{}TL\,t\sim{t}/{\tau_c}$ involves the
different time scale $\tau_c\sim1/(TL)$,
according to the physical argument given in section~\ref{sec:background}. 
As a consequence
Eq.~(\ref{StartingPoint}) indeed leads to~\cite{LudMir04}
\begin{equation}
  \label{LMlanguage}
    \Delta\tilde\sigma_n
  \sim - L_N\, \EXP{-\frac\pi8nL/L_c}
  \:,
\end{equation}
where 
\begin{equation}
  \label{eq:4}
  \boxed{
  L_c=\frac{L_N^{3/2}}{L^{1/2}}
  } 
  \:.
\end{equation}

\begin{figure}[!ht]
  \centering
  \includegraphics[scale=0.35]{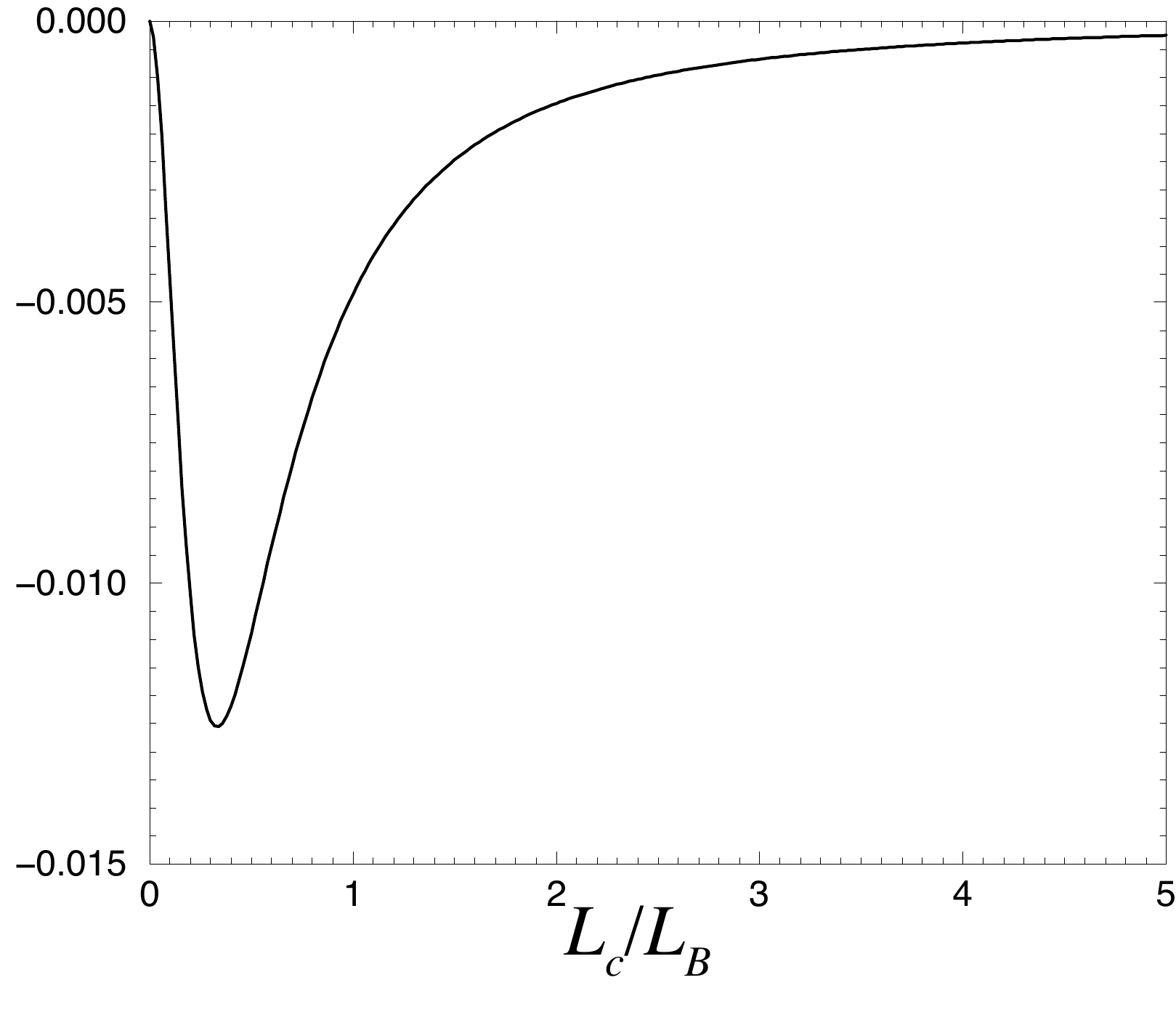}
  \caption{
           {\it 
             The relative difference ($1-\eta(x^2)/\sqrt{(\pi/8)^2+x^2}$)
             between the effective length
             $L_\varphi^\mathrm{osc}$, eq.~\eqref{LphiEff}, 
             and its approximation, eq.~\eqref{Combin},
             does not exceed 1.5\%.}}
  \label{fig:penetBfieldHarm}
\end{figure}

\vspace{0.25cm}

\mathversion{bold}
\noindent{\bf Phase coherence length~: $L_\varphi\propto{}T^{-1/3}$ or $L_\varphi\propto{}T^{-1/2}$~?--}
\mathversion{normal}
Note that the introduction of a new length scale~\cite{LudMir04} $L_c$
might appear arbitrary since the harmonics may be written uniquely in
terms~\cite{TexMon05b} of $L$ and $L_N$.
The difference between (\ref{LMTM}) and (\ref{LMlanguage}) is a
matter of convention and  may be related to the experimental
procedure.  
The usual method extracts
the phase coherence length from the analysis of MC harmonics. Then it
is natural to see how the winding number $n$ scales with the phase
coherence length, or more properly how the length $nL$ scales with 
$L_\varphi$ and therefore assume the
form~$\Delta\tilde\sigma_n\propto{}f(nL/L_\varphi)$. From
eq.~(\ref{LMlanguage}) we see that the function is simply the exponential,
$f(x)=\EXP{-x}$, with a {\it perimeter dependent} phase coherence length
$L_\varphi\to{L_c}\propto{}(TL)^{-1/2}$.
Another procedure may consist in studying the harmonics content as a
function of the perimeter $L$, that is for different samples.
The experiment is then analyzed with the form
$\Delta\tilde\sigma_n\propto{}f_n(L/L_\varphi)$. Eq.~(\ref{LMTM}) gives
$f_n(x)=\EXP{-nx^{3/2}}$ with the geometry independent phase coherence
length~$L_\varphi\to{L_N}\propto{}T^{-1/3}$. 

\vspace{0.25cm}

The temperature dependence $\Delta\sigma_n\sim\EXP{-L^{3/2}T^{1/2}}$ was
first predicted in Ref.~\cite{LudMir04} using instanton method (with a
different pre-exponential dependence) and studied in details in 
Ref.~\cite{TexMon05b} 
where the path integral 
\eqref{eq:CQ} was computed exactly for the
isolated ring. The effect of the connecting arms was clarified
in Ref.~\cite{Tex07b}. The fact that the pre-exponential factor is $L_N$ is
related to the fact that the smooth part of the MC, due to the penetration of
the field in the wire, probes the same length scale as in the infinite wire.

\vspace{0.25cm}

It is worth pointing the recent work~\cite{TreYevMarDelLer09} in which
the crossover to the 0d limit is studied in a ring weakly
connected. In this case the authors get a crossover from
$\tau_\varphi\propto{}T^{-1}$ (diffusive ring) to $\tau_\varphi\propto{}T^{-2}$ 
(ergodic) for temperature  below the Thouless energy. This latter
behaviour coincides with the result known for quantum 
dots in the same regime~\cite{SivImrAro94}.

\subsection{Penetration of the magnetic field in the wires of the
  ring}

Networks are made of wires of finite width $w$. The penetration of
the magnetic field in the wires is responsible for fluctuations of the
magnetic flux enclosed by trajectories with the same winding number
but different areas. 
In the weak magnetic field limit, this effect is described by introducing
an effective dephasing rate~\cite{AltAro81} 
\begin{equation}
  \gamma\to1/L_\mathcal{B}^2=\frac13\left(\frac{e\mathcal{B}w}{\hbar}\right)^2
  \:.
\end{equation}
The question of how to combine the two decoherence mechanisms ({\it
  models A \& B}) in  
the ring was discussed in Ref.~\cite{TexMon05b}. It was shown that the WL
correction of the ring presents the structure 
\begin{equation}
  \label{eq:RES2005}
  \Delta\tilde\sigma_n \simeq
  L_N\,
 \frac{\mathrm{Ai}(L_N^2/L_\mathcal{B}^2)}{\mathrm{Ai}'(L_N^2/L_\mathcal{B}^2)}
  \,\EXP{ -nL/L_\varphi^\mathrm{osc}(L_c,L_\mathcal{B}) }
\end{equation}
for $L_N\ll L$, with
\begin{equation}
  \label{LphiEff}
   L_\varphi^\mathrm{osc}(L_c,L_\mathcal{B}) 
  = \frac{L_c}{\eta(L_c^2/L_\mathcal{B}^2)}
  \:,
\end{equation}
where $\eta(x)=(\frac14+x)\arctan\frac1{2\sqrt{x}}+\frac{\sqrt{x}}2$. 

\vspace{0.25cm}

\noindent{\it Prefactor.--}
In eq.~\eqref{eq:RES2005}, the pre-exponential factor coincides with the result
obtained for an infinite wire \cite{AltAroKhm82}.  
The ratio of Airy functions can be approximated as~\cite{Pie00,AkkMon07}
$\frac{\mathrm{Ai}(x)}{\mathrm{Ai}'(x)}\simeq-(\frac12+x)^{-1/2}$ 
(figure~\ref{fig:penetBfield}). In other terms, we may write the zero harmonic
({\it i.e.} the result for the infinite wire) as 
\begin{equation}
  \Delta\tilde\sigma_0 \simeq -L_\varphi^\mathrm{env}
\end{equation}
where
\begin{equation}
  \label{LphiEnv}
    L_\varphi^\mathrm{env}
  =  \left(  \frac{1}{2L_N^2} + \frac1{L_\mathcal{B}^2} \right)^{-1/2}
  \:.
\end{equation}
This combination expresses that, in a wire of width $w$, the
penetration of the magnetic field provides the dominant cutoff when
typical trajectories enclose more than one quantum
flux~$L_\varphi{}w\mathcal{B}\gtrsim\phi_0$ (here $L_\varphi\sim{}L_N$
for trajectories with winding $n=0$). 

\vspace{0.25cm}

\noindent{\it Exponential damping.--}
In the exponential of eq.~\eqref{eq:RES2005}, the effective length
interpolates between $L_\varphi^\mathrm{osc}\simeq\frac8\pi{L_c}$ for
$L_\mathcal{B}\gg{L_c}$ and $L_\varphi^\mathrm{osc}\simeq{}L_\mathcal{B}$
for $L_\mathcal{B}\ll{L_c}$. Its overall behaviour
is well approximated by
\begin{equation}
  \label{Combin}
  L_\varphi^\mathrm{osc}
  \simeq 
  \left[ \left(\frac{\pi}{8L_c}\right)^2 + \frac1{L_\mathcal{B}^2} \right]^{-1/2}
  \hspace{-0.35cm}
  =  \left(  \frac{\pi^2L}{64L_N^3} + \frac1{L_\mathcal{B}^2} \right)^{-1/2}
\end{equation}
which differs with (\ref{LphiEff}) by less
than~1.5\% (figure~\ref{fig:penetBfieldHarm}).  
When $L_\mathcal{B}$ is the shortest length, the decay of AAS
oscillations $\sim\EXP{-nL/L_\mathcal{B}}$ can be understood from the fact
that modulations of the flux enclosed by trajectories with finite winding
become larger than the quantum flux $nLw\mathcal{B}\gtrsim\phi_0$.
Eq.~(\ref{Combin}) was 
used in the analysis of the recent
experiment~\cite{FerRowGueBouTexMon08}.

\vspace{0.25cm}

\begin{figure}[!ht]
  \centering
  \includegraphics[scale=0.425]{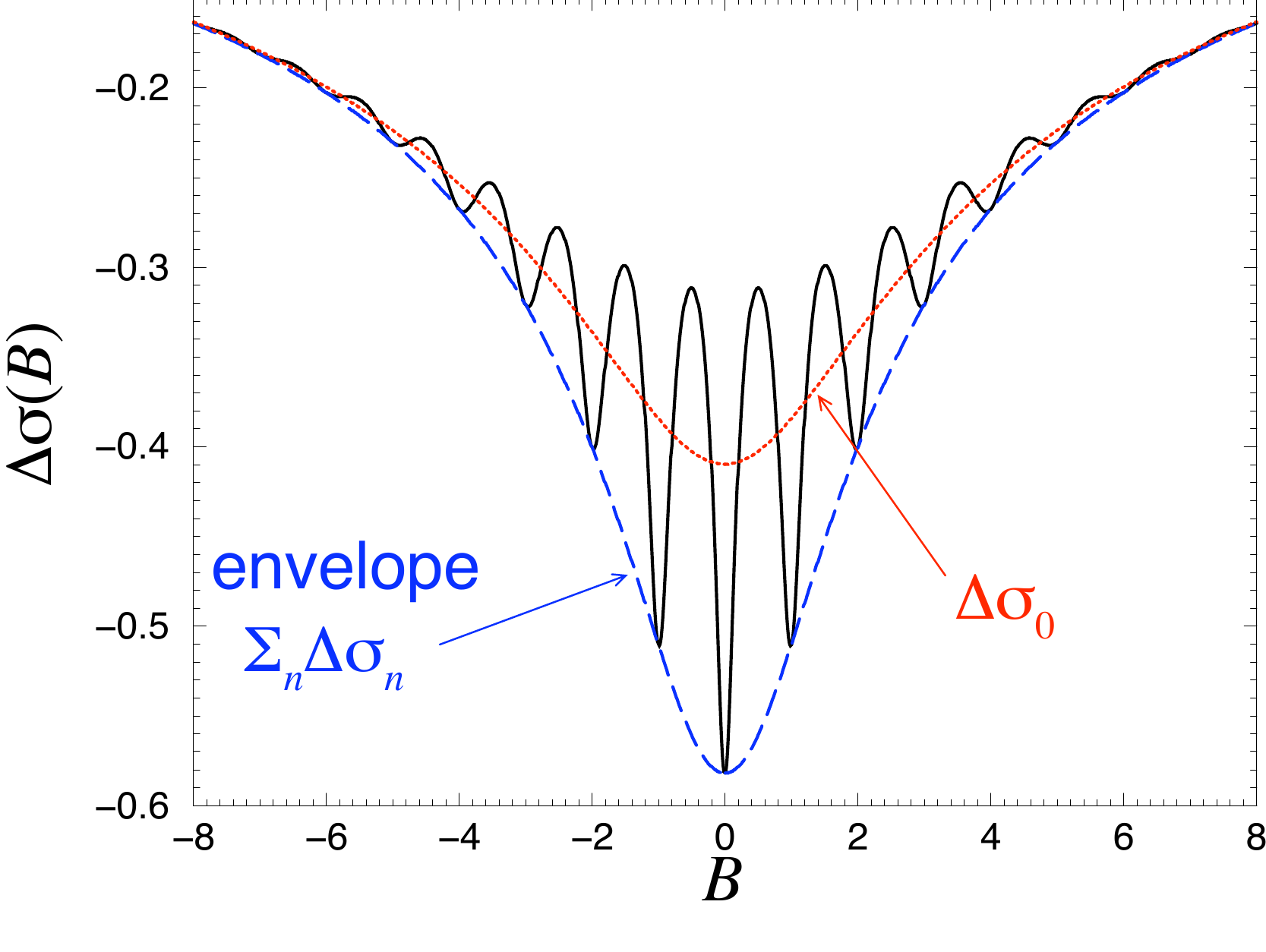}
  \caption{\it Typical shape of the MC curve of a network (here a
    chain of rings). Rapid oscillations are AAS oscillations. Damping
    of oscillations over large scale is due to the penetration of the
    magnetic field in the wires.  } 
  \label{fig:mc_chain}
\end{figure}

These two remarks show that the magnetic length $L_\mathcal{B}$ probes
two different length scales~:
in the pre-exponential factor $L_\mathcal{B}$ probes the Nyquist
length~$L_N\propto{T}^{-1/3}$, whereas in the ratio of harmonics
$\Delta\tilde\sigma_n/\Delta\tilde\sigma_0$, the magnetic length
$L_\mathcal{B}$ probes the length scale~$L_c\propto{T}^{-1/2}$.

\subsection{How to analyze MC experiments in networks}

In order to understand the implications of this remark, let us discuss
the structure of the typical MC curve of a network.
The following discussion applies to the case 
$L_N\lesssim{}L$ where \eqref{eq:RES2005} holds.
Figure~\ref{fig:mc_chain} represents a typical MC curve, here for a
chain of rings. It exhibits rapid AAS oscillations with a period given
by $\mathcal{B}_\mathrm{osc}\sim\phi_0/L^2$, superimposed with a
smooth variation over a scale 
$\mathcal{B}_\mathrm{damp}\sim\phi_0/wL_\varphi$.
The phase coherence length can be extracted either from the amplitude of
the oscillations or from the decay of the envelope
of the MC curve. 
Which $L_\varphi$ ($L_N$ or $L_c$) is obtained from such a curve~? 
%
%
\begin{figure}[htbp]
\begin{center}
\includegraphics[scale=0.85]{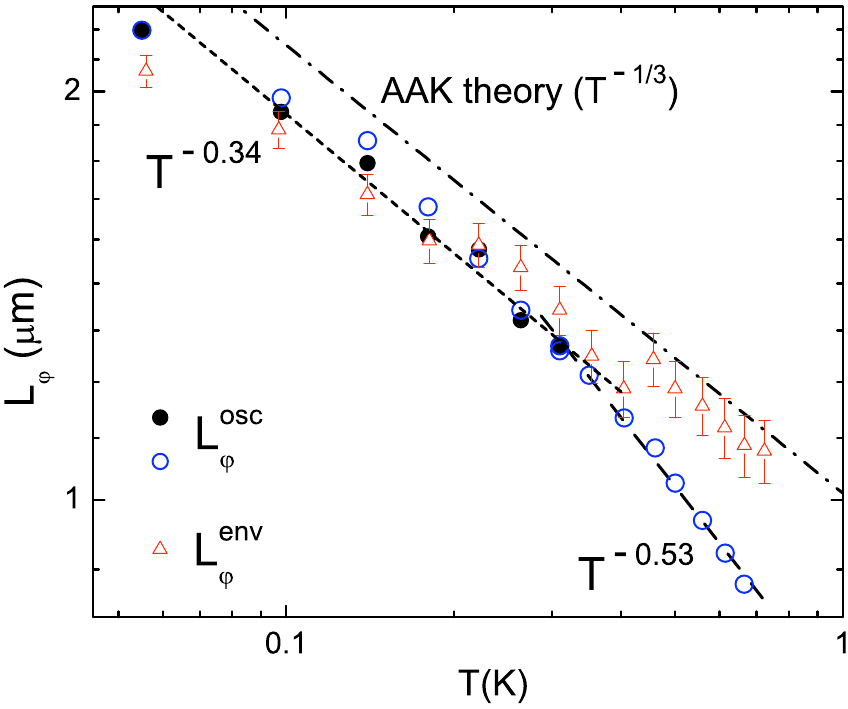}
\caption{{\it Phase coherence length as a function of the temperature
   obtained  
  from measurements realized on a large square
  network~\cite{FerRowGueBouTexMon08} of lattice 
  spacing $a=1\:\mu$m (perimeter $L=4\:\mu$m). 
  The length scale $L_\varphi\to{}L_\varphi^\mathrm{env}\propto{T}^{-1/3}$ 
  has been extracted from the smooth MC
  while the length $L_\varphi\to{}L_\varphi^\mathrm{osc}\propto{T}^{-1/2}$ 
  has been extracted from  ratio of harmonics. 
  (Note that 
   $L_\varphi^\mathrm{env}$ and
  $L_\varphi^\mathrm{osc}$ in Ref.~\cite{FerRowGueBouTexMon08} denote
  eqs.~\eqref{LphiEnv} and \eqref{Combin} for $L_\mathcal{B}=\infty$).}
  }
\label{fig:meydi}
\end{center}
\end{figure}
According to
\eqref{eq:RES2005} we see that $L_\varphi\to{}L_N$ in the
pre-exponential factor, which mostly dominates the smooth envelope,
while $L_\varphi\to{}L_c$ in the exponential
decay, which dominates the damping of the rapid oscillations. 
In order to decouple the two effects the analysis of the experiments of 
Ref.~\cite{FerRowGueBouTexMon08}
have been analyzed as follows~\cite{Fer04,FerAngRowGueBouTexMonMai04}~:
the Fourier transform of the MC curve 
$\Delta\sigma(\mathcal{B})$
presents broadened Fourier peaks due to the penetration of the magnetic
field in the wires. Integration of Fourier peaks eliminates this effect. 
Ratio of harmonics involve the length scale $L_c$.
The length $L_N$ was extracted from the smooth envelope 
$\Delta\tilde\sigma_\mathrm{env}=\sum_n\Delta\tilde\sigma_n\simeq\Delta\tilde\sigma_0$
for $L_N\ll{}L$.
The temperature dependence of the phase coherence length was extracted
in this way in Ref.~\cite{FerRowGueBouTexMon08}. Results are plotted on
figure~\ref{fig:meydi}, exhibiting clearly the two length scales in
the regime $L_N\ll{}L$.
We see that it is crucial to analyze the experiment in terms of the MC
harmonics~$\Delta\sigma_n$.

\vspace{0.25cm}

\noindent{\bf Isolated ring {\it vs} ring embedded in a network.--}
In transport experiments the ring is never isolated~: it is at least
connected to contacts through which current is injected. Moreover the samples
are often made of a large number of loops, in order to realize disorder
averaging. 
The results obtained for the isolated ring are fortunately
relevant to describe a more complex network of equivalent rings
(Figs.~\ref{fig:sn} \& \ref{fig:chains}) when the rings can be
considered as  {\it independent}, {\it i.e.} when interference phenomena
do not involve several rings~; this occurs when $L_\varphi\ll{L}$ (or
$L_N\ll{L}$), in practice in a high temperature regime.
This temperature dependence of harmonics is rather difficult to extract
from measurements since harmonics are suppressed exponentially.
This has been done only very recently in Ref.~\cite{FerRowGueBouTexMon08}.
Another difficulty is that the ``high temperature regime'' 
$L_N\ll{}L$ is in practice
quite narrow in these samples due to fact that electron-phonon
interaction dominates the decoherence above $1\:$K (in the sample of
Refs.~\cite{SchMalMaiTexMonSamBau07,Mal06} $L_N$ is much larger and
when $L_N\ll{}L$ the role of electron-eletron interaction is negligible).

It is an important issue to obtain the expression of the
WL correction for a broader temperature 
range, that is to study the regime $L\lesssim{}L_N$. 
This regime is reached in several experiments 
\cite{FerAngRowGueBouTexMonMai04,SchMalMaiTexMonSamBau07,Mal06}.
In this case diffusive
interfering trajectories responsible for AAS harmonics are not
constrained to remain 
inside a unique ring, but explore the surrounding network (see
figs.~\ref{fig:sn}, \ref{fig:loop5} \& \ref{fig:chains}). This
affects both the winding properties and the nature of decoherence. 
The MC oscillations are therefore network dependent. 
In the following sections we discuss the behaviour of the MC harmonics
in the limit $L\lesssim{}L_\varphi$ (or $L\lesssim{}L_N$) for
different networks~: a ring connected to long arms, a necklace of
rings and a large square network. The case of a long hollow cylinder will
also be discussed.


\section{The connected ring\label{sec:conring}}

\begin{figure}[htbp]
\begin{center}
\includegraphics{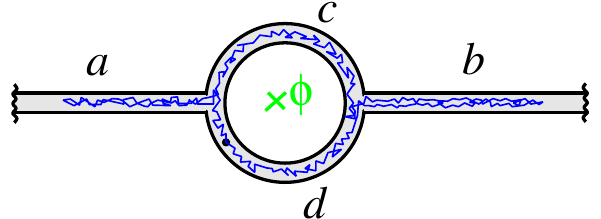}
\caption{{\it A ring connected to reservoirs through long wires $a$ and $b$.
  In the regime $L_\varphi\gtrsim{}L$, the WL correction is dominated
  by trajectories exploring the arms, as represented here.}}
\label{fig:loop5}
\end{center}
\end{figure}

In this section we consider the case of a single ring connected to
two wires supposed much longer than $L_\varphi$ (figure~\ref{fig:loop5}). This
problem has already  been considered in 
Refs.~\cite{TexMon05,TexMon05b}.

\subsection{Model A}

\noindent
$\bullet$ Let us consider first the case $L_\varphi\ll{}L$. 
The Cooperon is constructed in appendix~\ref{app:roa} (see
Ref.~\cite{TexMon05}) and we obtain
\begin{equation}
  \label{eq:SR1}
  \Delta\tilde\sigma_n(x\in\:\mbox{ring})
  \simeq -\left(\frac23\right)^{2n}L_\varphi\:\EXP{-nL/L_\varphi}
  \:,
\end{equation}
where $x$ is any position inside the ring far from the vertices (at
distance larger than $L_\varphi$).
\\
$\bullet$ Now we turn to the regime  $L_\varphi\gg{}L$. 
The Cooperon is uniform inside the ring and is computed in
appendix~\ref{app:roa}. We have~\cite{TexMon05}~:
\begin{align}
  \label{eq:SR2}
    \Delta\tilde\sigma_n(x)
     \simeq
    -\sqrt{\frac{L_\varphi L}{2}}\,
    \EXP{-n\sqrt{2L/L_\varphi}}
  \hspace{0.5cm} \mbox{for } L \ll L_\varphi 
\end{align}
where $x$ is any position inside the ring, or in the arms at a
distance to the ring smaller than $L_\varphi$.
We emphasize that this behaviour, quite different from (\ref{AAS}), is
due to the fact that the 
diffusive trajectories spend most of the time in the wires~\cite{TexMon05} 
(the distribution of the time spent by winding trajectories in the arm
was analyzed in Ref.~\cite{ComDesTex05}).

\vspace{0.25cm}

\begin{figure}[htbp]
\begin{center}
\includegraphics[scale=0.7]{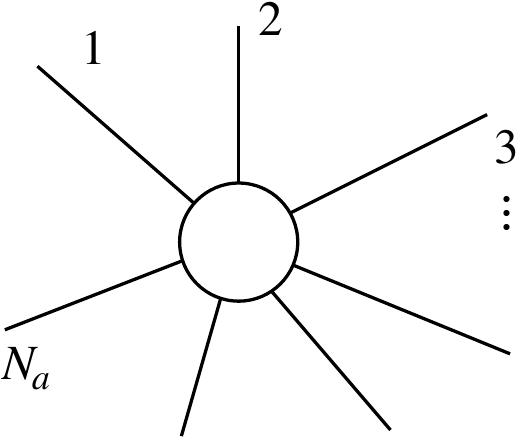}
\caption{ {\it A ring connected to $N_a$ wires.}}
\label{fig:loop7}
\end{center}
\end{figure}

\mathversion{bold}
\noindent{\bf Winding probability in a ring connected to $N_a$ arms.--}
\mathversion{normal}
We now derive the winding probability for a ring connected to $N_a$
infinitely long arms (figure~\ref{fig:loop7}) from the inverse Laplace
transform of the Cooperon $P_c^{(n)}(x,x)=-\frac12\Delta\tilde\sigma_n(x)$.
At small time, \eqref{eq:SR1} gives
\begin{equation}
   \mathcal{P}_n(x,x;t) \simeq\left(\frac23\right)^{N_an}
  \hspace{-0.15cm}
  \frac1{\sqrt{4\pi{t}}}\,\EXP{-\frac{(nL)^2}{4t}}
   \hspace{0.25cm} \mbox{for } t \ll L^2
  \:,
\end{equation}
where $x$ is inside the ring, far from a vertex (at distance larger
than~$\sqrt{t}$). 

At large time scales, $t\gg{}L^2$, the arms strongly modify the
winding properties around the ring~: the time dependence of the
typical winding number becomes subdiffusive $n_t\sim{t}^{1/4}$, to be
compared with the behaviour 
for the isolated ring $n_t\sim{t}^{1/2}$ reflected by eq.~(\ref{PnRing}). 
For a ring connected to $N_a$ infinite wires,
eq.~(\ref{PnRing}) is replaced by the probability~\cite{TexMon05} 
\begin{equation}
  \label{Pnconnectedring}
  \boxed{
  \mathcal{P}_n(x,x;t) \simeq \frac{\sqrt{L/N_a}}{2\,t^{3/4}}
  \ \Psi\!\left( \frac{n\sqrt{N_aL}}{t^{1/4}} \right)
  \hspace{0.5cm} \mbox{for } t\gg L^2
 } 
  \:,
\end{equation}
where $N_a$ is the number of arms  (as far as $x$ is inside the ring or at a
distance to the ring smaller than
$t^{1/2}\sim{}1/\sqrt\gamma=L_\varphi$, the Cooperon, or   
the corresponding probability is almost independent on $x$).
The function $\Psi(\xi)$, given by~\cite{TexMon05,ComDesTex05}
\begin{equation}
  \label{eq:DefPsi}
  \Psi(\xi) = \frac2\pi \re\left[
    \EXP{-\I\frac\pi4} \int_0^\infty\D{u}\,\sqrt{u}\,
    \EXP{-u^2-\sqrt{u}\,\xi\, \EXP{-\I\pi/4}}
  \right]
  \:,
\end{equation}
is studied in the appendix~\ref{app:fctPsi} and
plotted in the conclusion (figure~\ref{fig:QdeX}).

\vspace{0.25cm}

\noindent{\bf From the conductivity to the conductance.--}
In the geometry of figure~\ref{fig:loop5}, the conductance is not
simply related to the conductivity.
The classical conductance of the connected ring
is given by $g=\alpha_dN_c\ell_e/\mathcal{L}$ 
with $\mathcal{L}=l_a+l_{c\parallel d}+l_b$ where $l_i$ is the length
of the wire $i$. $l_{c\parallel d}^{-1}=l_c^{-1}+l_d^{-1}$ is the
equivalent length. From eq.~(\ref{Res2004})~:
\begin{align}
  \label{condweighted}
  \Delta g_n &= \frac1{(l_a+l_{c\parallel d}+l_b)^2}
  \\\nonumber
  &\times\left[
    \int_a+\frac{l_d^2}{(l_c+l_d)^2}\int_c
    +\frac{l_c^2}{(l_c+l_d)^2}\int_d + \int_b
  \right]
  \D x \, \Delta\tilde\sigma_n(x)
  \:.
\end{align}
The Cooperon  $\Delta\tilde\sigma_n(x)=-2P_c^{(n)}(x,x)$ has been
constructed for different positions of the coordinate $x$ in
appendix~\ref{app:roa}.   
Depending on the ratio $L_\varphi/L$, the WL correction
$\Delta{g}_n$ is dominated by different terms.
\\
$\bullet$ For $L_\varphi\ll{}L$, eq.~(\ref{PnXia}) shows that
harmonics of the   
Cooperon decay exponentially in the arms (figure~\ref{fig:cir})~;
inside the ring, the Cooperon (\ref{PnXir}) 
is almost uniform, apart for small variations near the nodes. Therefore
$\Delta{g}_n$ is dominated by integrals $\int_c$ and $\int_d$ in the
ring and we have
\begin{equation}
  \label{eq:cvc0}
  \Delta{g}_n\simeq
  \frac{l_{c\parallel d}}{\mathcal{L}^2}\Delta\tilde\sigma_n(x\in\:\mathrm{ring})
  \hspace{0.5cm} \mbox{for }  L_\varphi \ll L 
  \:,
\end{equation}
where $x$ is any position inside the ring far from the vertices (at
distance larger than $L_\varphi$).
\\
$\bullet$ For $L_\varphi\gg{}L$, using 
eqs.~(\ref{PnXia},\ref{PnXir}) we see that the terms $\int_c$ and
$\int_d$ bring a contribution 
proportional to the perimeter $L$ whereas the terms $\int_a$ and $\int_b$
bring larger contributions proportional to $L_\varphi$~:
$\int_{a}\D{x}\,P_c^{(n)}(x,x)\simeq\frac12L_\varphi{}P_c^{(n)}(0,0)$, 
therefore
\begin{equation}
  \label{eq:cvc}
  \Delta{g}_n \simeq
  \frac{L_\varphi}{\mathcal{L}^2}\Delta\tilde\sigma_n(x\in\:\mathrm{ring})  
  \hspace{0.5cm} \mbox{for } L \ll L_\varphi 
  \:.
\end{equation}
The general expression describing the crossover between
\eqref{eq:cvc0} and \eqref{eq:cvc} can be obtained easily using the
formalism of Ref.~\cite{TexMon04}.

\subsection{Model B}

We now compute the harmonics of the conductivity within {\it model B}.  
We have now to consider
eqs.~(\ref{StartingPoint},\ref{localFDT},\ref{Decoherence}). 
The function $W(x,x')$ has been constructed in Ref.~\cite{TexMon05b}.
In the limit $l_a,\,l_b\gg{L}$ and if $x$ and $x'$ belong
to the connecting wires for $x,\,x'$(i.e.~$L_N$)$\ll{}l_a,\,l_b$,
the function coincides with the one of the infinite
wire $W(x,x')\simeq\frac12|x-x'|$.
Therefore, since 
in the limit $L\ll L_N$  the
diffusive trajectories spend most of the time in the
wires~\cite{ComDesTex05} (figure~\ref{fig:loop5}), 
the dephasing mostly occurs in the wires and the relaxation of the phase 
coherence is similar to the one for the wire, 
eq.~(\ref{dephasingwire}), irrespectively of the winding~:
$\smean{ \EXP{\I\Phi_V[\mathcal{C}^n_t]} }_{V,\mathcal{C}^n_t}\to{}f_\mathrm{wire}(t/L_N^2)$.

\begin{figure}[!ht]
  \centering
  \includegraphics[scale=0.475]{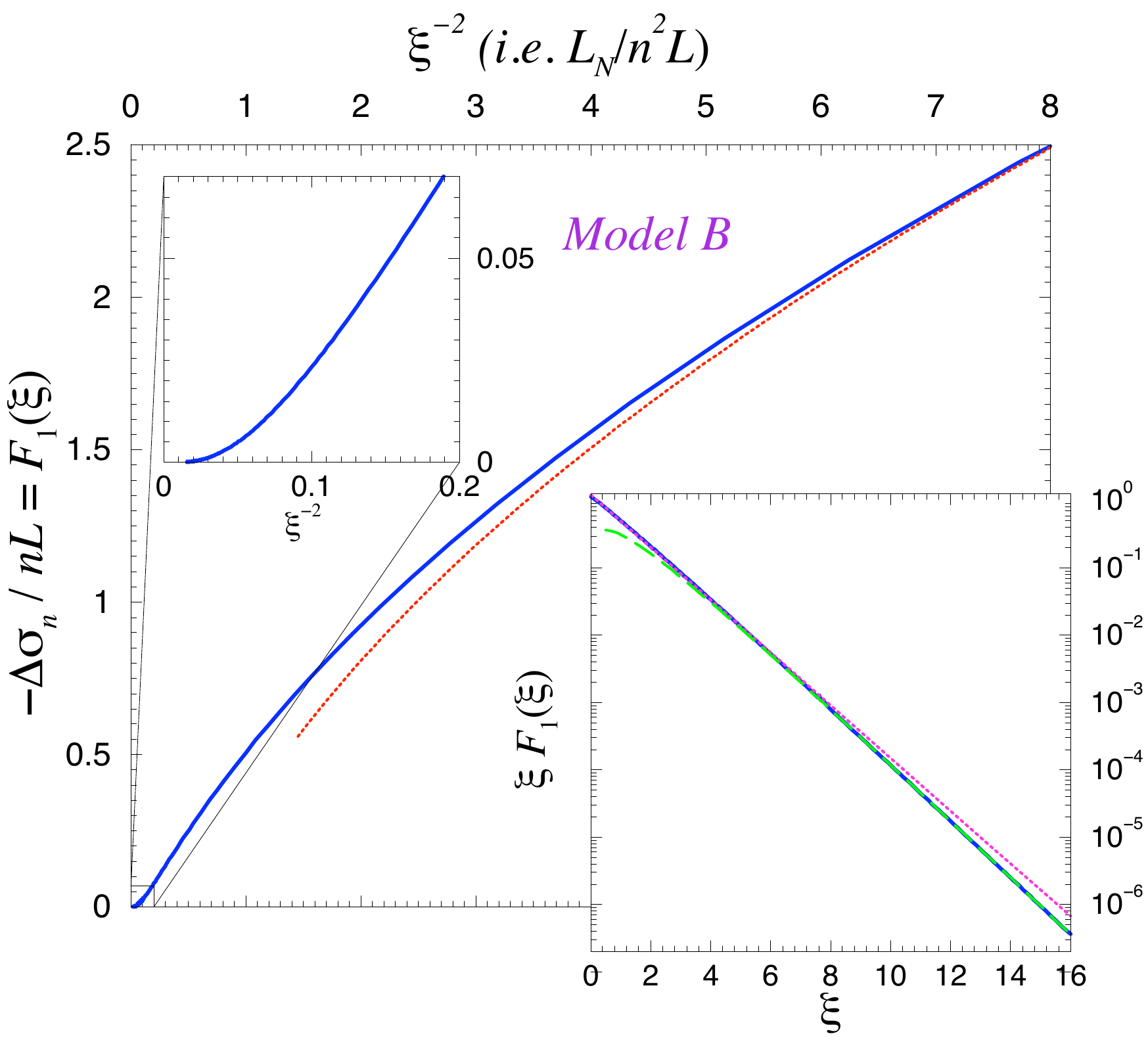}
  \caption{{\it $\Delta\tilde\sigma_n\propto{}F_1(\xi)$ 
   (continuous blue line) as a function of 
   $1/\xi^2\propto L_N/n^2L$. 
   Comparison with the limiting behaviour 
  $F_1(\xi\ll1)\simeq1.191/\xi-0.877$ (red dotted line).}
  Inset~:
  {\it $\xi\,F_1(\xi)$ in semi-log scale. 
    For moderate values of $\xi$, the function is well fitted by 
    $1.21\,\EXP{-0.9\xi}$ (magenta dots).
    Green dashed curve is the asymptotic expression, eq.~\eqref{Deltagnring2}.}
  }
  \label{fig:fctG1}
\end{figure}

Introducing (\ref{dephasingwire}) in (\ref{StartingPoint}) and
performing the change of variable $|u_m|t/\tau_N=v^2$, 
we obtain for the correction to the conductivity
\begin{align}
   \Delta\tilde\sigma_n(x)
   \simeq& -\sqrt{2\pi L_NL}\,
  \sum_{m=1}^\infty\frac1{|u_m|^{7/4}} \\\nonumber
  &\times\int_0^\infty\D v\,\sqrt{v}\:
   \Psi\!\left(|u_m|^{1/4}n\sqrt{\frac{2L}{v\,L_N}}\right)\,\EXP{-v^2}
  \:,
\end{align}
where $x$ is any position inside the ring.
Using \eqref{eq:DefPsi} we rewrite the double integral in polar
coordinates and perform integration over the radial coordinate. We find
\begin{align}
  \label{Deltagnring3}
  \boxed{
   \Delta\tilde\sigma_n(x)
   \simeq -L\,n\,\: F_1\!\left(n\sqrt{2L/L_N}\right)   
  }
\end{align}
with
\begin{align}
  \label{eq:fctG}
   F_1(\xi) 
  =\frac{2\sqrt2}{\xi}\,\sum_{m=1}^\infty\frac1{|u_m|^{7/4}}
  \,g\!\left(|u_m|^{1/4}\xi\right)
\end{align}
and 
\begin{align}
  g(\Lambda) 
  &= \re\left[
    \frac{\EXP{-\I\frac\pi4}}{4}
  \int_0^{\pi/2}\hspace{-0.25cm}\D\theta\,
  \sqrt{\sin2\theta}\,\EXP{-\Lambda\sqrt{\cotg\theta}\,\EXP{-\I\pi/4}}
  \right]
  \\
  \label{eq:Represg1}
   &=\frac1{\sqrt{2}}
  \re\left[
    \EXP{-\I\frac\pi4}
    \int_0^\infty\D t\,
    \frac{t^2\, \EXP{-\Lambda t\,\EXP{-\I\pi/4}}}{(t^4+1)^{3/2}}
  \right]
  \:.
\end{align}
A  convenient representation can be obtained by a rotation of 
$\pi/4$ of the axis of integration in the complex plane. We get
\begin{equation}
  \label{eq:Represg2}
    g(\Lambda)=\frac{\EXP{-\Lambda}}{\sqrt{2}}
  \int_0^\infty\D t\,
    \left(
       \frac{1}{8t^{3/2}}
      -\frac{(t+1)^2\,\EXP{-\Lambda t}}{[(t+1)^4-1]^{3/2}}
    \right)
  \:.
\end{equation}
The function $F_1(\xi)$ is plotted on figure~\ref{fig:fctG1}.
We now analyze the limiting behaviours.

We first consider the limit $L_N\ll n^2L$. 
Eq.~\eqref{eq:Represg2} gives
$g(\Lambda\gg1)\simeq\frac1{\sqrt2}\EXP{-\Lambda}\int_0^\infty\frac{\D{x}}{8x^{3/2}}(1-\EXP{-\Lambda{}x})=\frac1{4\sqrt2}\sqrt{\pi\Lambda}\,\EXP{-\Lambda}$,
therefore 
$F_1(\xi\gg1)\simeq\frac12|u_1|^{-13/8}\sqrt{\pi/\xi}\,\EXP{-|u_1|^{1/4}\xi}$ 
and
\begin{align}
  \label{Deltagnring2}
  &\boxed{
   \Delta\tilde\sigma_n(x)
   \simeq -\frac{L\,n\sqrt{\pi}}{2^{\frac54}|u_1|^{\frac{13}{8}}}
  \left(\frac{L_N}{n^2L}\right)^{\!1/4}\!\EXP{-\kappa_2n\sqrt{L/L_N}}
  }
  \\
  &\hspace{.5cm}
  \sim T^{-1/12}\EXP{ -n\,L^{1/2}T^{1/6} }
  \hspace{0.25cm} \mbox{for }  L\ll L_N  \ll n^2 L 
  \:,
\end{align}
where $\kappa_2=\sqrt{2}|u_1|^{1/4}\simeq1.421$.

For the lowest temperature $ n^2L\ll L_N$, 
eq.~\eqref{eq:Represg1}  gives
$g(0)=\frac1{4\sqrt\pi}\Gamma(3/4)^2$. Therefore
$F_1(\xi\ll1)\simeq{}A_1/\xi$
with 
$A_1=\frac12\sqrt{2/\pi}\,\Gamma(3/4)^2\sum_{m=1}^\infty|u_m|^{-7/4}\simeq1.191$
(the sum is $\sum_{m=1}^\infty|u_m|^{-7/4}\simeq1.989$)
\begin{align}
  \label{Deltagnring1}
  \boxed{
   \Delta\tilde\sigma_n(x)
   \simeq -A_1\sqrt{\frac{L_NL}{2}}
  \hspace{0.5cm} \mbox{for }  n^2 L \ll L_N
  }
  \:.
\end{align}

\vspace{0.25cm}

\noindent{\bf Comparison between models A \& B.--}
We have seen that for the wire, the MC obtained from the two models are
related through $L_\varphi\to\sqrt2\,L_N$ ({\it cf.}
section~\ref{sec:wr}).
It is tempting to look for a similar relation
for the connected ring in the limit $L\ll{}L_N$.

Let us compare the results for the two models of
decoherence. In the limit $n^2L\ll{}L_N$ the expressions
\eqref{Deltagnring1} is very close to \eqref{eq:SR2} because in this
case, the 
harmonics involve an integral over time of the function
$\smean{\EXP{\I\Phi}}_{V,\mathcal{C}_t}$. Therefore harmonics are
insensitive to the details of this function but only to the scale over
which it decays.
In the other limit $L_N\ll{}n^2L$, the calculation of the harmonics rather
involves the tail of the function
$\smean{\EXP{\I\Phi}}_{V,\mathcal{C}_t}$. 
Eq.~\eqref{Deltagnring2} presents an
exponential decays similar to eq.~\eqref{eq:SR2},  with a different
pre-exponential power law since the decay
$\smean{\EXP{\I\Phi}}_{V,\mathcal{C}_t}\propto\sqrt{t/\tau_N}\EXP{-|u_1|t/\tau_N}$
is different from the simple exponential decay $\EXP{-t/\tau_\varphi}$
for {\it model A}. 
The additional $\sqrt{t}$ in {\it model B} explains the different
pre-exponential terms
in eqs.~\eqref{eq:SR2} and~\eqref{Deltagnring2}.

Could we map the results of the two models through a simple
substitution of phase coherence length, as for the infinite wire~?
In the regime $L\ll{}L_N\ll{}n^2L$ we should compare the exponentials of  
\eqref{eq:SR2} and \eqref{Deltagnring2} what leads to
$L_\varphi\to{}L_N/\sqrt{|u_1|}\simeq0.99L_N$, however pre-exponential
factors cannot be matched, obviously.
In the regime $L_N\gg{}n^2L$ we rather compare the square roots
\eqref{eq:SR2} and \eqref{Deltagnring1} and therefore
$L_\varphi\to{}A_1^2L_N\simeq1.418L_N$.
Despite there is no unique simple substitution, we get in both cases 
$L_\varphi\sim{}L_N$.

\vspace{0.25cm}

\noindent{\bf Conductivity {\it vs} conductance.--}
We discuss the relation to the conductance.
In the regime $L_N\gg{}L$ discussed in this section, we
expect 
\begin{equation}
  \label{eq:cvcB}
    \Delta{g}_n\sim\frac{L_N}{\mathcal{L}^2}\Delta\tilde\sigma_n(x)
\end{equation}
for
$x\in$\:ring, for the same reason as for {\it model A},
eq.~\eqref{eq:cvc}, (the factor $L_N$ comes from the fact that the
integral \eqref{Res2004} is dominated by regions of typical size $L_N$
in the arms). 


\section{The chain of rings\label{sec:chain}}

\begin{figure}[!ht]
\centering
\includegraphics[scale=0.45]{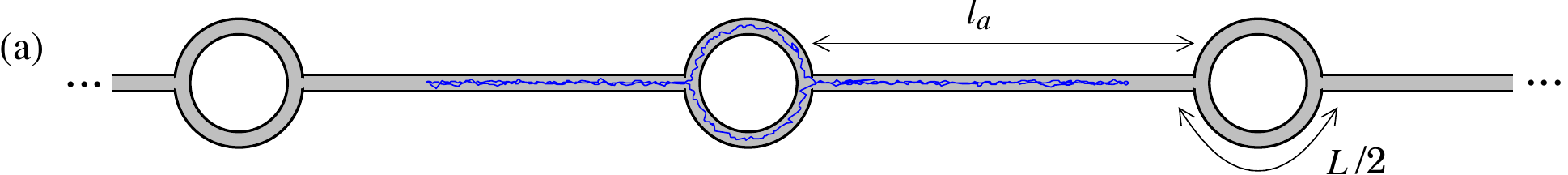}\\[0.25cm]
\includegraphics[scale=0.45]{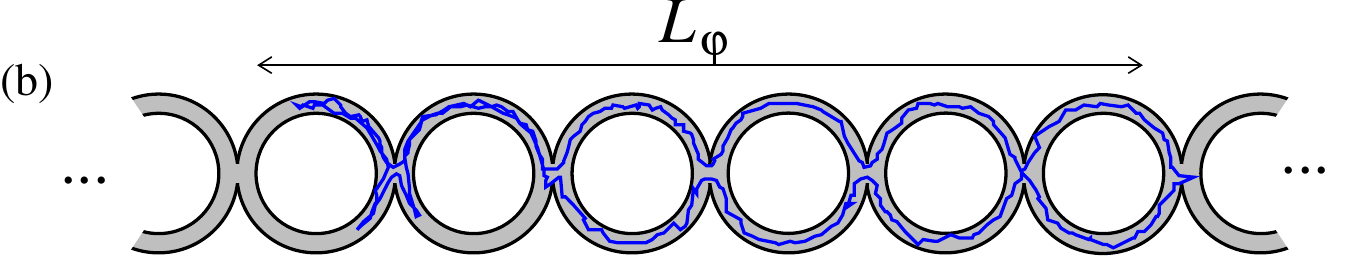}
\caption{\it 
  Chains of rings. If we consider the regime
  $l_a\gg{L_\varphi}($or\:$L_N)\gg{L}$ the rings of figure (a) can be
  considered as 
  independent. On the other hand, in the regime
  ${L_\varphi}($or\:$L_N)\gg{}l_a,\,{L}$ 
  interferences occur 
  between trajectories encircling several rings~; these latter cannot be
  anymore considered as independent in this case,
  as illustrated by the trajectory in the network (b).
  \label{fig:chains}}
\end{figure}

Let us consider now a chain of rings separated by arms of lengths $l_a$
(figure~\ref{fig:chains}.a).  
The case where the phase coherence length remains smaller than the arms
($L_\varphi\lesssim{}l_a$)
can be obtained from the results of the previous section since the
rings can be considered as independent. The conductance of a chain of
such $N_r$ rings  
is given by performing the substitution
$\frac1{\mathcal{L}^2}\to\frac{N_r}{[(N_r+1)l_a+N_rl_{c\parallel{}d}]^2}$ 
in the expression of $\Delta{g}_n$ for one ring,
eqs.~(\ref{eq:cvc0},\ref{eq:cvc},\ref{eq:cvcB}). 

When arms separating the rings are smaller than the phase coherence
length ($l_a\lesssim{}L_\varphi$), coherent trajectories enclose
magnetic fluxes in several rings.
In order to study how the MC harmonics are affected by this effect we
consider the limit when rings are directly attached to each other
(figure~\ref{fig:chains}.b). 

\subsection{Model A}
Considering the conductance of the  symmetric chain of
figure~\ref{fig:chains}.b,   
the weights of the wires involved in eq.~(\ref{Res2004}) are all
equal. This justifies 
a uniform integration of the Cooperon in the chain. In this case we can use the 
relation between the WL correction and the spectral determinant
\cite{Pas98,PasMon99,AkkComDesMonTex00} (appendix~\ref{app:spedet}).
The spectral determinant of the infinite chain is given in
appendix~\ref{app:necklace}. 
Averaging the Cooperon in the chain, 
$\Delta\tilde\sigma(\theta)\equiv\int_\mathrm{chain}\frac{\D{}x}{\mathrm{Vol}}\Delta\tilde\sigma(x,\theta)$, and
using (\ref{eq:PasMon},\ref{Schain}), we finally obtain~\cite{DouRam86}
\begin{widetext}
\begin{align}
  \label{eq:ChainMC}
  \Delta\tilde\sigma(\theta)
  = -\frac{L_\varphi}2
   \left(
   \coth(L/2L_\varphi) -\frac{2L_\varphi}{L}
   +\frac{\sinh(L/2L_\varphi)}{\sqrt{\cosh^2(L/2L_\varphi)-\cos^2(\theta/2)}}
   \right)
  \:,
\end{align}
where $\theta=4\pi\phi/\phi_0$ is the reduced flux per ring.
We now study the harmonics $n\neq0$~:
\begin{align}
  \Delta\tilde\sigma_n
  = - \frac{L_\varphi}{\sqrt2}\, \int_0^{2\pi}\frac{\D\theta}{2\pi}
  \frac{ \sinh(L/2L_\varphi)\, \EXP{\I n\theta} }
       { \sqrt{\cosh(L/L_\varphi)-\cos(\theta)} } 
  = - L_\varphi\, \sinh( L/2L_\varphi )  
  \oint_\mathrm{circle}\frac{\D z}{2\I\pi}
  \frac{z^{n-1}}{\sqrt{(\EXP{L/L_\varphi}-z)(z-\EXP{-L/L_\varphi})/z}}
  \:.
\end{align}
The integration in the complex plane is performed along the unit circle in the
clockwise direction.
The segment of the real axis $[0,\EXP{-L/L_\varphi}]$ is a branch cut. The
contour of integration is deformed to follow closely this segment. We obtain~:
\begin{align}
  \label{represinteg}
  \Delta\tilde\sigma_n &=
  - \frac{L}{2\pi}\,\frac{\sinh(L/2L_\varphi)}{L/2L_\varphi} \,
   \EXP{-(n-1/2)L/L_\varphi}  
   \int_0^1\D u\,
   \frac{(1-u)^{n-1/2}}{\sqrt{u^2+\varepsilon \, u}}
\end{align}
with $\varepsilon=\EXP{2L/L_\varphi}-1$.
We recognize the integral representation \eqref{eq:HyperIntRepres} of the 
hypergeometric function~\cite{gragra}~$F$~:
\begin{align}
  \label{ResultatExact}
\boxed{
  \Delta\tilde\sigma_n =
  - \frac{L}{2\pi}\,
  \frac{\sinh(L/2L_\varphi)}{L/2L_\varphi} \,
  \EXP{-(n+1/2)L/L_\varphi}\,
  B\!\left(\frac12,n+\frac12\right)\,
  F\!\left(\frac12,n+\frac12;n+1;\EXP{-2L/L_\varphi}\right)
}
  \:,
\end{align}
\end{widetext}
where $B(x,y)$ is the Euler $\beta$ function.

\vspace{0.25cm}

\noindent{\it Weakly coherent limit.--}
We consider the limit $L_\varphi\ll L$. 
Using  $F(\alpha,\beta;\gamma;\xi\to0)\to1$,
we obtain~:
\begin{equation}
  \label{necklace:unintersting}
  \Delta\tilde\sigma_n
  \simeq -\frac{(2n-1)!!}{2^{n+1}\,n!}\,
  L_\varphi\,\EXP{-nL/L_\varphi}
  \hspace{0.5cm}\mbox{ for }
  L_\varphi \ll L
  \:,
\end{equation}
a result reminiscent of the result of the isolated ring~(\ref{AAS}),
with a different prefactor originating from the probability to cross
the vertices of coordination number $4$ 
(note that $\frac{(2n-1)!!}{2^{n+1}\,n!}=\frac{\Gamma(n+1/2)}{2\sqrt{\pi}\,n!}\simeq\frac1{\sqrt{4\pi n}}$
for~$n\gg1$).

\vspace{0.25cm}

\noindent{\it Large coherence length.--}
In the opposite limit limit $L_\varphi\gg{L}$. 
Eq.~\eqref{eq:ExpF2} gives~:
\begin{align}
  \label{necklimit2}
  \Delta\tilde\sigma_n
   \simeq 
  - \sqrt{\frac{L_\varphi L}{8\pi n}}\,\EXP{-n L/L_\varphi}
    \hspace{0.25cm}\mbox{ for }   L \ll L_\varphi \ll nL
  \:.
\end{align}
We have recovered an exponential damping of the harmonics, reminiscent
of (\ref{AAS},\ref{necklace:unintersting}), but with a different $L_\varphi$
dependence of the pre-exponential factor.

On the other hand, for harmonics with $nL\ll{}L_\varphi$, 
the harmonics can be expanded by using eq.~\eqref{eq:ExpF}.
%
Let us introduce $b_n=\ln{n}-\psi(n+\frac12)+\psi(1)$.
These coefficients converge to a finite limit  at large $n$~:
$b_n=-$C$-\frac1{24n^2}+O(\frac1{n^4})$, where
C$=-\psi(1)=0.577215...$ is the Euler constant.  Finally we obtain 
\begin{equation}
  \label{necklimit1}
  \Delta\tilde\sigma_n  \simeq - \frac{L}{2\pi}\,
   \left[
     \ln(2L_\varphi/nL) + b_{n}  
   \right]
  \hspace{0.25cm}\mbox{ for }
 nL \ll  L_\varphi 
  \:.
\end{equation}

\begin{figure}[!ht]
  \centering
  \includegraphics[scale=0.9]{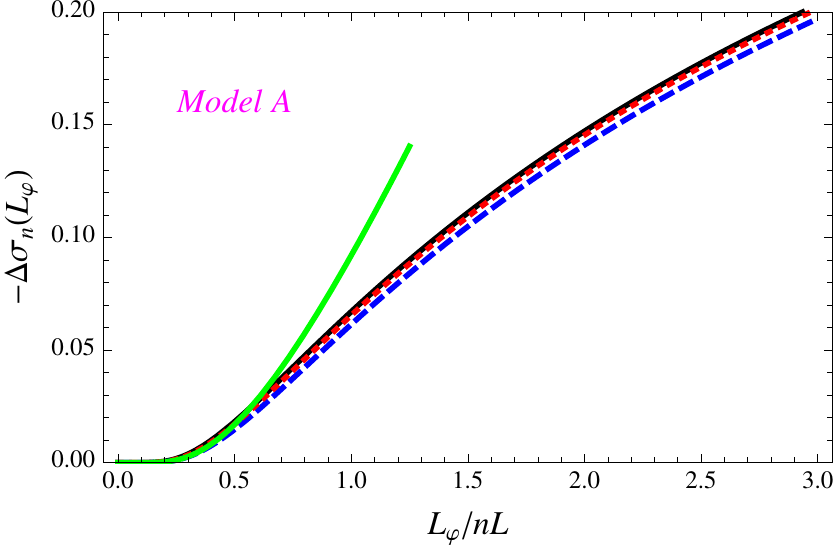}
  \caption{\it Comparison between exact result
    \eqref{ResultatExact} for
    $n=1$ (dashed blue line) and $2$ (dotted red line) and the
    approximation \eqref{eq:ApproxHarmChain} (black continuous line).  
    The interrupted green curve is \eqref{necklace:unintersting}.
    Even for small $n$ ($=2$)
    eq.~\eqref{eq:ApproxHarmChain} is a very good approximation of
    eq.~\eqref{ResultatExact}. 
     }
  \label{fig:chain_harm1and2}
\end{figure}

It is useful to remark that the expressions
\eqref{necklimit2} and \eqref{necklimit1} coincide with the limiting
behaviours of the modified Bessel function $K_0(z)$ for  large $n$
(the proof is given in appendix~\ref{app:RelSF})~: 
\begin{equation}
    \label{eq:ApproxHarmChain}
    \Delta\tilde\sigma_n  \simeq - \frac{L}{2\pi}\, K_0(nL/L_\varphi)
      \hspace{0.25cm}\mbox{ for } 
    L \ll  L_\varphi
  \:.
\end{equation}
Up to a factor $1/2$ interpreted below, this expression coincides with
the MC harmonics 
for a long hollow cylinder~\cite{AltAroSpi81}, eq.~\eqref{eq:AAScyl}
recalled in section~\ref{sec:cyl}.
We compare this approximation with the exact expression
\eqref{ResultatExact} on figure~\ref{fig:chain_harm1and2}. 
We see that the approximation
is already good for $n=1$, provided $L_\varphi\gtrsim L$.
The difference rapidly diminishes as $n$ increases.

\vspace{0.25cm}

\mathversion{bold}
\noindent{\bf Logarithmic divergence of the harmonics for $L_\varphi\to\infty$.--}
\mathversion{normal}
We see from eq.~(\ref{necklimit1}) that the harmonics are weakly 
dependent on $n$ (for $n\ll{}L_\varphi/L$). This logarithmic
behaviour reflects the singular behaviour 
$\Delta\tilde\sigma_\mathrm{osc}(\theta)\sim1/\theta$, with a cutoff at
$\theta\sim{L/L_\varphi}$~:
$\Delta\tilde\sigma_n\sim\int^{1/n}_{L/L_\varphi}\frac{\D\theta}{\theta}=\ln(L_\varphi/(nL))$.
The harmonics are therefore almost independent on $n$ as soon as $n$
is small enough compared to $L_\varphi/L$.
Note that in practice, this logarithmic divergence of the harmonics is
limited~: when the phase coherence length reaches the total length of
the chain $L_\varphi\sim{}N_rL$, harmonics reach a finite limit, 
$\Delta\tilde\sigma_n\simeq-\frac{L}{2\pi}\ln(N_r/n)$, due to the effect of 
boundaries (external contacts).


\vspace{0.25cm}

\noindent{\bf Winding probability.--}
We now extract the  probability
$\mathcal{P}_n(x,x;t)$ from these results.
First of all the behaviour
\eqref{necklace:unintersting} is related to 
\begin{equation}
  \label{necklace:unint2}
  \mathcal{P}_n(x,x;t) \simeq \frac{(2n-1)!!}{2^{n+1}\,n!}
  \frac{1}{\sqrt{4\pi t}}\,\EXP{-(nL)^2/4t}
  \hspace{0.5cm}\mbox{ for } t\ll L^2
  \:.
\end{equation}
We have recovered \eqref{PnRing} with an 
additional dimensionless factor coming from the
probability to cross the vertices of coordination number
$4$ (this factor can be understood when one writes the trace formula
for the heat kernel in the network~\cite{Rot83,AkkComDesMonTex00}).

The regime $L_\varphi\gtrsim{}L$ for the WL correction probes
the regime $t\gtrsim{}L^2$ for the winding probability.  
We use the approximation \eqref{eq:ApproxHarmChain} in order to perform the
inverse Laplace transform. Using  the integral
representation of the modified Bessel function~\cite{gragra}, we get~:
\begin{equation}
  \label{PnChain}
  \boxed{
  \mathcal{P}_n(x,x;t) \simeq \frac{L}{8\pi t}\,\EXP{-(nL)^2/4t}
  \hspace{0.5cm}\mbox{ for }  t \gg L^2
  }
  \:.
\end{equation}
We may check that (\ref{necklimit2},\ref{necklimit1}) coincide with
the limiting behaviours of this probability.
It is interesting to point that this probability is similar to the one found
for an infinitely long hollow cylinder, apart for a factor~$1/2$.
This additional factor can be understood from the fact that, starting
from a given ring, it is
equiprobable to return in one of its two arms.

Let us give a heuristic argument to recover roughly \eqref{PnChain},
that will be useful for the following. 
Arriving at a vertex, the diffusive particle equiprobably chooses one of the
four arms. Therefore it is equiprobable to wind a ring or not, while
diffusing along the chain. This
suggests that the winding probability is almost independent on $n$, up
to $n\sim\sqrt{t}/L$, the maximum number of rings explored for a time
$t$. 
This rough approximation would be
$\mathcal{P}_n(x,x;t)\sim\mathcal{N}_t\,\theta(\sqrt{t}/L-|n|)$.
The normalisation is estimated easily~:
since diffusion along the chain is one-dimensional, we expect 
$\sum_n\mathcal{P}_n(x,x;t)\sim1/\sqrt{t}$ so that
$\mathcal{P}_n(x,x;t)\sim{}L/t$ for $|n|\lesssim\sqrt{t}/L$ and $0$
otherwise.
This is a crude estimate of eq.~\eqref{PnChain}.

\subsection{Model B}
In order to compute the MC harmonics we first need to construct the
function $W$ entering in eq.~\eqref{localFDT}.
Following appendix~\ref{app:sdepn}, we introduce a coordinate
$\tilde{x}=(x,f)$ to locate a point in the chain (the continuous
variable $x$ measures the distance along the chain while the discrete
index $f\in\{u,d\}$ precise the arm, up or down).
If $\tilde{x}$ and $\tilde{x}'$ do not belong to the same ring we have
\begin{equation}
  W(\tilde{x},\tilde{x}') = \frac14|x-x'| = \frac12\,W_\mathrm{wire}(x,x')
  \:.
\end{equation}
Remembering that $W(x,x')$ is proportional to the resistance between
points $x$ and $x'$ this equation has a clear meaning~: when two
consecutive nodes are linked by two wires instead of one, the
resistance is diminished by a factor of $2$. In the limit $t\gg L_N^2$
the trajectories contributing to  
\begin{equation}
  \smean{\EXP{\I\Phi_V[\mathcal{C}^n_t]}}_{V,\mathcal{C}^n_t} 
  =\mean{
    \EXP{-\frac2{L_N^3}\int_0^t\D\tau\,W(x(\tau),x(t-\tau))}
  }_{\mathcal{C}^n_t} 
\end{equation}
are trajectories extending over distances $L_N\gg{L}$ along the chain.
In this case we can neglect the contributions to the integral where
the two arguments of $W(\tilde{x},\tilde{x}')$ are in the same ring.
We have seen that for $nL\lesssim\sqrt{t}$ the 
measure of the Brownian paths
weakly depends on $n$, therefore we expect that 
$
\smean{\EXP{\I\Phi}}_{V,\mathcal{C}^n_t}
\simeq\smean{\EXP{\I\Phi}}_{V,\mathcal{C}_t}
$, 
where the average of the {\it l.h.s.} is realized among Brownian
curves of definite winding 
whereas the average of the {\it r.h.s.} is among all Brownian
curves. The argument shows that the 
function describing decoherence corresponds to the result of the infinite wire
in which~$L_N^3\to2L_N^3$.
This factor $2$ stands from the ratio between the resistance of a wire
and of a chain of rings of the same length. Finally~:
\begin{equation}
  \smean{\EXP{\I\Phi_V[\mathcal{C}^n_t]}}_{V,\mathcal{C}^n_t} 
  \simeq 
  f_\mathrm{wire}(t/2^{2/3}\tau_N)
  \:.
\end{equation}
We can use (\ref{dephasingwire}) in order to 
compute 
$
\Delta\tilde\sigma_n\simeq-2
\int_{L^2}^\infty\D{t}\,
\frac{L\,\EXP{-(nL)^2/4t}}{8\pi{t}}\,f_\mathrm{wire}(t/2^{2/3}L_N^2)
$.
The lower cutoff takes into account the fact that expression of
$\mathcal{P}_n(x,x;t)$ is only valid for $t\gg{L}^2$. However, except if $n=0$, 
the cutoff is not important and can be replaced by $0$. Using 
(\ref{dephasingwire}) we obtain~:
\begin{equation}
  \label{eq:RESchain}
 \boxed{
  \Delta\tilde\sigma_n
  \simeq-L\, F_2\!\left(2^{-1/3}n{L}/{L_N} \right)
  } 
\end{equation}
where 
\begin{equation}
  F_2(\xi)=\frac14\,\sum_{m=1}^\infty\frac1{|u_m|^{3/2}}\EXP{-\sqrt{|u_m|}\,\xi} 
  \:.
\end{equation}

In the limit $L_N\ll nL$ the first term of the series dominates~:
\begin{align}
 & 
\label{eq:chainB1}
\boxed{
  \Delta\tilde\sigma_n 
  \simeq-\frac{L}{4|u_1|^{3/2}}\, \EXP{ -\kappa_3n{L}/{L_N} }
 }
  \\
  &\hspace{0.75cm}
  \sim \EXP{ -nL T^{1/3} }
  \hspace{0.5cm} \mbox{for }   L\ll L_N \ll nL \:,
\end{align}
where $\kappa_3=2^{-1/3}|u_1|^{1/2}\simeq0.801$.

\begin{figure}[!ht]
  \centering
  \includegraphics[scale=0.425]{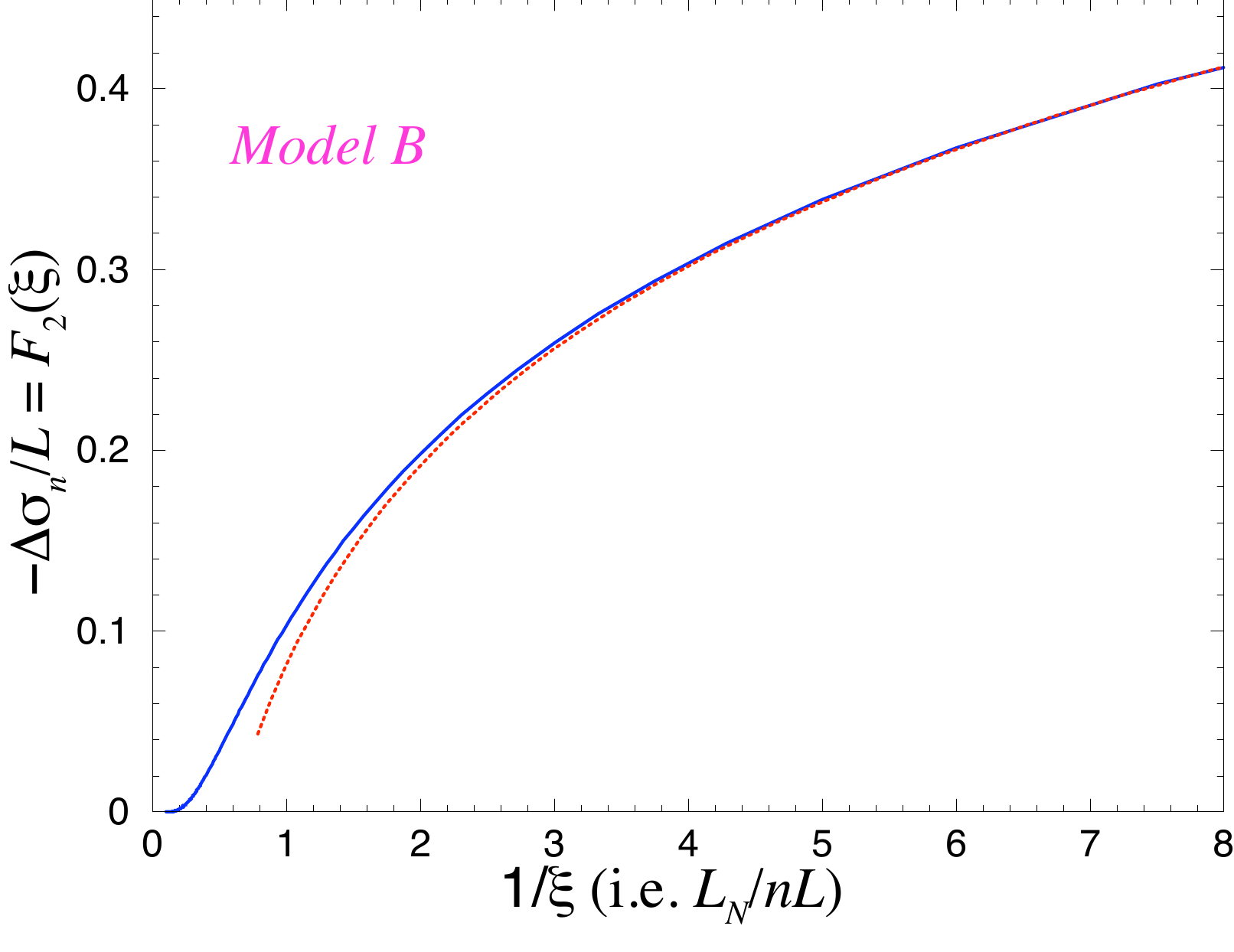}
  \caption{\it $\Delta\tilde\sigma_n\propto{}F_2(\xi)$ as a function of 
   $1/\xi\propto L_N/nL$ (blue continuous line). 
   Comparison with the limiting behaviour (red dotted line)
   $\frac1{2\pi}[\ln(1/\xi)+C_\mathrm{cyl}]$ with 
   $C_\mathrm{cyl}\simeq0.51$.}
  \label{fig:fctF2}
\end{figure}

In the opposite limit, for $nL\ll{}L_N$
($\xi\to0$), we can replace the sum by an integral and use the asymptotic
behaviour $u_m\simeq-[\frac{3\pi}2(m-\frac34)]^{2/3}$~:
\begin{align}
  &F_2(\xi) \underset{\xi\to0}{\simeq}
  \frac1{6\pi}\int_1^\infty\frac{\D{m}}{m}\,\EXP{-\xi[\frac{3\pi}4m]^{1/3}}
  \\
  &
  \simeq \frac1{6\pi}\int_1^{1/\xi^3}\frac{\D{m}}{m}+ \mathrm{cste}
  = \frac1{2\pi}\,[\ln(1/\xi) + C_\mathrm{cyl}]
  \:.
\end{align}
Finally we find a result similar to the one obtained for exponential
relaxation of phase coherence~:
\begin{equation}
  \label{eq:chainMBlimit}
  \boxed{
  \Delta\tilde\sigma_n 
  \simeq-\frac{L}{2\pi}\,
  \left[\ln(L_N/nL)+C_\mathrm{chain}\right]
  }
  \hspace{0.5cm} \mbox{for }  nL \ll L_N 
  \:.
\end{equation}
The constant is estimated numerically~: we find
$C_\mathrm{cyl}\simeq0.51$, hence $C_\mathrm{chain}\simeq0.74$. 
This result could also have been more simply obtained by noticing that
$f_\mathrm{wire}(t/2^{2/3}L_N^2)$ cuts the tail of $\mathcal{P}_n(x,x;t)$ on
a scale $L_N^2$~:
$\smean{\Delta\tilde\sigma_n}\simeq-2\int_{(nL)^2}^{L_N^2}\D{t}\,\frac{L}{8\pi t}$.

\vspace{0.25cm}

\noindent{\bf Comparison between models A \& B.--}
As we have done for the infinite wire and the connected ring, we
establish some correspondence between the results for the two models
when $L_N\gg{}L$. 
 
In the regime $L\ll{}L_N\ll{}n^2L$ the exponentials of  
\eqref{necklimit2} and \eqref{eq:chainB1} may be matched if
$L_\varphi\to{}L_N/\kappa_3\simeq1.25L_N$.

In the regime $L_N\gg{}n^2L$ the logarithmic behaviours 
\eqref{necklimit1} and \eqref{eq:chainMBlimit} coincide for
$L_\varphi\to1.87\,L_N$.

For not too large $n$, the two curves 
$ 
 \Delta\tilde\sigma_n^{(\mathrm{B})}(L_N)  \approx
  \Delta\tilde\sigma_n^{(\mathrm{A})}(L_\varphi\simeq1.87\,L_N)
$
are very close, apart for $L_N\ll{}L$ for
which there is a qualitative difference between \eqref{AAS} and~\eqref{LMTM}.

\vspace{0.25cm}

\noindent{\bf From the conductivity to the conductance.--}
The weights in eq.~\eqref{Res2004} are all equals and conductance is
related to an uniform integration of $\Delta\tilde\sigma(x)$ in the chain.
The dimensionless conductance is given
by~$\Delta{g}=\frac{4}{N_rL}\Delta\tilde\sigma$, where $N_r$ is the number
of rings of the chain.


\section{The square network}
\label{sec:sn}

The easiest way to realize disorder averaging experimentally is to use
networks with a large number of rings, like 2d networks 
(square~\cite{DolLicBis86,AroSha87,Fer04,FerAngRowGueBouTexMonMai04,SchMalMaiTexMonSamBau07,FerRowGueBouTexMon08}, 
honeycomb~\cite{PanChaRamGan84,PanChaRamGan85,AroSha87}, 
dice~\cite{Mal06,SchMalMaiTexMonSamBau07}). The ``high
temperature'' regime ($L_N\ll L$) is now well understood theoretically and 
experimentally
\cite{FerRowGueBouTexMon08}, but low temperature experimental results are
still unexplained~\cite{Mal06}. 
Therefore understanding the 
magnetoconductance of large networks when
decoherence is dominated by e-e interaction still deserves some clarification.
In this section we study the case of an infinite square network of lattice
spacing $a$ (figure~\ref{fig:sn}).

\subsection{Model A}
The weak localization correction was derived analytically by Dou\c{c}ot
\& Rammal (DR)
for rational fluxes $\theta_{p,q}=2\pi{}p/q$ with $p,\,q\in\NN$
(reduced flux is defined as $\frac{\theta}{2\pi}=2\phi/\phi_0$). They
obtained~\cite{DouRam86}  
\begin{widetext}
\begin{align}
  \label{DouRam86}
\boxed{
  \Delta\tilde\sigma(\theta_{p,q})
  =-\frac{L_\varphi}{2}
  \left[
    \coth\frac{a}{L_\varphi} -\frac{L_\varphi}{a} 
   + \frac{8\sinh\frac{a}{L_\varphi}}{\pi q}\,
      \frac{ P'_{p,q}(4\cosh\frac{a}{L_\varphi}) }
           { P_{p,q}(4\cosh\frac{a}{L_\varphi}) }\:
    \mathrm{K}\!\left(\frac4{P_{p,q}(4\cosh\frac{a}{L_\varphi})}\right)
  \right]
  }
  \:,
\end{align}
\end{widetext}
where $P_{p,q}(\varepsilon)$ is a polynomial of degree $q$ defined in
appendix~\ref{app:DouRam} where derivation of \eqref{DouRam86} is
recalled.   
$\mathrm{K}(x)$ 
is the elliptic integral of the first kind~\cite{gragra}.

\vspace{0.25cm}

\noindent{\it Weakly coherent network.--}
The harmonics are suppressed exponentially as
$\Delta\tilde\sigma_n\propto-L_\varphi\EXP{-4na/L_\varphi}$. 
Despite there is no close expression of the remaining dimensionless
$n$-dependent factor, a systematic expansion of the spectral
determinant can be written thanks to the trace formula of
Ref.~\cite{Rot83} (the first terms of this expansion are available in
Ref.~\cite{FerAngRowGueBouTexMonMai04}). 

\vspace{0.25cm}

\noindent{\it Large coherence length.--}
The rest of the section is devoted to the large coherence length
regime~$L_\varphi\gg{}a$.

\vspace{0.25cm}

\noindent{\bf Continuum limit.--}
In the limit of small flux, $\theta\ll1$, and large coherence length,
$L_\varphi\gg{}a$, the discrete character of the network disappears and
one should recover the results for the 2d plane in a uniform magnetic
field. 
Informations can be extracted from the study of this limit.

The zero field WL correction is obtained from eq.~(\ref{DouRam86})
with $p=q=1$, using $P_{1,1}(x)=-x$.  Using the expansion of the elliptic
integral~\cite{gragra,footnote9}, we
find~\cite{FerAngRowGueBouTexMonMai04,TexMon07c}~:  
\begin{equation}
  \label{eq:SNlim1}
  \Delta\tilde\sigma(0) \simeq-\frac{a}{\pi}\, 
  \left[ \ln(4L_\varphi/a)+\frac\pi6 \right]
  \:.
\end{equation}
This result is reminiscent of the WL correction of the film
\eqref{eq:Bergman}, but 
here, the cutoff at small scales is naturally provided by
the lattice spacing~$a$.

The limit of small fluxes is studied in details in
appendix~\ref{app:ConLimPlanNet}. Using that
$\theta=4\pi\phi/\phi_0=4\pi\mathcal{B}a^2/\phi_0$,
eq.~\eqref{eq:ConLimPlanNet} reads 
\begin{equation}
\label{eq:SnContLim}
  \boxed{
   \Delta\tilde\sigma(\theta\ll1) - \Delta\tilde\sigma(0)
  \simeq \frac{a}{2\pi}  
  \left[
    \psi\!\left(\frac12+\frac{a^2}{\theta L_\varphi^2}\right)
    -  \ln\left(\frac{a^2}{\theta L_\varphi^2}\right)
  \right]
  }
  \:.
\end{equation}
This expression gives a quadratic behaviour for small
flux~\cite{footnote14} 
\begin{equation}
    \label{snlimit2}
  \Delta\tilde\sigma(\theta) -
   \Delta\tilde\sigma(0) \simeq \frac{a}{48\pi}
  \left(\frac{\theta L_\varphi^2}{a^2}\right)^2
  \hspace{0.5cm}\mbox{ for } \theta \ll \frac{a^2}{L_\varphi^2}
\end{equation}
and a logarithmic behaviour for intermediate fluxes~\cite{footnote14}
\begin{equation}
     \label{snlimit1}
  \Delta\tilde\sigma(\theta)\simeq \frac{a}{2\pi} 
  \left[ \ln\theta + C_\mathrm{sn}\right]  
  \hspace{0.5cm}\mbox{ for } \frac{a^2}{L_\varphi^2}\ll\theta \ll 1 
  \:,
\end{equation}
where $C_\mathrm{sn}=-\mathrm{C}-3\ln2-\frac\pi3\simeq-3.704$.

\begin{figure}[htbp]
  \centering
  \includegraphics[scale=0.5]{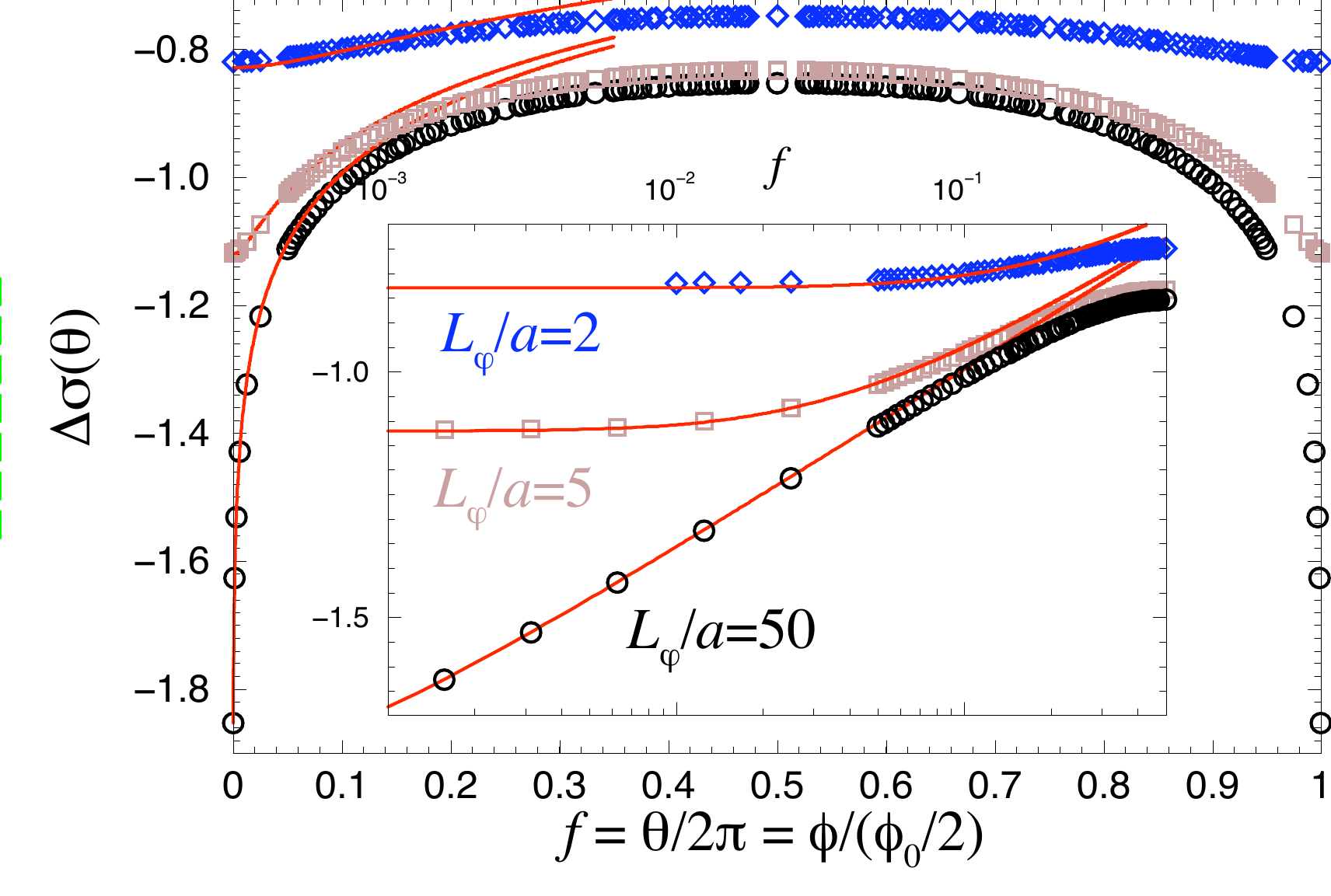}
  \caption{\it WL correction for $L_\varphi/a=2$ (blue diamonds),  
              $L_\varphi/a=5$ (brown squares) and $L_\varphi/a=50$
              (black circles).
              The red continuous lines correspond to the continuum limit, 
              eqs.~\eqref{eq:SNlim1} and \eqref{eq:SnContLim}.}
  \label{fig:wlsnBfield28}
\end{figure}

We now turn to the analysis of the MC harmonics.
A first simple remark allows to get the scaling of harmonics with time~:
the reduced flux $\theta$ is the variable conjugated to the
harmonic number $n$, therefore  the structure
$\Delta\tilde\sigma(\theta)=$fct$(a/\sqrt{\theta}L_\varphi)$
corresponds to
$\Delta\tilde\sigma_n=$fct$(\sqrt{n}a/L_\varphi)$. We now extract this
function.
Using the path integral formulation it is straightforward to get the
structure 
\begin{align}
    \Delta\tilde\sigma(\theta) - \Delta\tilde\sigma(0)
  &= -2\int_0^\infty\D t\,\mathcal{P}(x,x;t)
  \\\nonumber
  &\hspace{1cm}
  \times
  \left(
    \mean{ \EXP{\I\theta\mathcal{N}[\mathcal{C}_t]} }_{\mathcal{C}_t} - 1
  \right)\EXP{-t/L_\varphi^2}
\end{align}
where $\mathcal{N}[\mathcal{C}_t]$ is the winding number of the closed
trajectory. 
At large times the return probability coincides with the one of a plane,
$\mathcal{P}(x,x;t)\simeq\frac{a}{4\pi t}$~(appendix~\ref{app:Psn})~;
it can be obtained from  
eq.~\eqref{eq:SNlim1} thanks to an inverse Laplace transform. 
Using~\cite{gragra}
 $\int_0^\infty\D{}t\,\big(\frac1t-\frac{\lambda}{\sinh\lambda{}t}\big)\EXP{-\gamma{}t}=\psi\big(\frac12+\frac{\gamma}{2\lambda}\big)-\ln\big(\frac{\gamma}{2\lambda}\big)$,
we deduce that expression \eqref{eq:SnContLim} corresponds to
$
\smean{\EXP{\I\theta\mathcal{N}[\mathcal{C}_t]}}_{\mathcal{C}_t}
\simeq\frac{\theta\,t}{2a^2\sinh(\theta t/2a^2)}
$.
A Fourier transform~\cite{gragra} gives the distribution of the
winding number, plotted on figure~\ref{fig:QdeX}, 
\begin{equation}
    Q_t(\mathcal{N}) \simeq \frac{\pi a^2}{2t\,\cosh^2(\pi a^2\mathcal{N}/t)}
\end{equation}
We have recovered the well-known Levy law for the
distribution of the algebraic area  $\mathcal{A}=\mathcal{N}a^2$
enclosed by a planar Brownian 
motion~\cite{KhaWie88,Yor89,Dup89,ComDesOuv90}. 
For $t\gg a^2$ and $n\gg1$,
the return probability conditioned to 
 wind $n$ fluxes is therefore expected to behave as 
$\mathcal{P}_n(x,x;t)\simeq\mathcal{P}(x,x;t)Q_t(n)$~:
\begin{equation}
  \mathcal{P}_n(x,x;t) \simeq \frac{a^3}{8\,t^2}\,
  \frac{1}{\cosh^2(\pi na^2/t)}
\end{equation}
A Laplace transform gives the 
corresponding harmonics 
\begin{align}
  \label{resloidesaires}
  \boxed{
   \Delta\tilde\sigma_n  
  \simeq -\frac{a}{4\pi n}\, \mathcal{F}_3(\pi na^2/L_\varphi^2)
  }
\end{align}
where
\begin{equation}
    \mathcal{F}_3(\xi) = \int_0^\infty\D y\,\frac{\EXP{-\xi/y}}{\cosh^2y}
  \:.
\end{equation}
We extract the following limiting behaviours~:
\begin{align}
  \label{complimit1}
  \mathcal{F}_3(\xi)
  &\underset{ \xi\ll1}\simeq 1 - \xi \,\ln(\xi_0/\xi)
  \\
  \label{complimit2}
  &\underset{ \xi\gg1}\simeq \sqrt{4\pi} (2\xi)^{1/4}\,\EXP{-\sqrt{8\xi}}
  \:.
\end{align}
The constant $\xi_0$ is estimated numerically~:
$\xi_0\simeq1.239$. 
The tail of the distribution corresponds to 
\begin{equation}
  \label{eq:SNlimit1}
  \boxed{
  \Delta\tilde\sigma_n\simeq
  -\frac{a}{(2n)^{3/4}}\,
  \sqrt{\frac{a}{\pi L_\varphi}}\,
  \EXP{-\sqrt{8\pi n}\:a/L_\varphi}
   }
\end{equation}
for $L_\varphi \ll \sqrt{n}\,a$.
The saturation of the harmonics for $L_\varphi\to\infty$ is given by~:
\begin{equation}
  \label{eq:SNlimit2}
  \boxed{
  \Delta\tilde\sigma_n  \simeq
    -\frac{a}{4\pi n}\,
  \left[ 
   1 -\frac{\pi na^2}{L_\varphi^2}
        \ln\left(\frac{L_\varphi^2\xi_0}{\pi na^2}\right)
  \right]
  }
\end{equation}
for $\sqrt{n}\,a \ll L_\varphi$.

\vspace{0.25cm}

\mathversion{bold}
\noindent{\bf Harmonics reach a finite limit for
  $L_\varphi\to\infty$.--}
\mathversion{normal}
It is interesting to compare the MC of the chain and the MC of the 
square network. 
For the chain, the behaviour of the MC near zero flux 
$\Delta\tilde\sigma^\mathrm{(chain)}(\theta)\sim1/\theta+\mathrm{cste}$,
eq.~\eqref{eq:ChainMC},
is related to a weak logarithmic divergence of the harmonics 
$\Delta\tilde\sigma^\mathrm{(chain)}_n\sim\ln(L_\varphi/nL)$,
eq.~\eqref{necklimit1}.
For planar networks the WL correction presents a 
weaker divergence at zero magnetic field
$\Delta\tilde\sigma(\theta)\sim\ln\theta$. Therefore 
$\int_0^{2\pi}\D\theta\,\Delta\tilde\sigma(\theta)<\infty$ and the
harmonics reach  a finite value in the limit $L_\varphi\to\infty$.
Let us compute this value.
The singular behaviour 
$\Delta\tilde\sigma(\theta)\simeq\mathrm{Cste}+\frac{a}{2\pi}\ln|\theta|$ 
near zero flux is expected to dominate the harmonics behaviour
$
\Delta\tilde\sigma_n\sim
\int_0\D\theta\,\ln\theta\,\cos(n\theta)
$.
The typical scale over which $\ln\theta$ varies is $\theta\ln1/\theta$. 
If we define $\theta_c$ by $\theta_c\ln1/\theta_c=1/n$, we see that 
the integral is dominated by the interval $[0,\theta_c]$. 
We have $\theta_c\simeq1/(n\ln n)$, whence
$
\Delta\tilde\sigma_n\sim\int_0^{1/(n\ln n)}\D\theta\,\ln\theta\simeq-1/n
$. More precisely, we have obtained above~:
$
  \Delta\tilde\sigma_n \simeq
  -\frac{a}{4\pi n}
$ 
for $n\gg1$. 

\vspace{0.25cm}

\begin{figure}[htbp]
  \centering
  \includegraphics[scale=0.4]{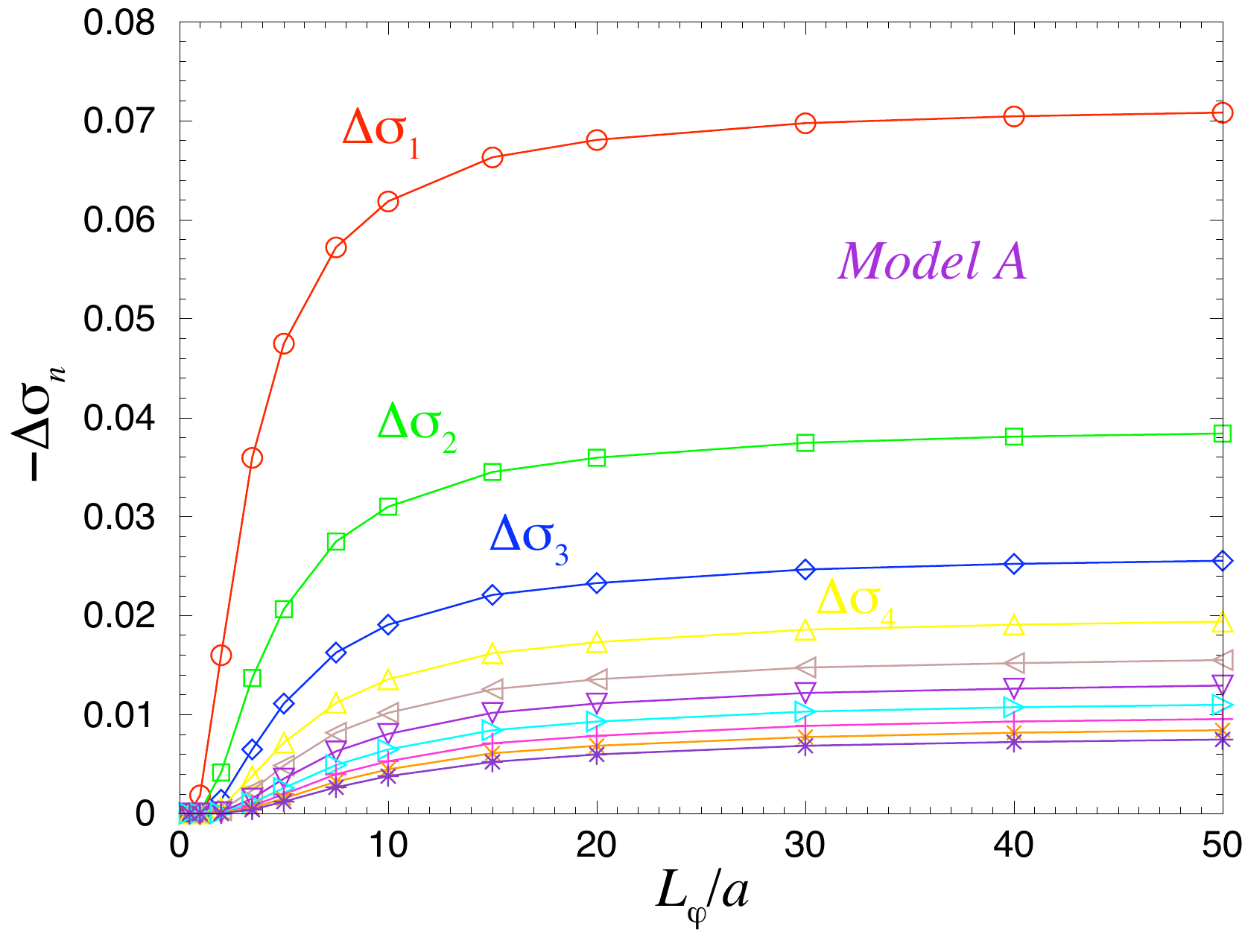} 
  \caption{\it Numerical calculation of ten first MC harmonics
    $\Delta\tilde\sigma_n$ for a square network, computed for $L_\varphi/a$
    between $0.5$ and $50$. }
  \label{fig:harm8}
\end{figure}

\noindent{\bf Numerical calculations.--}
The MC is computed numerically as a function of the reduced
flux~$\theta$ for rational fluxes $\theta=2\pi p/q$.
As recalled in appendix~\ref{app:DouRam} the computation of the MC 
is related to the study of the spectrum of
a tight binding Hamiltonian on a square lattice submitted to a
magnetic field, the so-called Hofstadter problem. For rational flux
$\theta=2\pi{}p/q$, this spectrum presents $q$ bands determined by the
polynomials $P_{p,q}(\varepsilon)$. For example, band edges correspond to
roots of $P_{p,q}(\varepsilon)=\pm4$.
For small $q$ (in practice we choose $\leq8$) the MC is computed by using
eq.~(\ref{DouRam86}). 
For  large $q$ (large number of Hofstadter bands), we use a more
efficient procedure and rather follow
Ref.~\cite{Mon89}~: we neglect the dispersion of Hofstadter 
bands, according to which eq.~(\ref{fctGreenHof}) reduces to 
$\frac1{N_xN_y}\,\tr{ \frac1{N(\gamma,\theta_{p,q})} }\simeq\:\frac1q
\sum_{r=1}^q\frac1{4\cosh(\sqrt\gamma\,a) + \bar\varepsilon_r}$,
where $\bar\varepsilon_r$ designates the position of the band.
The weak localization correction is represented on
figure~\ref{fig:wlsnBfield28} 
as a function of the reduced flux~$\theta$ for three values of the
ratio $L_\varphi/a$. As this latter increases, the
MC becomes sharper around zero flux, according to the above
discussion and the harmonic content becomes richer.
The MC is computed in this way for different values of the phase
coherence length ranging from $L_\varphi/a=0.5$ to $50$.
For each curve the first ten harmonics are extracted and plotted as a
function of $L_\varphi/a$ on figure~\ref{fig:harm8}.

\begin{figure}[htbp]
  \centering
  \includegraphics[scale=0.4]{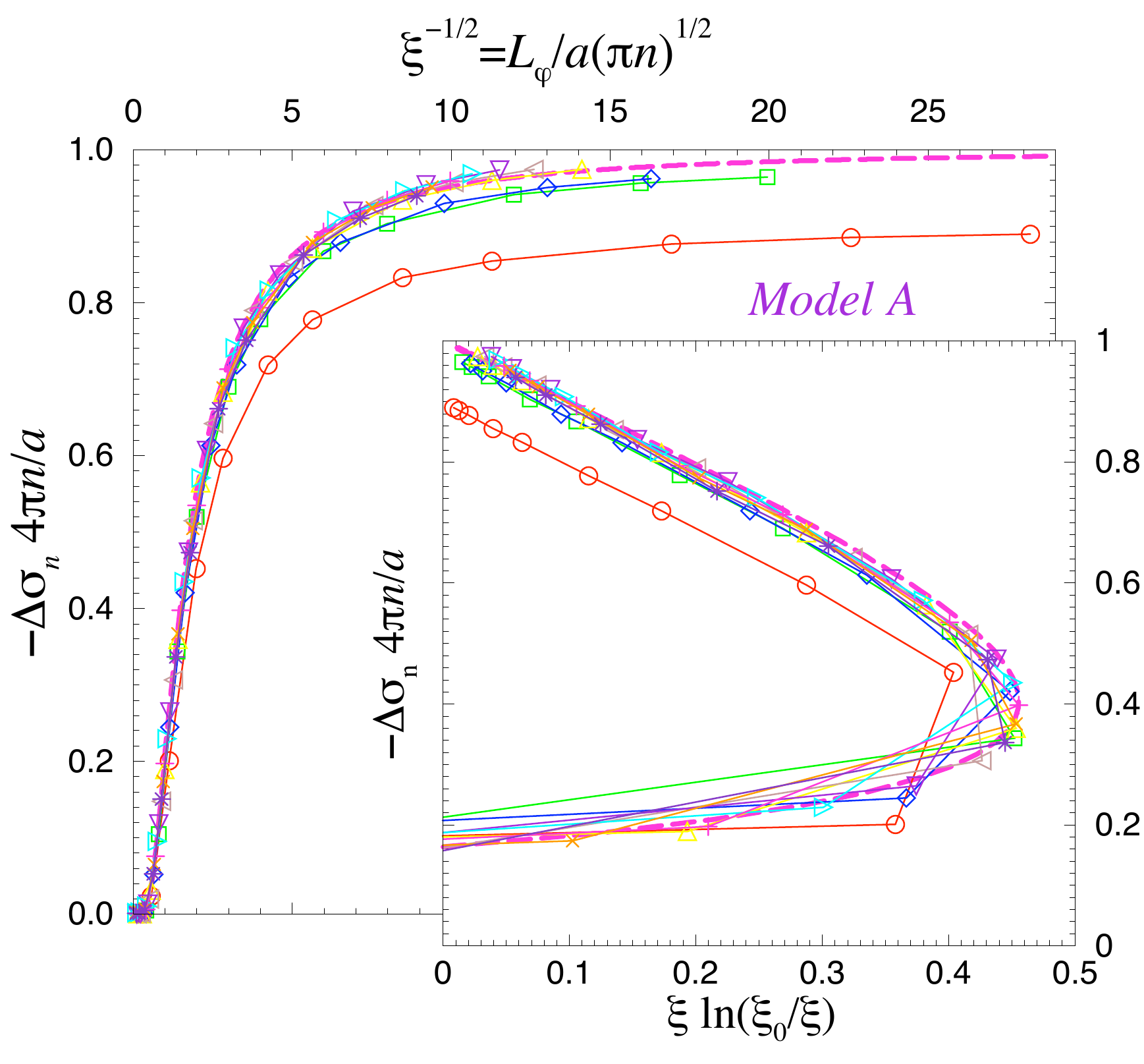} 
  \caption{{\it The harmonics represented on Fig.~\ref{fig:harm8}
    are plotted here as a function of the 
    variable~$\sqrt{\xi}=L_\varphi/a\sqrt{\pi n}$. The magenta dashed line
    corresponds to the continuum limit~:  
           function $\mathcal{F}_3(\xi)$. }
    Inset~: {\it Same functions as a function of $\xi\ln(\xi_0/\xi)$ with 
    $\xi_0\simeq1.239$. }
    }
  \label{fig:harm8b}
\end{figure}

In order to analyze the numerical results, we use the discussion of the
above paragraph on the continuum limit.
In the limit $L_\varphi\gg a$ we expect the scaling
$n_t\sim t/a^2$. On figure~\ref{fig:harm8b}, we plot
$\frac{4\pi{}n}{a}\Delta\tilde\sigma_n$ as a 
function of $1/\sqrt\xi=L_\varphi/a\sqrt{\pi{}n}$
(note that the 
scaling $n_t\sim t/a^2$ is only expected for $L_\varphi\gtrsim{a}$ when we
reach a ``two-dimensional limit''~; for $L_\varphi\lesssim{a}$ we rather
expect the scaling corresponding to the isolated ring $n_t\sim\sqrt{t}/a$).

After rescaling, all curves of figure~\ref{fig:harm8} collapse
onto each other  
as we can see on figure~\ref{fig:harm8b} (at least in the domain
$L_\varphi\gtrsim{a}$).  
Some  significant deviation from expression~\eqref{resloidesaires} 
occurs only for $n=1$.
In order to analyze the behaviour for largest $L_\varphi$ more precisely,
harmonics are re-plotted as funtions of the variable 
$\xi\ln(\xi_0/\xi)$ on the inset of figure~\ref{fig:harm8b}~:
we check the linear behaviour with this variable.
Surprisingly, 
the continuum limit can be considered as a very good approximation
already for $n\geq2$.

\vspace{0.5cm}

\noindent{\it Remark~: Brownian path/random walk.--}
We have shown that the distribution of the number $n\in\ZZ$ of cells enclosed
by a Brownian path in the square lattice is very close from the Levy
law describing the distribution of the algebraic area enclosed by a
planar Brownian motion (continuum limit) already for $n\geq2$.
It is interesting to point out that this remark also holds for the number
of cells enclosed by a {\it discrete} random walk jumping between the
different nodes of the square lattice~\cite{BelCamBarCla97,Des08}.

\subsection{Model B}

\subsubsection{The two-dimensional limit}
\label{sec:2dlimit}

\noindent{\bf The thin film.--}
Let us first recall some known results for
the plane (or thin film of thickness $b$)
\cite{AltAroKhm82,AltAro85,AleAltGer99}. In two dimensions the diffuson
presents a logarithmic behaviour. The function $W$ behaves in the same way, 
with a cutoff at small scales at the thermal length
$L_T$~\cite{footnote7,AltAroKhm82,AltAro85,footnote10}~:
$W(\vec{r},\vec{r}\,')=\frac1{2\pi}\ln(||\vec{r}-\vec{r}\,'||/L_T)$ for 
$||\vec{r}-\vec{r}\,'||\gtrsim{}L_T$.
Therefore the functional governing decoherence behaves as
\begin{equation}
  \label{eq60}
  \Gamma[\mathcal{C}_t]\, t\equiv\frac12\smean{\Phi_V[\mathcal{C}_t]^2}_V
  \sim
  \frac{2e^2T}{\sigma_0b}\,t\,\frac1{2\pi}\ln\frac{\sqrt{Dt}}{L_T}
  \:.
\end{equation}
We recognize the sheet resistance~\cite{footnote11}
$R_\square=1/(b\sigma_0)$ of the film 
of thickness $b$. 
The phase coherence (Nyquist) time is evaluated from
$\Gamma[\mathcal{C}_t]\,t\sim1$. 
We obtain the temperature dependence~\cite{AltAroKhm82,AltAro85,AleAltGer99}
\begin{equation}
  \label{NyquistFilm}
  \frac1{\tau_N^\mathrm{film}} = \frac{R_\square}{h/e^2}\,T\,
  \ln\left[\frac{h/e^2}{2R_\square}\right]
  \:,
\end{equation}
valid for $L_T\gg{}b$. This behaviour was observed experimentally 
for thin metallic film~\cite{EchThoGouBoz92} and
two-dimensional electron gas~\cite{EshEisKarPal06}.
We recall that 2d magnetoconductance is given
by~\cite{Ber84,AkkMon07,footnote12}~:
\begin{align}
  \label{eq:Bergman}
  \Delta\sigma^\mathrm{film}(\mathcal{B})
  &= \frac{2e^2}{h}\,\frac1{2\pi b}\nonumber\\
  &\times
  \left[
    \psi\!\left(\frac12+\frac{\tau_\mathcal{B}}{\tau_N^\mathrm{film}} \right)
    - \ln\left(\frac{\tau_\mathcal{B}}{\tau_e} \right)
  \right]
  +\mbox{cste}
\end{align}
where~$\tau_\mathcal{B}=\phi_0/(8\pi{D}\mathcal{B})$ and 
$\psi(z)$ is the Digamma function~\cite{footnote14}
(the additional factor $1/2$ in the Digamma function, compared to
\eqref{eq:SnContLim}, is explained in appendix~\ref{app:ConLimPlanNet}).
We may simply write
$\Delta\sigma\simeq-\frac{2e^2}{h}\,\frac1{2\pi{}b}\ln[\min{\tau_N^\mathrm{film}}{\tau_\mathcal{B}}/\tau_e]+$cste.
The small time cutoff $\tau_e$ in eq.~\eqref{eq:Bergman}
is introduced by hand to account for the fact
that the diffusion approximation only holds for
times~\cite{footnote13}~$t\gtrsim\tau_e$.

\vspace{0.25cm}

\noindent{\bf The square network.--}
For large time scale ($\tau_\varphi\sim{}t\gg{a}^2$) and small
magnetic fields (such that $\phi\ll\phi_0$) the result for the network should
coincide with the one for a plane.
In this case
the function $W$ entering the decoherence rate is
$W(x,x')\simeq\frac{a}{2\pi}\ln(||x-x'||/a)$ where $||x-x'||$ is the distance
between the two points of the network. The logarithmic behaviour is now cut
off naturally at the scale $a$. 
Because the function $W$ presents a smooth logarithmic behaviour, 
we extract the relevant time scale (phase coherence time) by following
the same lines as for the plane.
We write
\begin{align}
  \Gamma[\mathcal{C}_t]\, t
  &\equiv\frac12\smean{\Phi_V[\mathcal{C}_t]^2}_V
  \nonumber\\
  &\sim
  \frac{2D}{L_N^3}\,t\,\frac{a}{2\pi}\ln\frac{\sqrt{Dt}}{a}
  =\frac{2e^2T}{\sigma_0\sw}\,t\,\frac{a}{2\pi}\ln\frac{\sqrt{Dt}}{a}
\end{align}
where $\sw=wb$ is the section of the wires of width $w$ and thickness $b$ 
(figure~\ref{fig:sn}).
From this expression we extract a time scale
reminiscent of eq.~(\ref{NyquistFilm}) for the film~:
\begin{equation}
  \label{NyquistNetTau}
  \boxed{
  \frac1{\tau^\mathrm{net}_N} = 
  \frac{R_\square^\mathrm{net}}{h/e^2}\,T\,
  \ln\left[\frac{L_T^2}{a^2}\frac{h/e^2}{2R_\square^\mathrm{net}}\right]
  }
  \:.
\end{equation}
This result is valid for $L_T\lesssim{}a$. 
In the opposite limit $L_T\gtrsim{}a$, the cutoff in
the function $W$ should rather be \cite{footnote7} $L_T$ therefore
$1/\tau^\mathrm{net}_N=\frac{R_\square^\mathrm{net}}{h/e^2}\,T\,\ln\big[\frac{h/e^2}{2R_\square^\mathrm{net}}\big]$.
However this latter regime seems less relevant from the experimental
point of view~\cite{footnote4}.
The sheet resistance of the network is
\begin{equation}
  R_\square^\mathrm{net} = \frac{a}{wb\sigma_0}=\frac{a}{\sw\sigma_0} 
  = R_\square \frac{a}{w}
  \:.
\end{equation}
This characteristic time is reduced by a factor $w/a$, compared to the
Nyquist time (\ref{NyquistFilm}) obtained for a film of same thickness~:
$\tau^\mathrm{net}_N\sim\frac{w}{a}\tau^\mathrm{film}_N$.
We can also
introduce a Nyquist length for the network
$L_N^\mathrm{net}=\sqrt{D\tau^\mathrm{net}_N}$, related to the Nyquist length
of the wire $L_N$ by~\cite{footnote15}~:
\begin{equation}
  \label{NyquistNet}
  \boxed{
  \frac1{L_N^\mathrm{net}} 
  = \frac1{L_T} \,\sqrt{ \frac{R_\square^\mathrm{net}}{h/e^2}\, \ln\left[\frac{L_T^2}{a^2}\frac{h/e^2}{2R_\square^\mathrm{net}}\right] }
  = \sqrt{ \frac{3\,a}{2\pi L_N^3} \, \ln(L_N/a) } 
  }
  \:.
\end{equation}
We expect that the MC presents the logarithmic behaviour
$\Delta\tilde\sigma\simeq\frac{a}{2\pi}[\ln\theta+C_\mathrm{sn}]$ which is cut
off at very low  magnetic field~:
$\Delta\tilde\sigma\simeq-\frac{a}{\pi}\ln[\min{L_N^\mathrm{net}}{L_\mathcal{B}}/a]$,
where $L_\mathcal{B}=\sqrt{\phi_0/(4\pi\mathcal{B})}$ is the 2d cutoff.

\subsubsection{MC Harmonics}

In the network, the diffuson behaves logarithmically at large distances
$P_d(x,x')\simeq-\frac{a}{2\pi}\ln(||x-x'||/a)$. Therefore we expect that 
the relaxation of phase coherence is controlled by 
\begin{align}
  \frac12\smean{\Phi_V[\mathcal{C}_t]^2}_{V,\mathcal{C}_t}
  \sim \frac{a}{\pi L_N^3}\, t\,\ln(\sqrt{t}/a)
  \to \frac{3a}{2\pi L_N^3}\, t\,\ln(L_N/a)
  \:.
\end{align}
As for the plane we use the fact that the functional describing
decoherence weakly depends on trajectories since $W(x,x')\sim\ln||x-x'||$.
This suggests that the  result for
{\it model B} is given by performing, in the result for {\it model A},
the substitution 
\begin{equation}
  \frac{1}{L_\varphi^2}\to\frac{3a}{2\pi L_N^3}\ln(L_N/a)
  \hspace{0.5cm}\mbox{ {\it i.e.} }\hspace{0.5cm}
  L_\varphi \to L_N^\mathrm{net}
  \:,
\end{equation}
where the Nyquist length for the network is given by (\ref{NyquistNet}).
Using \eqref{eq:SNlimit1}, we get for the tail~:
\begin{align}
  \Delta\tilde\sigma_n  &
  \sim\left(L_N^\mathrm{net}\right)^{-1/2}\EXP{-\sqrt{8\pi n}\,a/L_N^\mathrm{net}}
  \\
  &  \sim
  L_N^{-3/4}\ln^{1/4}\big(\frac{L_N}{a}\big)\,\EXP{
    - \sqrt{n}\,(\frac{a}{L_N})^{3/2}
      \ln^{1/2}(\frac{L_N}{a})
   }
  \nonumber\\
  &\sim
  (T\ln1/T)^{1/4}\:\EXP{-n^{1/2}a^{3/2}(T\ln1/T)^{1/2}}
  \:.
\end{align}
A similar substitution in eq.~\eqref{eq:SNlimit2} leads to
\begin{equation}
  \Delta\tilde\sigma_n \sim -\frac{a}{4\pi n}
  \left[
     1 -  \frac{9na^3}{2L_N^3}\ln^2(L_N/a)
  \right]
\end{equation}
to describe the saturation of the harmonics at large $L_N/a$ (small
temperature).

We insist that since $\int_0\D\theta\,\Delta\sigma(\theta)<\infty$ the
harmonics reach a limit for $L_\varphi\to\infty$ (or $L_N\to\infty$)
which is independent on the decoherence mechanism.
In other terms the magnetoconductance curve $\Delta\tilde\sigma(\theta)$
reaches a limit apart in a very narrow region of width
$\delta\theta\sim{}(a/L_\varphi)^2$ around zero flux
(figure~\ref{fig:wlsnBfield28}). 


\section{The hollow cylinder}
\label{sec:cyl}

We have noticed that a network made of a large number of rings realizes
disorder averaging. Another natural way to realize this
averaging is to consider a long hollow cylinder of perimeter $L$ (longer than
$L_\varphi$) submitted to a magnetic field along its
axis~\cite{AltAroSpi81,ShaSha81,AltAroSpiShaSha82,AroSha87}. 
We study below how the original result of AAS~\cite{AltAroSpi81}
obtained within {\it model A} is
modified when decoherence is dominated by electron-electron interaction. 
In this section, it is natural to define the reduced dimensionless
conductivity $\tilde\sigma$ 
as~$\sigma=\frac{2e^2}{h\,b}\tilde\sigma$, where $b$ is the thickness
of the metallic film. 

\begin{figure}[htbp]
\begin{center}
\includegraphics[scale=1]{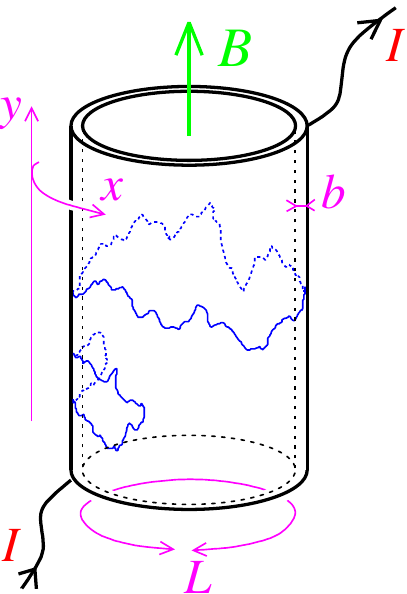}
\caption{\it A metallic film is deposited on an insulating wire.
  This allows to study quantum transport in a long hollow cylinder
  \cite{ShaSha81}. Two diffusive trajectories are represented 
  with winding $n=0$ and $n=1$.}
\label{fig:cyl}
\end{center}
\end{figure}

\subsection{Model A}

Let us first recall the well-known result for the weak localization
correction computed within {\it model A}~\cite{AltAroSpi81,AkkMon07}. 
We denote by $y\in\RR$ the coordinate along the axis of the cylinder and
$x\in[0,L]$ the coordinate in the perpendicular direction. 
The WL correction is written
as a path integral over Brownian paths
$\vec{r}(\tau)=(x(\tau),y(\tau))$ in the cylinder, where $x(\tau)$
describes a Brownian path on the circle and $y(\tau)$ on $\RR$
(figure~\ref{fig:cyl})~:
\begin{align}
  &\Delta\tilde\sigma_n^\mathrm{AAS} = -2\,
  \int_0^\infty \D{t}\, \EXP{-t/\tau_\varphi}
  \nonumber\\
  &\hspace{1.5cm}\times\int_{\vec{r}(0)=\vec{r}}^{\vec{r}(t)=\vec{r}}
  \hspace{-0.5cm}\mathcal{D}\vec{r}(\tau)\,
   \delta_{n,\mathcal{N}[x(\tau)]}\,
   \EXP{ -\int_0^t\D\tau\, \frac14\dot{\vec{r}}\,^2 }
  \\
  \label{eq:AAScyl}
 &=-2\,
  \int_0^\infty \D{t}\, \EXP{-t/\tau_\varphi}
  \frac{\EXP{-(nL)^2/4t}}{4\pi t}
  =-\frac1\pi \: K_0\!\left(\frac{nL}{L_\varphi}\right)
\end{align}
where $K_0(x)$ is a modified Bessel function.
Therefore~:
\begin{align}
  \label{AAScyl1}
  \Delta\tilde\sigma_n^\mathrm{AAS} 
  &\simeq -\sqrt{\frac{L_\varphi}{2\pi nL}}\, \EXP{-nL/L_\varphi}
  \hspace{0.25cm}\mbox{ for } L_\varphi\ll|nL| \\
  \label{AAScyl2}
  &\simeq -\frac1\pi \:\ln\left({L_\varphi}/nL\right)
  \hspace{0.75cm}\mbox{ for } |nL| \ll L_\varphi
  \:.
\end{align}
These results are very similar to the one obtained for the chain of
rings
(\ref{necklimit2},\ref{necklimit1},\ref{eq:ApproxHarmChain}).
This is due to the similar winding properties, what was already noticed
after eq.~(\ref{PnChain}).

\subsection{Model B : e-e interaction}

We have now to consider the path integral
\begin{align}
  \label{eq:PathIntCyl}
  \Delta\tilde\sigma_n &= -2\,
  \int_0^\infty \D{t}\,\\\nonumber
  &\times\int_{\vec{r}(0)=0}^{\vec{r}(t)=0}
  \hspace{-0.5cm}\mathcal{D}\vec{r}(\tau)\,
   \delta_{n,\mathcal{N}[x(\tau)]}\,
   \EXP{ -\int_0^t\D\tau\, 
         \big[ \frac14\dot{\vec{r}}\,^2 + 2e^2 T R_\square W(\vec{r}(\tau),0) \big]
       }
\end{align}
where we have used translation invariance along the two perpendicular
directions in order to deal with a path integral with action local in
time, in a similar way as for the ring~:
eq.~(\ref{Zenicerelation}). 

\subsubsection{The function $W$\label{sec:fctWcyl}}

The cylinder is
translation invariant in the two directions, therefore we may write
$W(\vec{r},\vec{r}\,')=W(\vec{r}-\vec{r}\,',0)$ with 
$W(\vec{r},0)=P_d(\vec{r}_c,0)-P_d(\vec{r},0)$, where 
$\vec{r}_c$ is a  short distance cutoff of order $L_T$ (we will see
that the direction of the vector $\vec{r}_c$ plays no role) 
\cite{footnote7}.

In order to avoid the divergent contribution of the zero mode of the
Laplace operator, we start by considering the solution of
$(\gamma-\Delta)\,P=\delta$~: 
\begin{align}
  P(\vec{r},0)
  &= \frac1L\sum_{n\in\ZZ} \int_{-\infty}^{+\infty}\frac{\D k}{2\pi}\,
  \frac{\EXP{2\I\pi n x/L+\I ky}}{\gamma + (\frac{2\pi n}{L})^2+k^2}
  \\
  &= \frac1{2L}\sum_{n\in\ZZ} 
  \frac1{\sqrt{\gamma + (\frac{2\pi n}{L})^2}}
  \EXP{2\I\pi n x/L-\sqrt{\gamma + (\frac{2\pi n}{L})^2}\,|y|}
  \:.
\end{align}
Next we take the limit $\gamma\to0$ in 
\begin{align}
  \label{eq75}
  W(\vec{r},0)
  &= \lim_{\gamma\to0}\left[ P(\vec{r}_c,0)-P(\vec{r},0)\right]
  \\
  &= \frac{|y|}{2L} 
  + \frac1{2\pi}\sum_{n=1}^\infty
  \frac1n
  \left(
    \cos\frac{2\pi nx_c}{L}\:\EXP{-\frac{2\pi ny_c}{L}}
   \right.
  \nonumber\\
  &\hspace{2.5cm}
  \left.  -\cos\frac{2\pi nx}{L}\:\EXP{-\frac{2\pi n|y|}{L}}
  \right) 
  \:.
\end{align}
We finally obtain
\begin{align}
 \label{fctWcyl}
 \boxed{
  W(\vec{r},0) = \frac{|y|}{2L} + \frac1{2\pi}
  \re\Big[
  \ln\left( 
    \frac{ 1-\EXP{-\frac{2\pi}{L}(\I x+|y|) } }{2\pi L_T/L}
  \right)
  \Big]
 }
\end{align}
where we used that $||\vec{r}_c||=L_T\ll{L}$. We can check that this
expression reproduces known 
results in two limits~: for $|y|\gg{L}$, we recover the 1d form
$W\simeq|y|/(2L)$. For $|y|\ll{L}$ we obtain the 2d result
$W\simeq\frac1{2\pi}\ln(||\vec{r}||/L_T)$.

\subsubsection{Harmonics}

The first term of eq.~\eqref{fctWcyl} originates from the 1d motion
along the cylinder. To this 1d motion, we can associate a 1d Nyquist
time similar to the one obtained for the wire, eq.~\eqref{nyquist},
\begin{align}
  \label{TauCyl1}
  \boxed{
    \frac1{\tau_{N,1d}} 
    = \left( \frac{e^2R_\square\sqrt{D}T}{L} \right)^{2/3} 
    = \left( \frac{e^2\sqrt{D}T}{\sigma_0bL} \right)^{2/3} 
  }
\end{align}
that coincides with (\ref{nyquist}) in which the section is taken
as~$\sw=bL$. 

We consider first the high temperature limit
$L\gg{}L_{N,1d}=\sqrt{D\tau_{N,1d}}=(\frac{\sigma_0DbL}{e^2T})^{1/3}$.
We remark that for the harmonic $n=0$, trajectories very
unlikely wind around the cylinder and we can use
$W\simeq\frac1{2\pi}\ln(||\vec{r}||/L_T)$ for $y\ll{}L$. Therefore the 
calculation of the path integral \eqref{eq:PathIntCyl} corresponds to
the one for the film, 
$\Delta\sigma_0\simeq\Delta\sigma^\mathrm{film}$,
with the time scale~(\ref{NyquistFilm}).

\begin{figure}[!ht]
  \centering
  \includegraphics[scale=0.7]{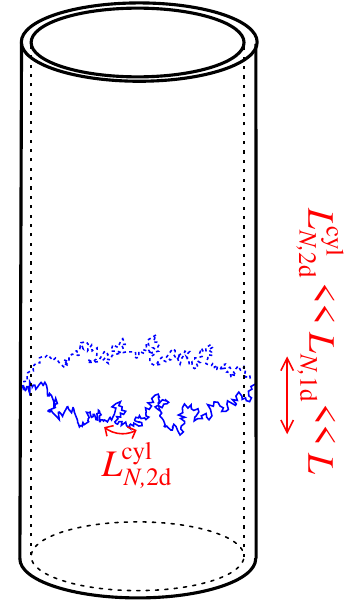}
  \hspace{0.5cm}
  \includegraphics[scale=0.7]{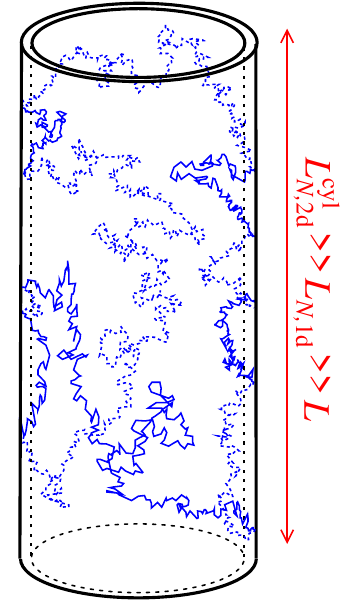}
  \caption{\it Trajectories contributing to first harmonic for ``high
    temperature'' (left) and ``low temperature'' (right).}
  \label{fig:cyl_sketch1}
\end{figure}

Next we consider non zero harmonics $n\neq0$.
In this case the trajectories have a small extension along
the wire $|y|\lesssim{}L_{N,1d}\ll{}L$ and we can neglect the $|y|$ in the
exponential in eq.~(\ref{fctWcyl}) 
(see figure~\ref{fig:cyl_sketch1}). Therefore we perform the substitution~: 
\begin{equation}
  W(\vec{r},0) \to \frac{|y|}{2L} + \widetilde W(x)
\end{equation}
with
\begin{equation}
  \widetilde W(x) = \frac1{2\pi}
  \ln\frac{ |\sin(\pi x/L)| }{\pi L_T/L}
\end{equation}
This approximation allows us to factorize the path integral as~:
\begin{widetext}
\begin{align}
   \Delta\tilde\sigma_n \simeq -2\,
  \int_0^\infty \D{t}\,
   \int_{x(0)=0}^{x(t)=0}\mathcal{D}x(\tau)\,
   \delta_{n,\mathcal{N}[x(\tau)]}\,   
   \EXP{ -\int_0^t\D\tau\, 
         \big[ \frac14\dot{x}^2 
               + 2e^2 T R_\square\widetilde W(x) \big]
       }
   \int_{y(0)=0}^{y(t)=0}\mathcal{D}y(\tau)\,
   \EXP{ -\int_0^t\D\tau\, 
         \big[ \frac14\dot{y}^2 + \frac{|y|}{L_{N,1d}^3} \big]
       }
  \:.
\end{align}
\end{widetext}
The first path integral runs over trajectories encircling the
cylinder. Therefore we can replace $\widetilde{W}(x)$ by its average
$\int_0^L\frac{\D{x}}{L}\widetilde{W}(x)$. 
This approximation is justified by the fact that $\widetilde{W}$ has
only a logarithmic dependence. 
This simplify the calculation by substituting the functional by a constant~:
\begin{equation}
  2e^2 T R_\square\int_0^t\D\tau\,\widetilde{W}(x(\tau)) 
  \to \frac{t}{\tau_{N,2d}^\mathrm{cyl}}
  \:,
\end{equation}
where we have introduced the time scale
\begin{align}
   \label{TauCyl2}
 \boxed{
    \frac1{\tau_{N,2d}^\mathrm{cyl}} 
   = \frac{R_\square}{h/e^2}\,T\,\ln\left(\frac{L^2}{L_T^2}\right)
  }
  \:.
\end{align}
This time is reminiscent of the Nyquist time for the film,
eq.~\eqref{NyquistFilm}, but the two times differ by the argument of
the logarithm.

The second path integral precisely coincides with the one for a wire~:
$\frac1{\sqrt{4\pi t}}\smean{\EXP{\I\Phi}}_{V,\mathcal{C}_t}$ given by
(\ref{dephasingwire}). 
Finally
\begin{align}
  \Delta\tilde\sigma_n &\simeq -2\,
  \int_0^\infty \D{t}\,
  \frac1{\sqrt{4\pi{t}}}\,\EXP{-\frac{(nL)^2}{4t}} 
  \EXP{-t/\tau_{N,2d}^\mathrm{cyl}}\:
  \nonumber\\
   & 
    \times\frac1{\sqrt{4\pi t}}
  \sqrt{\frac{\pi t}{\tau_{N,1d}}} 
  \sum_{m=1}^\infty\frac1{|u_m|}\EXP{-|u_m|t/\tau_{N,1d}} 
 \:,
\end{align}
which leads to the series (for $n\neq0$)~:
\begin{align}
  \label{eq:REScyl}
  \boxed{
    \Delta\tilde\sigma_n  \simeq -\frac{1}{2\sqrt{\tau_{N,1d}}}
    \sum_{m=1}^\infty\frac{\sqrt{\tau_m}}{|u_m|}
    \EXP{-nL/\sqrt{\tau_m}}
  }
\end{align}
where the times $\tau_m$ are defined as
\begin{equation}
  \label{SetOfTimes} 
  \frac1{\tau_m} \eqdef 
  \frac{|u_m|}{ \tau_{N,1d} } + \frac1{ \tau_{N,2d}^\mathrm{cyl}}
  \:,
\end{equation}
(we recall that $u_m$'s are zeros of Airy function Ai$'$).

This expression assumes that $L\gg{}L_{N,1d}$.
We show that \eqref{eq:REScyl} it
 also  valid for the other regime $L\ll{}L_{N,1d}$~: in this
case the path integral 
runs over trajectories such that $|y|\gg{}L$ 
(see figure~\ref{fig:cyl_sketch1}), therefore, in \eqref{fctWcyl}, the
exponential damping suppresses the $x$ and $y$ dependence in the
logarithmic of $W$ what leads to
the same conclusion for the two regimes since  
$W\to\frac{|y|}{2L}+\frac1{2\pi}\ln(L/2\pi{}L_T)$.

In order to analyze the two limiting cases into more details it is
convenient to relate the two times as~:
\begin{equation}
  \frac{ \tau_{N,1d} }{ \tau_{N,2d}^\mathrm{cyl} } = \frac1\pi 
  \, \left(\frac{L}{L_{N,1d}}\right)\,
  \ln \left(\frac{L}{L_T}\right)
  \:.
\end{equation}
Since the two lengths ${L}_{N,2d}^\mathrm{cyl}$ and $L_{N,1d}$ are related,
we just have to consider two different regimes.
As it is clear from eq.~\eqref{SetOfTimes}, the harmonics are always
controlled by the smallest scale among ${L}_{N,2d}^\mathrm{cyl}$ and~$L_{N,1d}$.

\vspace{0.25cm}

\noindent$\bullet$ {\it High temperature} $L_{N,1d}\ll{}L$ (then
${L}_{N,2d}^\mathrm{cyl}\ll{}L_{N,1d}$).--  
The WL correction is dominated by non winding trajectories in this case
\begin{equation}
  \Delta\tilde\sigma  \simeq \Delta\tilde\sigma_0  \simeq
  -\frac1{2\pi} \,
   \ln\left( \frac{\tau_{N}^\mathrm{film}}{\tau_e} \right) + \mbox{cste}
\end{equation}
involves the 2d Nyquist time  (\ref{NyquistFilm}).
Considering the oscillating part of the MC, only the first term of the series 
dominates. The harmonics are governed by the smallest length among
${L}_{N,2d}^\mathrm{cyl}$ and~$L_{N,1d}$~:
\begin{equation}
  \boxed{
  \Delta\tilde\sigma_n  \simeq -\frac{ {L}_{N,2d}^\mathrm{cyl} }{ 2|u_1|\,L_{N,1d} }
  \EXP{-n L/L_{N,2d}^\mathrm{cyl}}
  }
  \sim \EXP{-n\,L\, (T\ln T)^{1/2}}
\end{equation}
for $L_{N,1d} \ll L$.
Note that this result is reminiscent of the result (\ref{LMTM}) for a ring
$\Delta\tilde\sigma_n^\mathrm{ring}\sim\EXP{-nL^{3/2}T^{1/2}}$~: up to some
logarithmic correction it presents a similar $T^{1/2}$ in the exponential for
the similar reason (related to potential fluctuations seen by winding
trajectories). However the $L$ dependence differs from the one of the
ring.

\vspace{0.25cm}

\noindent$\bullet$ {\it Low temperature} $L\ll{}L_{N,1d}$ (then
$L_{N,1d}\ll{L}_{N,2d}^\mathrm{cyl}$).--
The harmonics involve $L_{N,1d}$, the smallest length among
${L}_{N,2d}^\mathrm{cyl}$ and $L_{N,1d}$. Eq.~\eqref{eq:REScyl}
coincides with the one obtained for the chain of rings
\eqref{eq:RESchain}~:
\begin{equation}
  \label{eq:6}
  \Delta\tilde\sigma_n\simeq-2\,F_2(nL/L_{N,1d})
  \:.
\end{equation}
For intermediate coherence length,  $L \ll L_{N,1d} \ll nL$,
eq.~\eqref{eq:REScyl} gives  
\begin{equation}
  \boxed{
   \Delta\tilde\sigma_n  \simeq -\frac{1}{|u_1|^{3/2}}
  \EXP{-|u_1|^{1/2}nL/L_{N,1d}}
  }\sim\EXP{-nLT^{1/3}}
  \:.
\end{equation}

For largest phase coherence length $nL \ll L_{N,1d}$,  
we may use the form derived in section~\ref{sec:chain}
\begin{equation}
  \boxed{
  \Delta\tilde\sigma_n  \simeq -\frac1\pi\left[
  \ln\left( \frac{L_{N,1d}}{nL} \right) + C_\mathrm{cyl}
  \right]
  }
\end{equation}
where the constant, $C_\mathrm{cyl}\simeq0.51$, was introduced in
section~\ref{sec:chain}. 

\vspace{0.25cm}

\noindent{\bf Discussion.--}
It is worth emphasizing the similarity between the results for the
cylinder and  for the networks.

In section~\ref{sec:wr} we have seen that the MC harmonics
of a weakly coherent ring probe two length scales $L_N\propto{T}^{-1/3}$
or~$L_c\propto{T}^{-1/2}$. For the lowest temperatures the surrounding
network matters and another length scale emerges~: the MC of the
square network involves a unique time scale
$1/\tau_N^\mathrm{net}\sim{T}\ln{1/T}$, eq.~(\ref{NyquistNet}),
reminiscent of the 2d Nyquist time~(\ref{NyquistFilm}).

For a cylinder the MC also probes several time scales. At high
temperature the zero harmonic related to non-winding trajectories
probes the 2d Nyquist time
$\frac1{\tau_N^\mathrm{film}}=\frac{R_\square}{h/e^2}\,T\,\ln[{h/e^2}{2R_\square}]$
whereas the nonzero harmonics probe the time
$\frac1{\tau_{N,2d}^\mathrm{cyl}}=\frac{R_\square}{h/e^2}\,T\,\ln[{L^2}/{L_T^2}]$.
The main dependence of the
corresponding length ${L}_{N,2d}^\mathrm{cyl}\propto{}T^{-1/2}$ has the same
origin as for a single ring and reflects that winding trajectories
feel fluctuations of the potential over length scale given by the
perimeter (figure~\ref{fig:cyl_sketch1}, left).
For lower temperature, trajectories diffuse along the cylinder over
length scale much larger than the perimeter and the WL correction is
controlled by a unique length $L_{N,1d}$, corresponding to the 
usual 1d Nyquist time~$\tau_{N,1d}\propto{T}^{-2/3}$.



\section{Conclusion}

We have studied the weak localization correction in metallic networks
and in a hollow cylinder.
This study relies on a detailed analysis of the winding
properties of closed Brownian trajectories in these systems. 
We now summarize our results.

We first recall the behaviour of the probability to return to the
starting point after a time $t$ for trajectories conditioned to wind $n$
rings. In the short time limit $t\ll{}L^2$, we have
$\mathcal{P}_n(x,x;t)\simeq{}p_n\frac1{\sqrt{4\pi t}}\EXP{-(nL)^2/4t}$
where $p_n$ depends on the network~: $p_n=1$ for the isolated ring, 
$p_n=\big(\frac23\big)^{nN_a}$ for the ring connected to $N_a$ long
wires and $p_n=\frac{(2n-1)!!}{2^{n+1}n!}$ in the chain of rings.
For the square network, there is no close expression but a
systematic expansion may be found in
Ref.~\cite{FerAngRowGueBouTexMonMai04} with the trace
formula of Ref.~\cite{Rot83}.  

\begin{table}[htdp]
\begin{center}
\begin{tabular}{llll}
  Network & $\mathcal{P}(x,x;t)$ 
          & $Q_t(n)$
          & $q(x)$ \\[0.2cm]
\hline
  \diagram{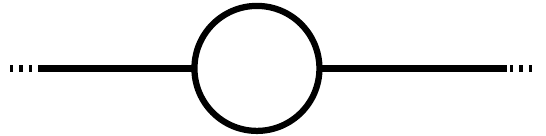}{0.3}{-0.25cm}
    & $\frac1{N_a\sqrt{\pi t}}$ 
    & $\frac{\sqrt{N_aL}}{(4\pi t)^{1/4}}\,q\big(\frac{n\sqrt{N_aL}}{(4\pi t)^{1/4}}\big)$
    & $\frac{\pi^{3/4}}{\sqrt{2}}\Psi((4\pi)^{1/4}x)$ \\[0.2cm]
  \diagram{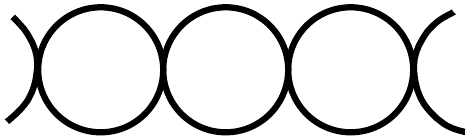}{0.3}{-0.25cm}
    & $\frac1{4\sqrt{\pi t}}$ 
    & $\frac{L}{\sqrt{2t}}\,q\big(\frac{nL}{\sqrt{2t}}\big)$ 
    & $\frac1{\sqrt{2\pi}}\EXP{-\frac12x^2}$ \\[0.2cm]
  \diagram{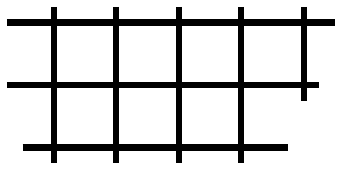}{0.3}{-0.25cm}
    & $\frac{a}{4\pi t}$
    & $\frac{2\sqrt3a^2}{t}\,q\big(\frac{n2\sqrt3a^2}{t}\big)$
    & $\frac{\pi}{4\sqrt3}\frac1{\cosh^2(\frac{\pi x}{2\sqrt3})}$\\
\hline
\end{tabular}
\end{center}
\caption{\it 
  Return probability and distribution
  $Q_t(n)=\frac{\mathcal{P}_n(x,x;t)}{\mathcal{P}(x,x;t)}$ of the winding
  number in the large time limit $t\gg{}L^2$. The function $\Psi(\xi)$ is
  defined in eq.~\eqref{eq:DefPsi} } 
  \label{table:QdeX}
\end{table}%

\begin{figure}[!ht]
  \centering
  \includegraphics[scale=0.45]{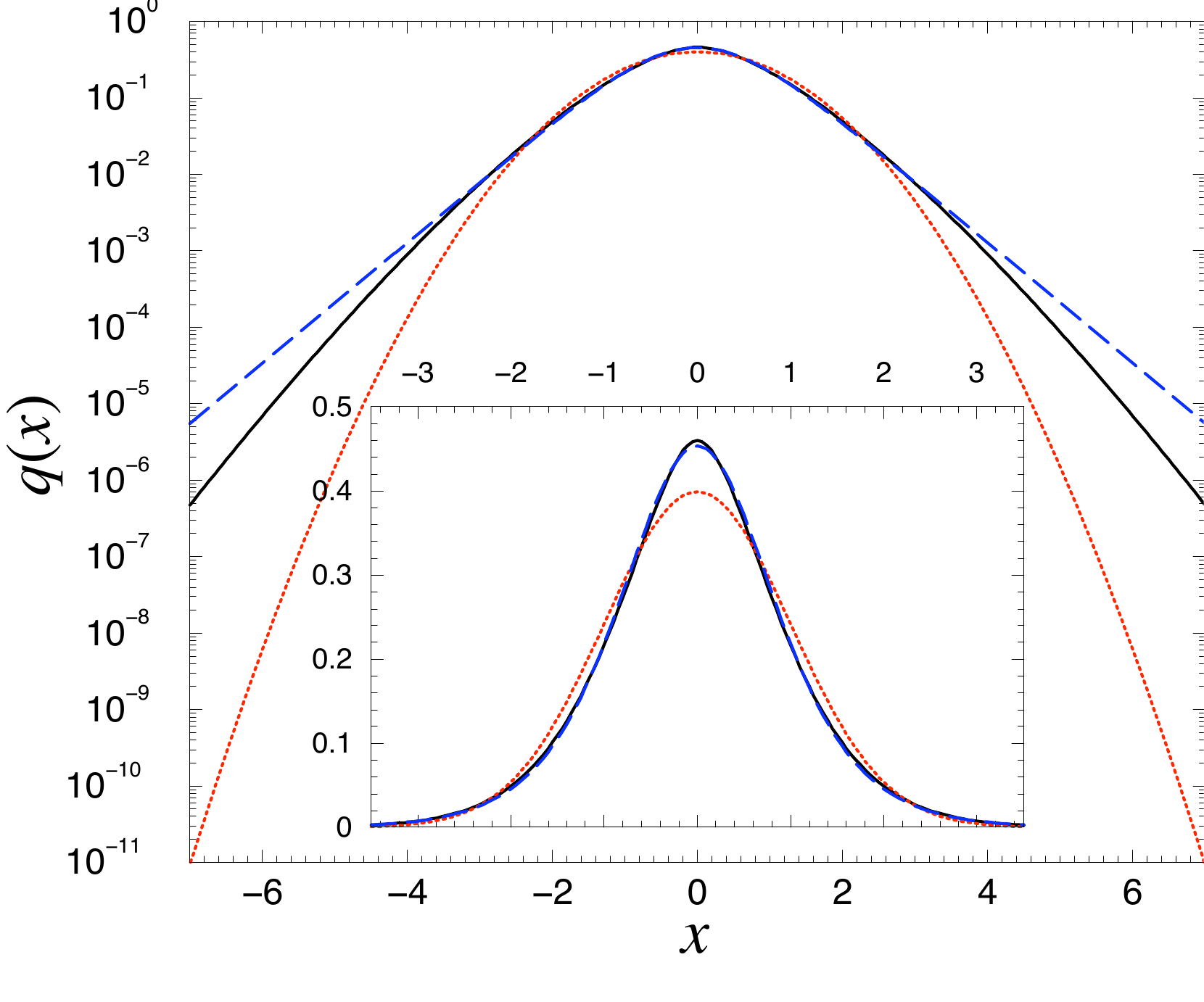}
  \caption{{\it Function $q(x)$ for the connected ring (black
    continuous line), the
    chain of rings (red dotted line) and square network (blue dashed
    line) in semilog scale.}
    Inset~: {\it Same functions in linear scale.} }
  \label{fig:QdeX}
\end{figure}

At large times $t\gg{}L^2$ the typical winding number scales as 
$n_t\sim{}(t/L^2)^\alpha$, where $\alpha$ is a network dependent
exponent. 
Introducing the return probability
$\mathcal{P}(x,x;t)=\sum_n\mathcal{P}_n(x,x;t)$, we may write the
winding probability as 
\begin{equation}
  \label{eq:Resume}
  \mathcal{P}_n(x,x;t) \simeq 
  \frac{\mathcal{P}(x,x;t)}{c_2\,(t/L^2)^\alpha}
  \:q\!\left(\frac{n}{c_2\,(t/L^2)^\alpha}\right)
  \:,
\end{equation}
where $\int\D x\,q(x)=1$. The dimensionless number $c_2$ ensures that 
$\int\D x\,x^2\,q(x)=1$.
Since $\mathcal{P}(x,x;t)\sim{}t^{-d/2}$ where $d$ is the
effective dimensionality of the network, we may also write
\begin{equation}
  \label{FormeResumee}
    \mathcal{P}_n(x,x;t) \sim \frac1{t^{\alpha+d/2}}\,
    q\!\left(\frac{n}{t^\alpha}\right)
\end{equation}
($L$ may be re-introduced by dimensional analysis).
The function $q(x)$ is
given for the various networks in the table~\ref{table:QdeX},
and represented on figure~\ref{fig:QdeX}.
Surprinsingly the functions for the connected ring and the plane are
very close~; they only differ in the wings when functions are
exponentially small.

\vspace{0.25cm}

We have analyzed in details the harmonics of the magnetoconductance
oscillations obtained when decoherence is described by a simple
exponential relaxation ({\it model A}). In the limit of large coherence length
compared to the perimeter of the rings, the scaling of the harmonics
can be easily understood from the Laplace transform 
$\Delta\tilde\sigma_n\sim\int_0^\infty\D{}t\,\mathcal{P}_n(x,x;t)\,\EXP{-t/L_\varphi^2}$. We
see that the time scale coincides with $t\sim{}L_\varphi^2$.
We deduce from \eqref{FormeResumee} that harmonics are of the form
\begin{equation}
  \label{FormeResumee2}
   \Delta\tilde\sigma_n\sim \frac1{L_\varphi^{d-2+2\alpha}}\:
    \Phi\!\left(\frac{n}{L_\varphi^{2\alpha}}\right)
  \:,
\end{equation}
where $\Phi(x)$ is a dimensionless network dependent function (the
perimeter $L$ is easily reintroduced by reminding that
$\Delta\tilde\sigma$ has dimension of a length).
We may then summarize for each geometry~:

\begin{itemize}

\item For the isolated ring ($d=0$, $\alpha=1/2$), the form of the harmonics
  $\Delta\tilde\sigma_n\sim{}L_\varphi\Phi_\mathrm{i.r}(nL/L_\varphi)$
  is related to the scaling $n_t\sim t^{1/2}$.

\item For the connected ring ($d=1$, $\alpha=1/4$)~: 
  $\Delta\tilde\sigma_n\sim\sqrt{L_\varphi L}\,\Phi_\mathrm{c.r}(n\sqrt{L/L_\varphi})$
  can be understood from $n_t\sim t^{1/4}$.
  
\item For the chain of rings  ($d=1$, $\alpha=1/2$)~:
  $\Delta\tilde\sigma_n\sim\Phi_\mathrm{chain}(nL/L_\varphi)$ reflects 
  $n_t\sim t^{1/2}$.

\item For the square network ($d=2$, $\alpha=1$) 
 $\Delta\tilde\sigma_n\sim \frac{L}{n}\,\tilde\Phi_\mathrm{s.n}(nL^2/L_\varphi^2)$
  originates from $n_t\sim t$ (here the harmonics were not written
  exactly under the form \eqref{FormeResumee2}, but in terms of
  the function $\tilde\Phi(x)=x\Phi(x)$ in order to emphasize that
  harmonics reach a finite value for $L_\varphi\to\infty$).

\end{itemize}
The precise behaviours for the harmonics $\Delta{}g_n$ are summarized in
table~\ref{table:summary} (we recall that
$\Delta{}g_n\sim\Delta\sigma_n$ apart for the connected ring for which
$\Delta{}g_n\sim\frac{L_\varphi}{l_a}\Delta\sigma_n$ where $l_a$ is
the length of the connecting wires).


\vspace{0.25cm}

\begin{widetext}
\begin{table}[htdp]
\begin{center}
\begin{tabular}{lllll}
    & {\it Model A} (exp. relax.)  &
    & {\it Model B} (e-e inter.) & \\
  \hline\hline
  & & {\it Regime} $L_\varphi\ll L$ & &
  \\
  \hline
  \diagram{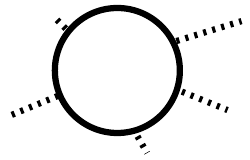}{0.5}{-0.5cm}
  &
  $L_\varphi\,\EXP{-nL/L_\varphi}$ & 
  & $L_N\,\EXP{-\kappa_1n(L/L_N)^{3/2}}$ & for $L_{N}\ll L$
  \\
  \diagram{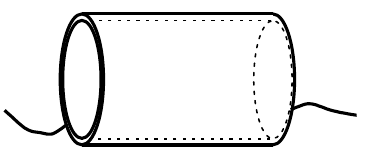}{0.45}{-0.25cm}
    & $L_\varphi^{1/2}\EXP{-nL/L_\varphi}$ & 
    & $\frac{L_{N,2d}^\mathrm{cyl}}{L_{N,1d}}\EXP{-nL/L_{N,2d}^\mathrm{cyl}}$
    & for $L_{N,2d}^\mathrm{cyl}\ll L_{N,1d}\ll L$
  \\
  \hline\hline
  & & {\it Regime}  $L_\varphi\gg L$ & &
  \\
  \hline
  \diagram{cartoon_ring_co}{0.45}{-0.25cm}
    & $L_\varphi^{3/2}$ & for $n^2\ll L_\varphi/L$ 
    & $L_N^{3/2}$        & for $n^2\ll L_N/L$ \\
    & $L_\varphi^{3/2}\EXP{-n\sqrt{2L/L_\varphi}}$ & for $n^2\gg L_\varphi/L$ 
    & $L_N^{5/4}\EXP{-\kappa_2n\sqrt{L/L_N}}$ & for $n^2\gg L_N/L$ \\
  \hline
  \diagram{cartoon_chain}{0.45}{-0.25cm}
    & $\ln(L_\varphi/nL)$ & for $n\ll L_\varphi/L$ 
    & $\ln(L_N/nL)$ & for $n\ll L_N/L$ \\
    & $L_\varphi^{1/2}\EXP{-nL/L_\varphi}$ & for $n\gg L_\varphi/L$
    & $\EXP{-\kappa_3nL/L_N}$ & for $n\gg L_N/L$ \\
  \hline
  \diagram{cartoon_sn}{0.5}{-0.5cm}
    & 
    $
    \frac{1}{n}\,\big[ 
      1 -\frac{\pi nL^2}{16L_\varphi^2}
      \ln(\frac{L_\varphi^2}{nL^2})
    \big]
    $ 
    &  for $\sqrt{n}\ll L_\varphi/L$
    & \ \hspace{1.5cm} idem  for & $L_\varphi\to L_N^\mathrm{net}$ \\[-0.1cm]
    & $L_\varphi^{-1/2}\EXP{-\sqrt{\frac\pi2n}\,L/L_\varphi}$ &  for $\sqrt{n}\gg L_\varphi/L$
    &  \ \hspace{1.5cm} idem  for & $L_\varphi\to L_N^\mathrm{net}$  \\
  \hline
  \diagram{cartoon_cyl}{0.45}{-0.5cm}
    & $\ln(L_\varphi/nL)$ & for $n\ll L_\varphi/L$ 
    & $\ln(L_{N,1d}/nL)$ & for $n\ll L_{N,1d}/L$  \\[-0.2cm]
    & $L_\varphi^{1/2}\EXP{-nL/L_\varphi}$ & for $n\gg L_\varphi/L$ 
    & $\EXP{-\kappa_4nL/L_{N,1d}}$
    &   for $n\gg L_{N,1d}/L$  \\
  \hline\hline
\end{tabular}
\end{center}
\caption{\it Harmonics of MC 
  $\Delta{g}_n/g$ for different networks.
  In the high temperature regime $L_\varphi\ll L$,  winding trajectories 
  cannot explore more than a single ring and harmonics do not depend
  on the network. Dimensionless constants are
  $\kappa_1=\pi^2/8\simeq1.234$, 
  $\kappa_2=\sqrt{2}|u_1|^{1/4}\simeq1.421$,
  $\kappa_3=2^{-1/3}|u_1|^{1/2}\simeq0.801$ and  
  $\kappa_4=|u_1|^{1/2}\simeq1.009$.
  The various Nyquist lengths are $L_N\sim{}T^{-1/3}$, 
  $L_N^\mathrm{net}\sim(T\ln1/T)^{-1/2}$ and  
  ${L}_{N,2d}^\mathrm{cyl}\sim(T\ln{T})^{-1/2}$.
  }
\label{table:summary}
\end{table}%
\end{widetext}

For each situation we have also discussed the effect of decoherence
due to electron-electron interaction ({\it model B}), 
the dominant phase breaking
mechanism at low temperature. As recalled at the begining of the
paper, this mechanism requires a refined description~: the
simple exponential decay of phase coherence is replaced by a
functional of the trajectories, eqs.~(\ref{localFDT},\ref{Decoherence}).
In networks of quasi-1d wires the decoherence due to e-e
interaction is controlled by the Nyquist length $L_N\propto{T}^{-1/3}$.

In the ``high temperature'' limit $L_N\ll{}L$ the fact that
trajectories with finite winding number and trajectories with winding
$n=0$ do probe different length scales is responsible for the
emergence of two length scales $L_N\propto{T}^{-1/3}$ and
$L_c\propto{T}^{-1/2}$ 
(or $L_N^\mathrm{film}\propto{T}^{-1/2}$ and
$L_{N,2d}^\mathrm{cyl}\propto(T\ln{}T)^{-1/2}$ for the cylinder). 
The {\it models A \& B} give different dependences in the phase coherence
length~:
$\Delta\tilde\sigma^{(A)}\sim-L_\varphi\EXP{-nL/L_\varphi}$
and 
$\Delta\tilde\sigma^{(B)}\sim-L_N\EXP{-n(L/L_N)^{3/2}}$
The exponential decay of harmonics is almost independent on the
network. 

\begin{figure}[!ht]
  \centering
  \includegraphics[scale=0.8]{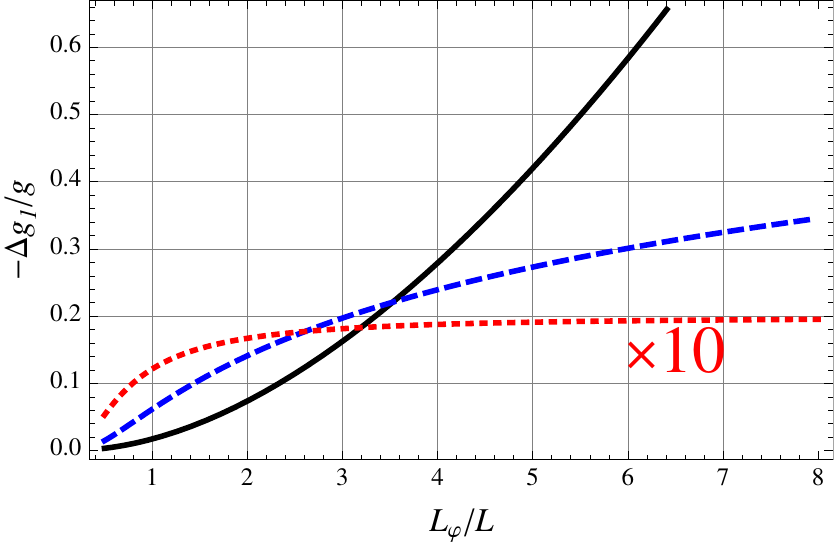}
  \\[0.5cm]
  \includegraphics[scale=0.8]{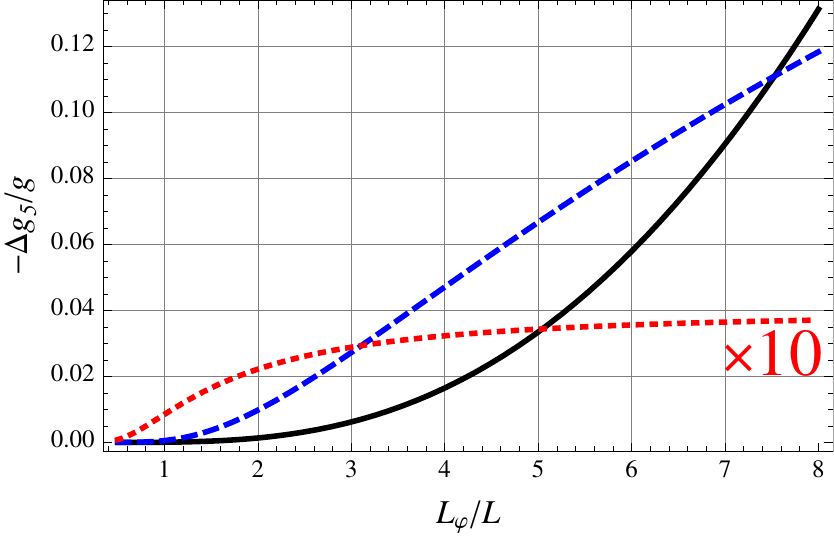}
  \caption{\it Comparison between harmonics $\Delta{}g_n/g$ for 
    different networks in the low temperature regime 
   $L_\varphi\gtrsim L$ with $n=1$ and $n=5$. 
   For the connected ring (continuous black line) we choose
   $l_a/L=10$. 
   The blue dashed line corresponds to the chain of rings.
   The result for the square network (dotted red line) has been multiplied
   by a factor 
   $10$ for visibility.}
  \label{fig:compnet}
\end{figure}

In the ``low temperature'' limit  $L_N\gg{}L$, all trajectories probe
the same typical scale, irrespectively of the winding. However this
length scale depends on the geometry~:
$L_N\propto{}T^{-1/3}$ for the chains of rings and the hollow cylinder, and 
$L_N^\mathrm{net}=(\frac{2\pi{}L_N^3}{3a})^{1/2}\ln^{-1/2}(L_N/a)\sim(T\ln1/T)^{-1/2}$
for the square network.
As a function of the phase coherence length, {\it models A \& B} predict
harmonics of similar form strongly network dependent. 
We have compared harmonics as a function of the phase coherence length for
the different networks on figure~\ref{fig:compnet} (for {\it model~A}).

All results are summarized in table~\ref{table:summary}.
We have plotted the WL correction to conductances for the three
different networks on figure~\ref{fig:compnet}.

\vspace{0.25cm}

An experimental verification of these predictions would be interesting
and would confirm our understanding of decoherence due to
electron-electron  interaction in complex geometries.
In particular an interesting and clear experimental test would be to
compare the MC harmonics for the chain of rings for independent rings and
coherent rings (networks of figure~\ref{fig:chains}) in the ``low
temperature'' limit $L_N\gg L$.
The experimental  analysis of the MC harmonics
for the square network in this limit seems more difficult due to the
fact that harmonics reach a value independent on the phase
breaking mechanism. Therefore, contrarily to the chains of
rings, the MC harmonics for the square network are less sensitive to
the model of  decoherence for large phase coherence length.

\section*{Acknowledgements}

We thank
Christopher B\"auerle, H\'el\`ene Bouchiat, Markus B\"uttiker, Richard
Deblock, Jean Desbois, Meydi Ferrier, Sophie Gu\'eron, Alberto Rosso
and Laurent Saminadayar for stimulating discussions.


\begin{appendix}

\section{A useful property of winding Brownian trajectories\label{app:winding}}

The difficulty for computing the path integral \eqref{eq:CQ}
lies in the time nonlocality of the action. In this appendix we show
how it is possible to get rid of time nonlocality in certain cases, as
explained in Refs.~\cite{ComDesTex05,TexMon05b}. 
For that purpose we demonstrate the identity 
\begin{align}
  \label{Zenicerelation}
  &\int_{x(0)=x}^{x(t)=x}\hspace{-0.5cm}\mathcal{D}x\,
  \delta_{n,\mathcal{N}[x]}\,
  \EXP{-\frac12\int_0^t\D\tau\,\dot
    x^2-\int_0^t\D\tau\,V(x(\tau)-x(t-\tau))}
  \nonumber\\
  &=  \int_{x(0)=0}^{x(t)=0}\hspace{-0.5cm}\mathcal{D}x\,
   \delta_{n,\mathcal{N}[x]}\,
  \EXP{-\frac12\int_0^t\D\tau\,\dot x^2-\int_0^t\D\tau\,V(x(\tau))}
\end{align}
where $x(\tau)$ is a Brownian path on the circle (here identified with
the interval $[0,1]$). The identity 
is valid for any symmetric and periodic function~:
$V(-x)=V(x)$ and $V(x+n)=V(x)$ for $n\in\ZZ$.

Demonstration for $n=0$ was given in
Ref.~\cite{TexMon05b,ComDesTex05}, where we pointed that, for a
Brownian bridge 
on $\RR$ $(x(\tau),\,0\leq\tau\leq{t}\,|\,x(0)=x(t)=0)$, we have
the following equality in law~\cite{footnote16}~:
\begin{equation}
  x(\tau) - x(t-\tau) \eqlaw x(2\tau)
  \hspace{0.5cm} \mbox{for }
  0\leq\tau\leq t/2
  \:.
\end{equation}
The proof lies on the fact that we can relate the bridge to a free
Brownian motion (Wiener process) 
$(W(\tau),\,0\leq\tau\leq{t},\,W(0)=0)$~:
$x(\tau) \eqlaw W(\tau) - \frac{\tau}{t}W(t)$.

Here we generalize this relation when $x(\tau)$ lives on the circle
$[0,1]$ and when we constraint the winding number.
Let us unfold the ring in order to work on $\RR$. A close path 
winding $n$ times around the ring is related to the following path
living on the real axis~:
$(x_n(\tau),\,0\leq\tau\leq{t}\,|\,\,x_n(0)=0\,;\,x_n(t)=n)$
that can be writen as
\begin{equation}
  x_n(\tau) \eqlaw W(\tau) + \frac{\tau}{t}(n-W(t))
  \eqlaw x_0(\tau) +n\frac{\tau}{t}
\end{equation}
$x_0(\tau)$ is the Brownian bridge. It is now easy to show
that~\cite{footnote17} 
\begin{equation}
  x_n(\tau) - x_n(t-\tau) \eqlaw x_n(2\tau) -n
  \hspace{0.5cm} \mbox{for }
  0\leq\tau\leq t/2
  \:.
\end{equation}
Since $x_n(\tau) - x_n(t-\tau)$ is argument of the periodic function,
the integer shift can be forgotten.
The symmetry $V(x)=V(-x)$
ensures the equality of contributions of intervals $\int_0^{t/2}$ and
$\int_{t/2}^t$. It follows that
\begin{equation}
  \label{eq:94}
  \boxed{
  \int_0^t\D\tau\,V(x_n(\tau) - x_n(t-\tau))
  \eqlaw
  \int_0^t\D\tau\,V(x_n(\tau))
  }
\end{equation}
which demonstrates eq.~(\ref{Zenicerelation}).

\vspace{0.25cm}

\noindent{\it Infinite wire~:}
Using (\ref{Zenicerelation}), we see that the path
integral (\ref{PathIntInfWire}) involves an action local in time
\begin{align}
   P_c(x,x) &\equiv
   -\frac12\Delta\tilde\sigma(x) = \int_0^\infty\D{t}\,\EXP{-\gamma t}
\nonumber\\
  &\times
   \int_{x(0)=0}^{x(t)=0}\hspace{-0.5cm}\mathcal{D}x(\tau)\,
   \EXP{ -\int_0^t\D\tau\,\big[\frac14\dot{x}^2
         + \frac{1}{L_N^3}\,|x(\tau)|\big]}
  \:,
\end{align}
that can now be computed.
We obtain 
$P_c(0,0)
=-L_N\,\frac{\mathrm{Ai}(\gamma L_N^2)}{2\,\mathrm{Ai}'(\gamma L_N^2)}$
derived in  Ref.~\cite{AltAroKhm82} (numerical factors are incorrect
in this reference). 

\vspace{0.25cm}

\noindent{\it Isolated ring~:}
The  function $W(x,x')$ is given by
eq.~\eqref{Wring}. The path integral \eqref{eq:CQ} can be rewritten as 
\begin{align}
    \int_0^\infty
   \hspace{-0.25cm}\D{t}\,\EXP{-\gamma t}
   \int_{x(0)=0}^{x(t)=0}\hspace{-0.5cm}\mathcal{D}x\,
  \delta_{n,\mathcal{N}[x]}\,
   \EXP{ -\int_0^t\D\tau\,\big[\frac{\dot{x}^2}4
         + \frac{|x|}{L_N^3}\,\big(1-\frac{|x|}{L}\big)\big]}
\end{align}
that can be expressed in terms of Hermite functions~\cite{TexMon05b}.


\mathversion{bold}
\section{\label{app:fctPsi} The function $\Psi(\xi)$}
\mathversion{normal}

We analyze several properties of the function \eqref{eq:DefPsi}, that
we rewrite
\begin{equation}
  \label{eq:DefPsi2}
  \Psi(\xi) = \frac4\pi \Lambda^{3/4} \re\left[
    \EXP{-\I\frac\pi4} \int_{\RR^+}\D{z}\,z^2\,
    \EXP{-\Lambda\,\varphi(z)}
  \right]
  \:,
\end{equation}
where $\varphi(z)=z^4+4z\EXP{-\I\pi/4}$ and $\Lambda=(\xi/4)^{4/3}$.
The value of the function at the origin is
$\Psi(0)=\frac{\Gamma(3/4)}{\pi\sqrt2}$.

\begin{figure}[!ht]
  \centering
  \includegraphics[scale=1.2]{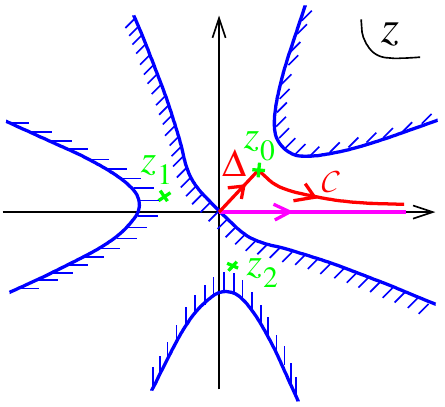}
  \caption{\it Appropriate contour deformation in order to estimate
  \eqref{eq:DefPsi2}.
  The dashed area corresponds to the region where
  $\re[\varphi(z)]<0$.}
  \label{fig:contourPsi}
\end{figure}

The asymptotic behaviour for $\xi\gg1$ may be studied by the steepest
descent method. $\varphi'(z)=0$ has three solutions
$z_n=\EXP{\I\pi/4+2\I{}n\pi/3}$, with $n=0,\,1,\,2$
(figure~\ref{fig:contourPsi}).  
An appropriate contour deformation in the complex plane  of the
variable $z$ must remain in the region where $\re[\varphi(z)]>0$. This
domain can be easily determined by performing a rotation
$z=w\EXP{\I\pi/4}$~: writing $w=u+\I{}v$ we have
$\re[\varphi(z)]=-u^4+6u^2v^2-v^4+4u$ that vanishes for 
$v=\pm\sqrt{3u^2\pm2\sqrt{2u^4+u}}$. The domain where
$\re[\varphi(z)]>0$ is represented on figure~\ref{fig:contourPsi}.
This shows that the contour can only visit $z_0=\EXP{\I\pi/4}$.
The integration over $\RR^+$ is replaced by integration over the
segment $\Delta$ from the origin to $z_0$ and the contour
$\mathcal{C}$ issuing from $z_0$ and going to infinity
(figure~\ref{fig:contourPsi}). 
Noticing that $\int_\Delta\D{z}\,z^2\,\EXP{-\Lambda\,\varphi(z)}$ is
purely imaginary, we are left with the contribution of the contour
$\mathcal{C}$ only. We now use the steepest descent method
$\Psi(\xi)=\frac4\pi\Lambda^{3/4}\re[\EXP{-\I\frac\pi4}\int_\mathcal{C}\D{z}\,z^2\,\EXP{-\Lambda\,\varphi(z)}]\simeq\frac4\pi\Lambda^{3/4}\re[\EXP{-\I\frac\pi4}\frac12\sqrt{\frac{2\pi}{\Lambda\varphi''(z_0)}}\EXP{-\Lambda\,\varphi(z_0)}]$,
where the $1/2$ is due to the fact that the contour issues from the
stationary point, hence
\begin{equation}
    \Psi(\xi\gg1)\simeq\frac2{\sqrt{6\pi}}(\xi/4)^{1/3}\EXP{-3(\xi/4)^{4/3}}
\end{equation}
(note that a factor $1/2$ is missing in Ref.~\cite{TexMon05}).

Finally the relation to the function $q(x)$ introduced in the
conclusion requires the two integrals
$\int_0^\infty\D\xi\,\Psi(\xi)=1/\sqrt\pi$ and
$\int_0^\infty\D\xi\,\xi^2\,\Psi(\xi)=2$.

\mathversion{bold}
\section{\label{app:RelSF} Hypergeometric function $F(\frac12,n+\frac12;n+1;\xi)$}
\mathversion{normal}

This appendix is devoted to the study of the hypergeometric function
$F(\frac12,n+\frac12;n+1;\xi)$.
Our starting point is the integral representation~\cite{gragra}
\begin{align}
  \label{eq:HyperIntRepres}
  &
  B\!\left(\frac12,n+\frac12\right)
  \,F\!\left(\frac12,n+\frac12;n+1;\xi\right) 
\nonumber\\
  &=
  \int_0^1\D t\,\frac{t^{n-1/2}}{\sqrt{(1-t)(1-\xi t)}}
  \:.
\end{align}
We recall that the hypergeometric function is regular at the origin
$F(\alpha,\beta;\gamma;0)=1$. 
Note that the Euler $\beta$ funtion
$B(\frac12,n+\frac12)={\sqrt\pi\,\Gamma(n+\frac12)}/{n!}$ is well
approximated by $B(\frac12,n+\frac12)\simeq\sqrt{\pi/n}$ in the large
$n$ limit. 

In order to analyze the behaviour of the hypergeometric function for
$\xi\to1$ we rewrite the integral of eq.~\eqref{eq:HyperIntRepres} as 
\begin{align}
  \label{eq:B3}
  &
  \int_0^1\D t\,\frac{(1-t)^{n-1/2}}{\sqrt{t(1-\xi+\xi t)}}
  \\\nonumber
  &=\frac1{\sqrt\xi}\left(
     \int_0^1\D t\,
     \frac{1}{\sqrt{t^2+\varepsilon\,t}}
     +\int_0^1\D t\,\frac{(1-t)^{n-1/2}-1}{\sqrt{t^2+\varepsilon\,t}}
   \right)
\end{align}
where $\varepsilon=1/\xi-1$. The first integral is
$2\argsinh(1/\sqrt\varepsilon)=\ln(4/\varepsilon)+O(\varepsilon)$.
The second integral reaches a finite limit for $\varepsilon\to0$,
expressed in terms of the Digamma function~\cite{gragra}
$\int_0^1\D{u}\,\frac{u^{n-1/2}-1}{1-u}=\psi(1)-\psi(n+\frac12)$. 
We can show that correction to this constant is of order 
$\varepsilon\ln(\varepsilon)$, therefore
\begin{align}
  \label{eq:ExpF}
   & F\!\left(\frac12,n+\frac12;n+1;\xi\right) \underset{\xi\to1}{=}
   \frac{1}{B\!\left(\frac12,n+\frac12\right)}
  \\\nonumber
  &\times\left[
   \ln\left(\frac4{1-\xi}\right)+\psi(1)-\psi\!\left(n+\frac12\right)
   +O(\varepsilon\ln\varepsilon)
  \right]
  \:.
\end{align}
The behaviour \eqref{eq:ExpF} only holds for $n$ not too large,
$n\ll(1-\xi)^{-1}$. In the opposite case $n\gg(1-\xi)^{-1}\gg1$, the
factor $(1-t)^{n-1/2}\simeq\EXP{-(n-1/2)t}$ in eq.~\eqref{eq:B3}
selects an interval of
width $1/n\ll(1-\xi)$ and we can neglect the quadratic term
$\xi t^2$ below the square root. Therefore eq.~\eqref{eq:B3}
is
$\simeq\int_0^\infty\D{t}\,\frac1{\sqrt{(1-\xi)t}}\EXP{-(n-1/2)t}$. Finally
\begin{equation}
  \label{eq:ExpF2}
    F\!\left(\frac12,n+\frac12;n+1;\xi\right) \underset{\xi\to1}{\simeq}
   \frac1{\sqrt{1-\xi}}
\end{equation}
for $n\gg(1-\xi)^{-1}\gg1$.

\begin{figure}[!ht]
  \centering
  \includegraphics[scale=0.8]{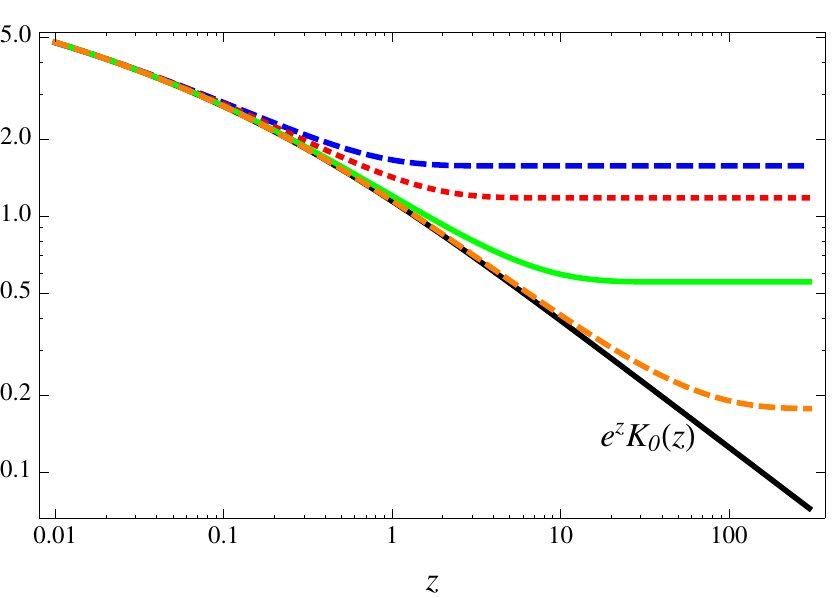}
  \caption{\it Comparison between the two sides of equation
    \eqref{eq:ApproxHypergeo} for $n=1$, $2$, $10$ and~$100$.
    } 
  \label{fig:chain_hypergeo}
\end{figure}

We now prove a useful relation between the hypergeometric function and the
MacDonald function (modified Bessel function).
If $\xi=\EXP{-\frac{2z}{n}}\simeq1-2z/n$, the integral \eqref{eq:B3} may be
rewritten as 
\begin{align}
  \int_0^n\frac{\D t}{n}
  \frac{\left(1-\frac{t}{n}\right)^{n-1/2}}
       {\sqrt{\frac{t}{n}\left(\frac{2z}{n}+\frac{t}{n}\right)}}  
  \underset{n\to\infty}{\longrightarrow}
  \int_0^\infty\D t\,
  \frac{\EXP{-t}}{\sqrt{t(2z+t)}}  
  \:.
\end{align}
We recognize an integral representation of the MacDonald
function~\cite{gragra}. 
Therefore, for $z\ll n$, we can write
\begin{equation}
  \label{eq:ApproxHypergeo}
  B\!\left(\frac12,n+\frac12\right)\,
  F\!\left(\frac12,n+\frac12;n+1;\EXP{-\frac{2z}{n}}\right)
  \simeq \EXP{z}\,K_0(z) 
  \:.
\end{equation}
The {\it r.h.s.} describes the crossover between \eqref{eq:ExpF} and
\eqref{eq:ExpF2}.
%
%
We compare the two sides of equation \eqref{eq:ApproxHypergeo} for
different values of $n$ on figure~\ref{fig:chain_hypergeo}.


\section{Laplace equation in networks~: spectral determinant}
\label{app:spedet}

In this appendix we introduce an important tool, the spectral determinant,
used to study some properties of the equation
\begin{equation}
  \label{EqP}
  (\gamma -\Delta)P(x,x') = \delta(x-x')
\end{equation}
in networks. 

The spectral determinant is formally defined as
$S(\gamma)=\det(\gamma-\Delta)=\prod_n(\gamma+E_n)$ where $\{E_n\}$ is
the spectrum of the Laplace operator $-\Delta$ (in the presence of a
magnetic field, $\Delta\to[\nabla-2\I eA(x)]^2$).  Despite this
operator acts in a space of infinite dimension, the spectral
determinant can be related to the determinant of a finite size matrix,
of dimension equal to the number of vertices. This
matrix encodes all informations on the network (topology, lengths of
the wires, magnetic field, boundary conditions describing connections
to reservoirs). Let us label vertices with 
greek letters. $l_\ab$ designates the length of the wire $(\ab)$ and
$\theta_\ab$ the circulation of the vector potential along the wire.
The topology is encoded in the adjacency matrix~: $a_\ab=1$ if
$\alpha$ and $\beta$ are linked by a wire, $a_\ab=0$ otherwise.
We consider the case where Laplace operator acts on functions $\varphi(x)$
({\it i}) continuous at the vertices satisfying 
({\it ii}) $\sum_\beta{}a_\ab\varphi_\ab'(0)=\lambda_\alpha\,\varphi_\alpha$
where $\varphi_\ab(x)$ designates the component of the function on the wire
$(\ab)$ and $\varphi_\alpha$ its value at the vertex. Self-adjointness
of the Laplace  
operator is ensured if $\lambda_\alpha\in\RR$ (more details may be 
found in Refs.~\cite{AkkComDesMonTex00,ComDesTex05}).
$\lambda_\alpha=\infty$ corresponds to Dirichlet boundary condition at
the vertex 
and describe the case where $\alpha$ touches a reservoir through which 
current is injected in the network. 
$\lambda_\alpha=0$ for internal vertices.
The interest of mixed  boundary conditions (finite $\lambda_\alpha$)
is illustrated in appendix~\ref{app:sdepn}.
We introduce the matrix
\begin{align}
  \label{MatM}
  \mathcal{M}_\ab &= \delta_\ab
  \left(
   \lambda_\alpha
   + \sqrt\gamma\sum_\mu a_{\alpha\mu}\coth\sqrt\gamma l_{\alpha\mu}
  \right)
  \nonumber\\
  &\hspace{0.5cm}
  -a_\ab \,\sqrt\gamma \frac{\EXP{-\I\theta_\ab}}{\sinh\sqrt\gamma l_\ab}
  \:,
\end{align}
where the $a_{\alpha\mu}$ constrains the sum to run over neighbouring
vertices. Then~\cite{PasMon99,AkkComDesMonTex00}
\begin{equation}
  \label{eq:SpeDet}
  \boxed{
  S(\gamma) = \prod_{(\ab)}\frac{\sinh\sqrt\gamma l_\ab}{\sqrt\gamma}\:
  \det\mathcal{M}
  }
\end{equation}
where the product runs over all wires. 
Despite the spectral determinant encodes the spectral information, it is
also possible to extract some local information, like $P(x,x)$, by small
modifications of the matrix. This has been used in
Ref.~\cite{TexMon05} and is briefly discussed in appendix~\ref{app:sdepn}.

It is useful to remark that the matrix $\mathcal{M}$ can be used to express
$P(x,x')$ when $x$ and $x'$ coincides with nodes (this is always possible to
introduce a vertex anywhere without changing the properties of the
network)~:
\begin{equation}
  \boxed{
  \label{remark}
  P(\alpha,\beta) = \left(\mathcal{M}^{-1}\right)_{\alpha\beta}
  }
  \:.
\end{equation}

\vspace{0.25cm}

\noindent{\it WL correction in regular networks.--}
In large regular networks connected in such a way that currents are uniformly
distributed in the wires, we can assume that weights attributed to the
wires of the networks 
in eq.~(\ref{Res2004}) are equal. In this case, a uniform integration of the 
Cooperon $P_c(x,x)=\bra{x}\frac1{\gamma-\Delta}\ket{x}$
in the network leads to a meaningful quantity (relevant
experimentally). 
The Cooperon integrated uniformly is directly related to the spectral
determinant 
\cite{Pas98,AkkComDesMonTex00}
\begin{equation}
  \label{eq:PasMon}
  \int_\mathrm{network}\D{}x\: \Delta\tilde\sigma(x)
  =-2\, \derivp{}{\gamma}\ln S(\gamma)
  \:.
\end{equation}
This equation provides a very efficient way for calculating the WL
correction in arbitrary networks, when uniform integration of Cooperon
is justified.

\vspace{0.25cm}

\noindent{\it WL correction in arbitrary networks.--}
In the most general case, eq.~\eqref{Res2004} requires to construct
the Cooperon in each wire.
A general expression was provided in \cite{TexMon04} however it is
useful to notice that $P_c(x,x)$ can also be obtained from a spectral
determinant for a modified boundary condition at point $x$. It was
shown in Refs.~\cite{TexMon05,ComDesTex05} that if we introduce mixed
boundary conditions with a parameter $\lambda_x$ at $x$, then 
\begin{equation}
  \label{eq:7}
  P_c(x,x)=\derivp{}{\lambda_x}\ln{S}^{(\lambda_x)}(\gamma)\big|_{\lambda_x=0}
  \:.
\end{equation}


\section{Classical resistance/conductance\label{app:classresist}}

We calculate the resistance between two vertices of an arbitrary
network. We consider a network of wires of lengths $l_{\mu\nu}$ with same
sections $\sw$.  In this case the conductance of the wire $\mu\nu$ is
given by $\sigma_0\sw/l_{\mu\nu}$.
We introduce the matrix 
\begin{equation}
    (\mathcal{M}_0)_{\mu\nu} = \delta_{\mu\nu}
     \sum_\rho \frac{a_{\mu\rho}}{l_{\mu\rho}} - \frac{a_{\mu\nu}}{l_{\mu\nu}}
\end{equation}
whose matrix elements coincide with the conductances of the wires (up to the
factor $\sigma_0\sw$).
This matrix coincides with the matrix (\ref{MatM}) if all fluxes are
set to zero and the limit $\gamma\to0$ is taken and moreover with
$\lambda_\mu=0$, $\forall\:\mu$ for an isolated network.

\begin{figure}[!ht]
  \centering
  \includegraphics[scale=0.35]{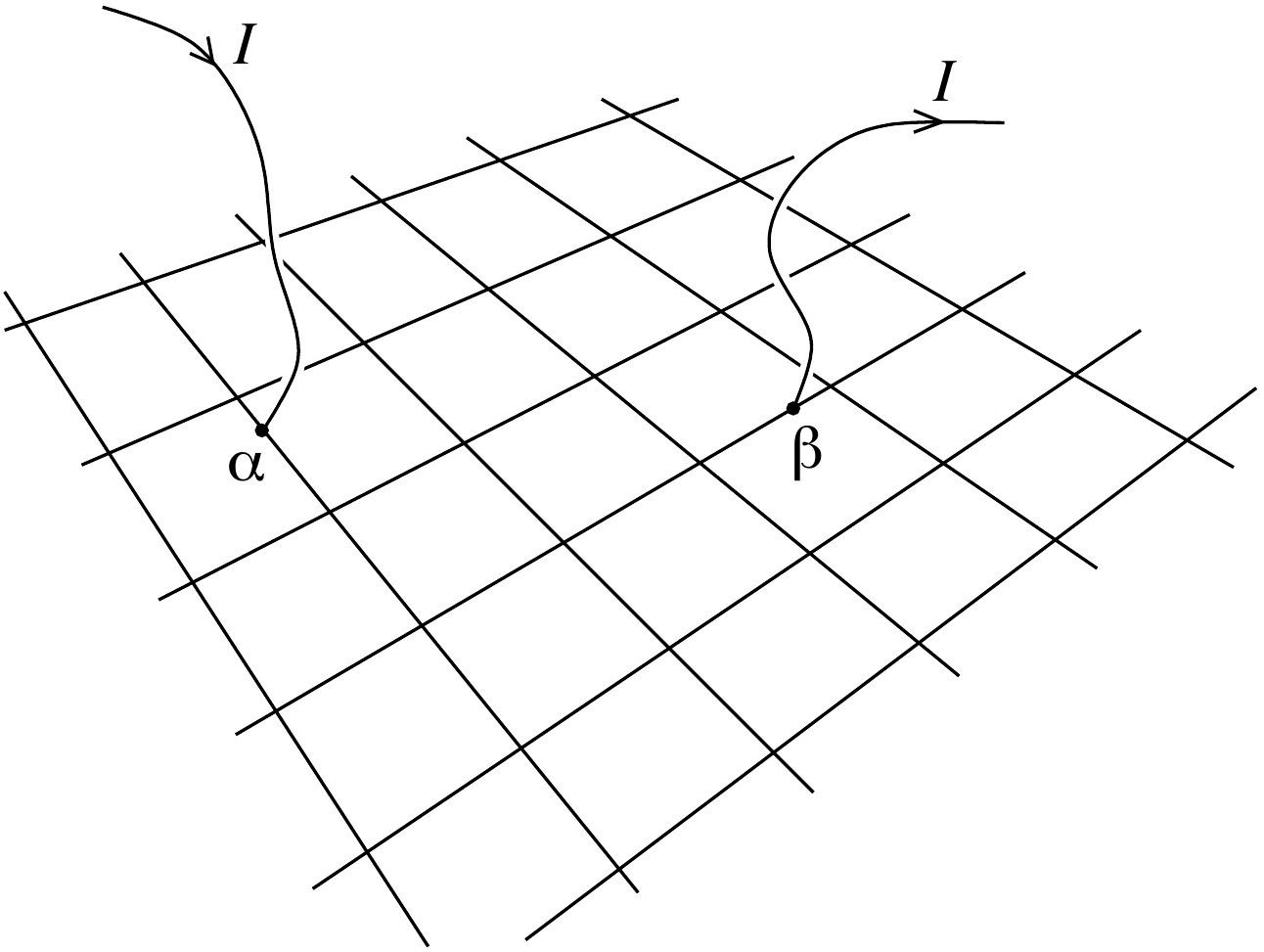}
  \caption{\it Injection of current in a network (here a regular
    square network).} 
  \label{fig:rixix}
\end{figure}

We now consider the situation where we inject a current at the vertex
$\alpha$. This current exits at vertex $\beta$ (see figure~\ref{fig:rixix}).
If we denote by $V_\mu$ the potential at $\mu$, Kirchhoff law at vertex $\mu$
takes the form
\begin{equation}
  \sigma_0 \sw \sum_\nu(\mathcal{M}_0)_{\mu\nu} V_\nu
  = I \, \left[ \delta_{\mu,\alpha}-\delta_{\mu,\beta} \right]
  \:.
\end{equation}
Potential is therefore given by~:
\begin{equation}
  \label{potentials}
  V_\mu = \frac{I}{\sigma_0\sw}
  \left[ (\mathcal{M}^{-1}_0)_{\mu\alpha}-(\mathcal{M}^{-1}_0)_{\mu\beta} \right]
  \:.
\end{equation}
Note that the matrix $\mathcal{M}_0$ is not inversible~; it is
explained below how to give a precise meaning to this expression.
We define the resistance between points $\alpha$ and $\beta$ as 
$\mathcal{R}_\ab=(V_\alpha-V_\beta)/I$. Therefore~:
\begin{align}
  \label{resistnet}
  \mathcal{R}_\ab= \frac2{\sigma_0\sw}
   \left[
     \frac{(\mathcal{M}^{-1}_0)_{\alpha\alpha}+(\mathcal{M}^{-1}_0)_{\beta\beta}}2
     - (\mathcal{M}^{-1}_0)_\ab 
   \right]
  \:.
\end{align}
Using eq.~(\ref{remark}) we see that we can express the resistance in
terms of the solution $P_d$ of the equation $-\Delta{}P_d=\delta$~:
\begin{align}
  \label{RqGilles}
  &\boxed{
  \mathcal{R}(x,x') = \frac2{\sigma_0 s}
   \left[
     \frac{P_d(x,x)+P_d(x',x')}2 - P_d(x,x')
   \right]
  }
  \nonumber\\
  &\hspace{1.5cm}= \frac2{\sigma_0 s}\,W(x,x')
  \:.
\end{align}
This demonstrates that the function $W(x,x')$ defined by \eqref{eq:defW}
is indeed the equivalent resistance between points $x$ and~$x'$.

\vspace{0.25cm}

\noindent{\it Remark~: $\mathcal{M}_0$ is not inversible.--}
It is easy to check that 
\begin{equation}
  \label{eq:ccgi}
  \sum_\mu(\mathcal{M}_0)_{\mu\nu}=0
  \:.
\end{equation}
Kernel of the matrix is the vector $(1,1,\cdots,1)$. Physically 
eq.~\eqref{eq:ccgi}
ensures~: ({\it i}) that sum of all currents arriving at vertex $\nu$
is zero, ({\it ii}) currents are zero if all potentials are equal
(equilibrium).  This problem can be overcome easily by noticing that
only {\it differences} of inverse matrix elements have appeared, in
eqs.~(\ref{potentials}) and~(\ref{resistnet}). We can always inverse
the matrix $\mathcal{M}_0$ in the space orthogonal to the vector
$(1,1,\cdots,1)$ and compute differences of such matrix elements. This
is how eqs.~(\ref{potentials},\ref{resistnet},\ref{RqGilles}) must be
understood.

In practice, an easier way to compute such differences is to regularize the
calculation by computing the inverse of matrix $\mathcal{M}$ for finite $\gamma$
(or at least one finite $\lambda_\alpha$) and take the limit $\gamma\to0$ (or
$\lambda_\alpha\to0$) after having computed the difference of inverse matrix
elements~:
\begin{equation}
  \label{eq105}
  (\mathcal{M}^{-1}_0)_{\mu\alpha}-(\mathcal{M}^{-1}_0)_{\mu\beta}
  = \lim_{\gamma\to0}
  \left[
  \mathcal{M}^{-1}_{\mu\alpha} - \mathcal{M}^{-1}_{\mu\beta}
  \right]
  \:.
\end{equation}
Note that $\det\mathcal{M}\neq0$ for $\gamma\in\RR^{+*}$ since
$\mathrm{Spec}(\Delta)\subset\RR^-$. 

This point is related to the fact that was already mentioned in the continuum limit
in order to construct the function $W(x,x')$ (in section~\ref{sec:fctWcyl} or
in appendix~\ref{app:sdepn})~: eq.~(\ref{eq105}) is the analogue of
eq.~(\ref{eq75}). In 
the continuum this problem is related to the fact that the Laplace
operator is not inversible in the space of functions satisfying Neumann
boundary conditions corresponding to an isolated conductor.

\vspace{0.25cm}

\noindent{\it Example~: function $W$ in an isolated ring.--}
The relation between the function $W$ and the resistance may be used in order
to construct easily $W$. Let us consider the case of a ring of
perimeter $L$. When the ring is connected at two wires at $x$ and $x'$, the
resistance $\mathcal{R}(x,x')$ corresponds to the one of two wires of lengths
$|x-x'|$ and $L-|x-x'|$ put in parallel. We straightforwardly recover
the function obtained in Ref.~\cite{TexMon05b}~:
\begin{align}
  W_\mathrm{ring}(x,x') 
  &= \frac12\,\frac1{\frac1{|x-x'|}+\frac1{L-|x-x'|}}
  \\
  \label{Wring}
  &= \frac12\,|x-x'|\left(1-\frac{|x-x'|}L\right)
  \:.
\end{align}

Note that if we consider a ring connected at reservoirs through arms of
finite length, the function $W(x,x')$ inside the ring is modified
\cite{LudMir04,TexMon05b}, since its construction needs to consider a four
terminal device. In the limit of infinitely long arms we can neglect currents
flowing through the leads and we recover the result of the isolated
ring~(appendix of Ref.~\cite{TexMon05b}).

\vspace{0.25cm}

\begin{figure}[!ht]
  \centering
  \includegraphics[scale=0.6]{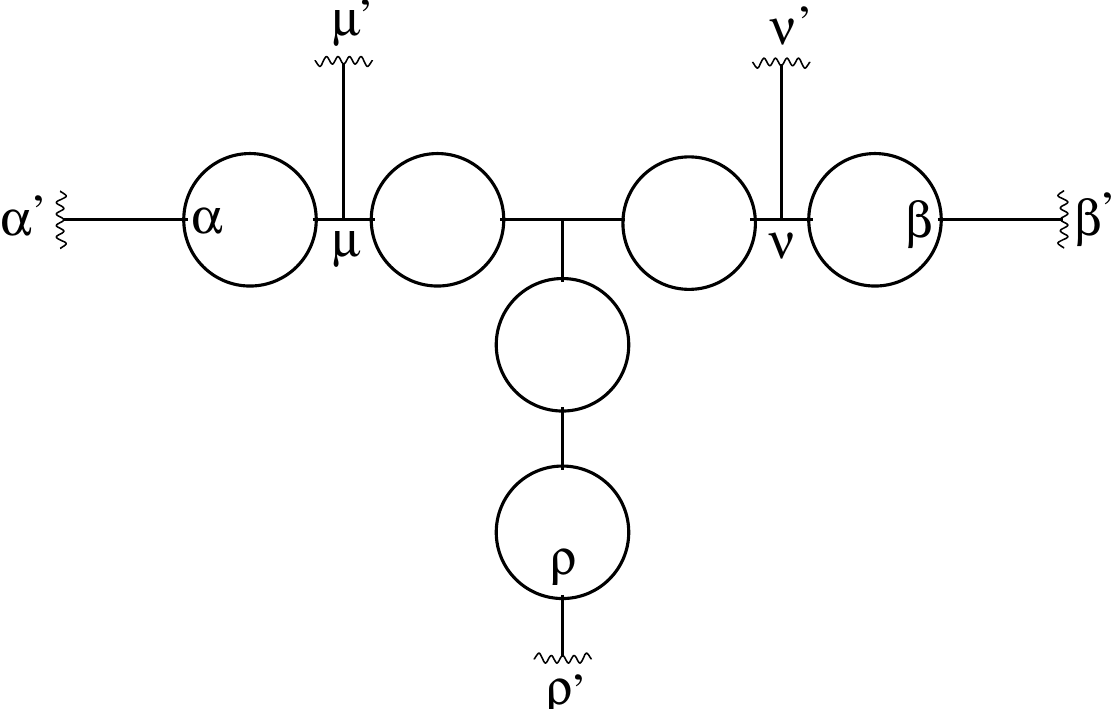}
  \caption{\it A multiterminal network. Wavy lines represent contacts
    through which current is injected. Contacts correspond to vertices
  with primed labels.}
  \label{fig:multiterm}
\end{figure}

\noindent{\it Classical conductance.--}
It is interesting to compare the formula obtained for the resistance
with the one obtained for the conductance matrix of a multiterminal
network~\cite{TexMon04}. 
We first stress that we consider two different situations. 
The eqs.~(\ref{resistnet},\ref{RqGilles}) give the {\it potentials}
when the {\it currents} injected at vertex $\alpha$ and extracted at vertex
$\beta$ are fixed (figure~\ref{fig:rixix}).
On the other hand, a conductance matrix allows to determine the {\it
  currents} through contacts when the {\it potentials} at these
contacts are fixed (figure~\ref{fig:multiterm}).

We consider a multiterminal network,
connected to external contacts through which current is injected at a
set of vertices indicated with 
primes~:  $\alpha'$, $\beta'$,... 
The connection is accounted for by introducing parameters
$\lambda_{\alpha'}=\infty$ added to the matrix $\mathcal{M}_0$, that now
become inversible. These parameters describe Dirichlet conditions for the
diffusion equation and permit to inverse the Laplace operator. 
The transport through the network is characterized by a conductance
matrix whose elements are given by~\cite{TexMon04}
\begin{align}
  \label{eq:Cond2004}
  \mathcal{G}_{\alpha'\beta'} = - \frac{\sigma_0\sw}{l_{\alpha\alpha'}l_{\beta\beta'}}
  (\mathcal{M}^{-1}_0)_\ab
  =- \frac{\sigma_0\sw}{l_{\alpha\alpha'}l_{\beta\beta'}}\,P_d(\alpha,\beta)
  \:.
\end{align}
At first sight diffuson $P_d(\alpha,\beta)$ ({\it i.e.} inverse matrix
element 
$(\mathcal{M}^{-1}_0)_\ab$) is related to both the conductance and the 
resistance.
Note however that the matrix $\mathcal{M}_0$ (and the corresponding
diffuson $P_d(x,x')$) in eq.~\eqref{eq:Cond2004} does account for the
boundary conditions (Dirichlet boundary conditions at contacts, {\it
  i.e.} primed vertices) whereas $\mathcal{M}_0$ (and $P_d(x,x')$) in
eqs.~(\ref{resistnet},\ref{RqGilles}) describe the isolated network.


\section{Solution of the diffusion equation in some particular networks\label{app:sdepn}}

We consider the solution of the diffusion eq.~(\ref{EqP})
for the networks studied in this article. 

\subsection{The ring with one or several arms\label{app:roa}}

We consider a ring of perimeter $L$ attached to an arm of length $b$
connected to a reservoir ({\it i.e.} with Dirichlet 
boundary condition at its end) and submitted to a magnetic field. 
The spectral determinant is~\cite{TexMon05}~:
\begin{align}
  \label{eq:Scr}
  S(\gamma) = \frac{2}{\sqrt\gamma} \sinh\sqrt\gamma b 
  \left[\cosh\sqrt\gamma L_\mathrm{eff}-\cos\theta\right]
\end{align}
where the effective perimeter is given by
\begin{align}
  \cosh\sqrt\gamma L_\mathrm{eff} = \cosh\sqrt\gamma L
  +\frac12\coth\sqrt\gamma b\sinh\sqrt\gamma L
  \:.
\end{align}
A systematic way for obtaining the spectral determinant of two
subgraphs glued at one vertex from the spectral determinants of the
subgraphs has been derived in Ref.~\cite{Tex08}. This allow to recover 
easily eq.~\eqref{eq:Scr}.

\vspace{0.25cm}

\begin{figure}[htbp]
  \centering
  \includegraphics{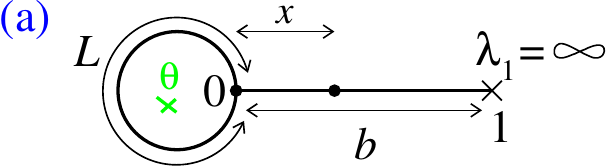}
  \vspace{0.5cm}
  \\
  \includegraphics{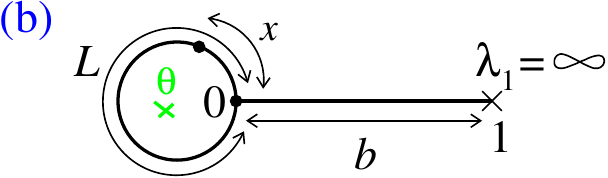}
  \caption{\it A ring with one wire. We choose a Dirichlet boundary condition
    at the vertex $1$ for simplicity ($\lambda_1=\infty$).}
  \label{fig:ouvrebouteille4}
\end{figure}

\noindent$\bullet$
Introducing mixed boundary conditions at the node (vertex $0$) we easily obtain 
$S^{(\lambda_0)}(\gamma)=\frac{\lambda_0}\gamma\sinh\sqrt\gamma{}L\sinh\sqrt\gamma{}b+S(\gamma)$
from \eqref{eq:SpeDet}.
Using eq.~\eqref{eq:7}, and performing a Fourier transform, we get the
Cooperon 
$P_c^{(n)}(0,0)=-\frac12\Delta\tilde\sigma_n(0)$ at the 
node~\cite{TexMon05} 
\begin{align}
  \label{CoopATN}
  P_c^{(n)}(0,0) = \frac1{2\sqrt\gamma}\,
  \frac{\sinh\sqrt\gamma L}{\sinh\sqrt\gamma L_\mathrm{eff}}\,
  \EXP{-n\sqrt\gamma L_\mathrm{eff}}
  \:.
\end{align}

In the weakly coherent limit we find
$L_\mathrm{eff}\simeq{}L+L_\varphi\ln(3/2)$, whence
$P_c^{(n)}(0,0)\simeq\frac1{2\sqrt\gamma}\,\big(\frac23\big)^{n+1}\EXP{-n\sqrt\gamma{}L}$.

In the limit $\sqrt\gamma L\ll1$ the  effective perimeter is 
$L_\mathrm{eff}\simeq\gamma^{-1/4}L^{1/2}$.
We have~\cite{TexMon05,ComDesTex05}~:
\begin{align}
  P_c^{(n)}(0,0)
  \label{LaplPnRL}
  \simeq \frac{\sqrt{L}}{2\gamma^{1/4}} \,\EXP{-n\sqrt{L}\gamma^{1/4}} 
  \:.
\end{align}

\vspace{0.25cm}

\noindent$\bullet$
In order to calculate the harmonics of the Cooperon in the arm
($x$ is the distance from the ring, see 
figure~\ref{fig:ouvrebouteille4}.a), we have to consider the spectral
determinant for the graph with mixed boundary conditions at $x$, with
parameter $\lambda_x$~:
\begin{widetext}
\begin{align}
  S^{(\lambda_x)}(\gamma) = S(\gamma) +
  \frac{\lambda_x}{\gamma}
  \bigg[
    \sinh L \frac{\sinh^2(b-x)}{\sinh b}
  + 2\sinh x\sinh(b-x)
      \left(\cosh L_\mathrm{eff}-\cos\theta\right)
  \bigg]
\end{align}
(for shorter notations we have omitted $\sqrt\gamma$ in hyperbolic functions).
From eq.~\eqref{eq:7} we obtain~:
\begin{align}
  \label{PnXia}
  \boxed{
  P_c^{(n)}(x,x)
  = \delta_{n,0}\frac1{\sqrt\gamma}\,
  \frac{\sinh\sqrt\gamma x\sinh\sqrt\gamma(b-x)}{\sinh\sqrt\gamma b}
  +\left(\frac{\sinh\sqrt\gamma(b-x)}{\sinh\sqrt\gamma b}\right)^2\,
   P_c^{(n)}(0,0) 
  }
\end{align}
for $x\in$\:arm (figure~\ref{fig:ouvrebouteille4}.a).
We recognize the first term as the result obtained for a wire of
length $b$ connected to reservoirs ({\it i.e.} with Dirichlet
boundaries). In
the limit $b\to\infty$ we have $P_c^{(n)}(x,x)\simeq
P_c^{(n)}(0,0)\,\EXP{-2\sqrt\gamma x}$ (for $n\neq0$).

\vspace{0.25cm}

\noindent$\bullet$
If $x$ is inside the ring (figure~\ref{fig:ouvrebouteille4}.b)
the modified spectral determinant  reads~:
\begin{align}
  S^{(\lambda_x)}(\gamma) = S(\gamma) +
  \frac{\lambda_x}{\gamma}
  \left[
    \sinh x\sinh(L-x)\cosh b + \sinh b \sinh L
  \right]
\end{align}
and from eq.~\eqref{eq:7} the Cooperon is therefore
\begin{align}
  \label{PnXir}
  \boxed{
   P_c^{(n)}(x,x)
   = \left[
     1 + \coth\sqrt\gamma b 
     \frac{\sinh\sqrt\gamma x\sinh\sqrt\gamma(L-x)}{\sinh\sqrt\gamma L}
   \right] \, P_c^{(n)}(0,0)
  }
\end{align}
\end{widetext}
for $x\in$\:ring (figure~\ref{fig:ouvrebouteille4}.b).

\begin{figure}[!ht]
  \centering
  \includegraphics[scale=0.6]{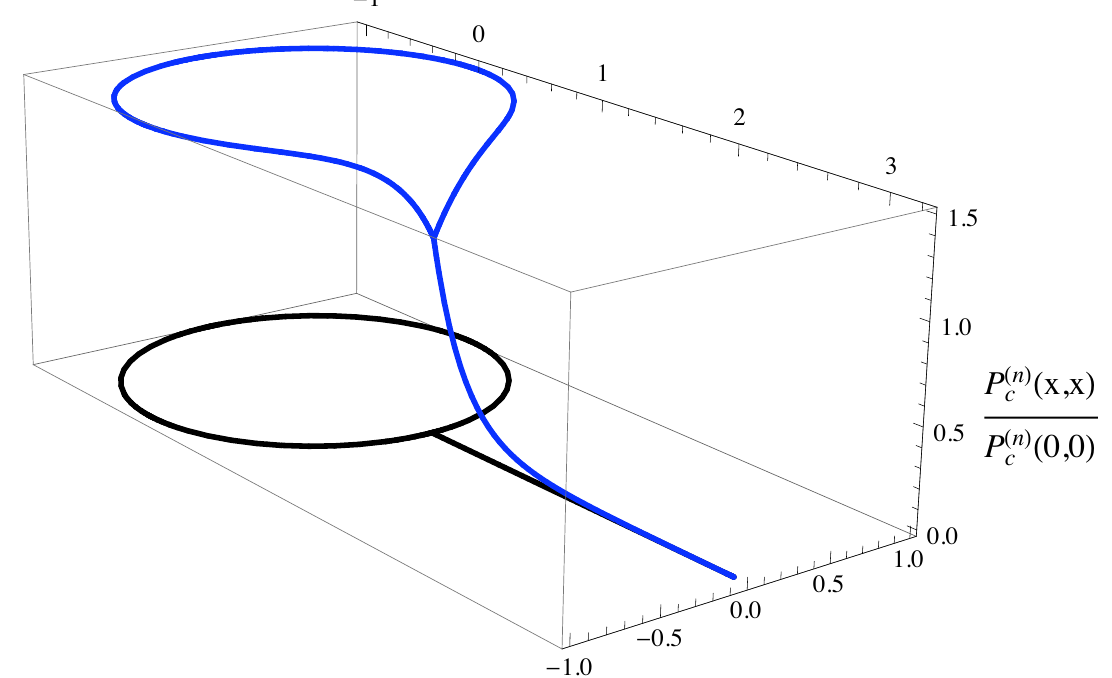}
  \caption{\it $x$ dependence of the Cooperon $P_c^{(n)}(x,x)$ in the
    connected ring for $L_\varphi\ll L$.}
  \label{fig:cir}
\end{figure}

If $L_\varphi\ll L$ we have
$P_c^{(n)}(x,x)\simeq[\frac32-\frac12\EXP{-2\sqrt\gamma x}]P_c^{(n)}(0,0)$ for
$x<L/2$. In the bulk (for $x$ and $L-x\gg L_\varphi$) we have 
$P_c^{(n)}(x,x)\simeq\frac32P_c^{(n)}(0,0)$
(figure~\ref{fig:cir}), where the factor $\frac32$ corresponds to the
ratio of coordination numbers at $0$ and at~$x$.

In the opposite limit $L\ll L_\varphi\ll b$ the Cooperon is
homogeneous inside the ring, as expected.

\vspace{0.25cm}

\noindent{\it From one to $N_a$ arms.--}
In the regime $L_\varphi\ll{}L$, we have seen that the presence of one arm
is responsible for factor 
$(\frac23)^n$ originating from the $n$ crossings of the vertex. We immediatly
deduce that the Cooperon in the ring is 
$P_c^{(n)}(x,x)\simeq\frac1{2\sqrt\gamma}\,\big(\frac23\big)^{nN_a}\EXP{-n\sqrt\gamma{}L}$.

In order to study the  regime  $L_\varphi\gg{}L$, 
instead of considering the network of figure~\ref{fig:loop7} we
discuss the case where all arms are attached at the same point
in the ring. The calculation is more simple in this case.
The two situations were studied in Ref.~\cite{TexMon05} where it was
shown that the Cooperons for the two networks only slightly differ in
the regime 
$L_\varphi\ll{}L$ and are equal in the regime $L_\varphi\gg{}L$ of
interest now.
The effective length is now given by
$\cosh\sqrt\gamma{}L_\mathrm{eff}=\cosh\sqrt\gamma{}L+\frac{N_a}{2}\sinh\sqrt\gamma{}L\,\coth\sqrt\gamma\,b$.
The structure \eqref{CoopATN}  still holds. 
When $L\ll{}L_\varphi\ll{}b$ we find
$L_\mathrm{eff}\simeq\gamma^{-1/4}\sqrt{N_aL}$, therefore
$P_c^{(n)}(x,x)\simeq\frac12\sqrt{L_\varphi{}L/N_a}\,\EXP{-n\sqrt{N_aL/L_\varphi}}$, 
whose inverse Laplace transform leads to eq.~(\ref{Pnconnectedring}).
This $N_a$ dependence may be more simply obtained by noticing that,
given the winding probability $\mathcal{P}_n(x,x;t)$ for one arm, the
one for $N_a$ arms is obtained thanks to the substitution $n\to nN_a$
and~$L\to{}L/N_a$.

\subsection{The necklace of rings \label{app:necklace}}

We analyze the solution of the diffusion equation 
in a chain of $N_r$ identical rings of perimeter $L$. Rings are attached 
in such a way that the two arms joining two vertices are symmetric. 
The chain is closed in order to form a necklace for
simplicity~; as soon as the total length is smaller than $L_\varphi$, the
results are insensitive to boundary conditions~: periodic (isolated necklace) or
Dirichlet (chain connected to external contacts). Let us label the vertices 
joining consecutive rings with greek letters $\alpha,\,\beta\in\{1,\cdots,N_r\}$. 
The matrix introduced in appendix~\ref{app:spedet} has the simple form
$
\mathcal{M}_\ab=\frac{2\sqrt\gamma}{\sinh(\sqrt{\gamma}L/2)}
\big[ \delta_\ab\,2\cosh(\sqrt{\gamma}L/2)-a_\ab\,\cos(\theta/2)\big]
$
where $\theta$ is the reduced flux per ring.
With our convention the adjacency matrix reads
$a_\ab=\delta_{\alpha,\beta+1}+\delta_{\alpha,\beta-1}$. Its spectrum
of eigenvalues is $2\cos(2n\pi/N_r)$ with $n\in\{1,\cdots,N_r\}$, therefore
the spectral determinant reads~:
\begin{align}
  S(\gamma) &= \left(\frac{4\sinh(\sqrt{\gamma}L/2)}{\sqrt{\gamma}}\right)^{N_r}
  \nonumber\\
  &\times
  \prod_{n=1}^{N_r}
  \left(
    \cosh(\sqrt{\gamma}L/2)-\cos(\theta/2)\cos\frac{2\pi n}{N_r}
  \right)
  \:.
\end{align}
We replace the product by a sum by considering the logarithm. 
The sum can then be computed in the limit $N_r\to\infty$~:
\begin{widetext}
\begin{align}
  \label{Schain}
   S(\gamma) \underset{N_r\to\infty}{=} 
   \left(
     \frac{2|\cos(\theta/2)|\,\sinh(\sqrt{\gamma}L/2)}{\sqrt{\gamma}}
     \: \EXP{ \argcosh\left(\frac{ \cosh(\sqrt\gamma L/2)}{|\cos(\theta/2)|} \right) }
   \right)^{N_r}
   \:.
\end{align}

We now consider $P(x,x')$ when arguments coincide with vertices~:
\begin{align}
\label{Pchaine}
  P(\alpha,\beta) = \left(\mathcal{M}\right)_{\alpha\beta}^{-1}
  =\frac{\sinh\sqrt\gamma L/2}{N_r 4\sqrt\gamma}
  \sum_{n=1}^{N_r}
  \frac{ \EXP{\frac{2\I\pi n}{N_r}(\alpha-\beta)} }
       { \cosh\sqrt\gamma L/2 - \cos\frac{2\pi n}{N_r} }
  \underset{N_r\to\infty}{=}
  \frac1{4\sqrt\gamma}\,\EXP{-|\alpha-\beta|\sqrt\gamma L/2}
  \:.
\end{align}
\end{widetext}

\begin{figure}[htbp]
  \centering
  \includegraphics{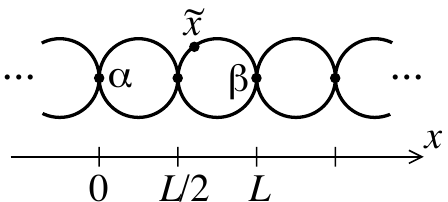}
  \caption{\it Infinite chain of rings.}
  \label{fig:necklace}
\end{figure}

We introduce the notation $\tilde{x}\equiv(x,f)$ to locate points in the
chain, where $x$ is the coordinate along the axis (see figure)~; in the $n$-th
ring $x\in[nL/2,(n+1)L/2]$. The discrete index $f\in\{u,d\}$ indicates
whether the point is in the upper branch or the lower branch.

If $\tilde{x}$ and $\tilde{x}'$ are arbitrary positions we can consider
several situations.

\noindent$\bullet$ $\tilde{x}$ and $\tilde{x}'$ belong to different rings.
The Cooperon is
\begin{equation}
  P(\tilde{x},\tilde{x}') 
  \simeq \frac1{4\sqrt\gamma}\,\EXP{-\sqrt\gamma|x-x'|}
  \:,
\end{equation}
that is {\it half} of the result for the infinite line.

\vspace{0.25cm}

%

\noindent$\bullet$ $\tilde{x}$ and $\tilde{x}'$ belong to the same ring.
\begin{equation}
  P(\tilde{x},\tilde{x}') 
  \underset{\gamma\to0}{\simeq} \frac1{4\sqrt\gamma}
  \:.
\end{equation}

We conclude that in time representation
\begin{align}
  \label{eq:PdexxtChain}
  \boxed{
  \mathcal{P}(\tilde{x},\tilde{x}';t) 
  \underset{t\gg L^2}{\simeq}
  \frac12\,\frac1{\sqrt{4\pi t}}\,\EXP{-(x-x')^2/4t}
  }
\end{align}
(the relation is exact when $\tilde{x}$ and $\tilde{x}'$ belong to different
rings). The $1/2$ corresponds to the probability to end in one of the two arms
of the rings. It ensures normalisation~:
$
\int\D\tilde{x}\,\mathcal{P}(\tilde{x},\tilde{x}';t)
\equiv
\sum_{f=u,d}\int\D{x}\,\mathcal{P}(\tilde{x},\tilde{x}';t)=1
$.

\vspace{0.25cm}

\mathversion{bold}
\noindent{\bf The function $W(x,x')$.--}
\mathversion{normal}
Expression (\ref{Pchaine}) can be used to construct $P_d(x,x')$ by taking the
limit $\gamma\to0$. We find for the function entering into description of
dephasing~:
\begin{align}
  \label{Wnecklace1}
  W(x,x') = \frac14|x-x'| = \frac12 W_\mathrm{wire}
\end{align}
for $x,\,x'$ in different rings, and
\begin{align}
  \label{Wnecklace2}
  W(x,x')= \frac12|x-x'|\left(1-\frac{|x-x'|}{L}\right) = W_\mathrm{ring}
\end{align}
for $x,\,x'\in$\:same ring.
In (\ref{Wnecklace1}) positions are measured along an axis along the chain~; the
result is {\it half} of the result for a wire.
As it has been already noticed this result can be understood thanks to the
relation~(\ref{RqGilles})~: when the number of wires between each node is
doubled, the resistance is divided by a factor of $2$.
In (\ref{Wnecklace2}) coordinates are relative to a unique axis inside the ring
(it is understood that the expression is periodic)~;
the result is exactly the result obtained for an isolated ring. Once again
interpretation is easy~: for an infinitely long necklace, when two external
wires are plugged inside a ring, no current can flow out of the ring and
resistance $\mathcal{R}(x,x')$ is not affected by the remaining rings.

\subsection{The square network \label{app:Psn}}

We construct the solution of the diffusion equation in the square
network. We use \eqref{remark} to express 
$P(x,x')$ when the two coordinates coincide with nodes of the
network. In this paragraph, nodes are labelled with a couple of integers
$\alpha\to(n,m)\equiv\vec{R}$. The matrix $\mathcal{M}$ has the
structure 
\begin{equation}
  \left(\mathcal{M}\right)_{\vec{R},\vec{R}\,'}  
  = \frac{\sqrt\gamma}{\sinh\sqrt\gamma a}
  \left[
    4\,\cosh\sqrt\gamma a\: \delta_{\vec{R},\vec{R}\,'} 
    -a_{\vec{R},\vec{R}\,'} 
  \right]
  \:.
\end{equation}
This matrix is easily inverted
\begin{align}
  \label{eq:E19}
  \left(\mathcal{M}^{-1}\right)_{\vec{R},\vec{R}\,'}  
  &= \frac{\sinh\sqrt\gamma a}{2\sqrt\gamma}
  \\ \nonumber
  & \times
  \int_\mathrm{ZdB}\frac{\D\vec Q}{(2\pi)^2}
  \frac{\EXP{\I \vec Q\cdot (\vec{R} - \vec{R}\,')}}
       {2\cosh\sqrt\gamma a-\cos Q_x-\cos Q_y} 
  \:,
\end{align}
where integral runs over wavevectors of the Brillouin zone.
If we are interested in $\mathcal{P}(x,x';t)$ in the large time limit,
$t\gg{}a^2$, this corresponds to consider $\gamma{}a^2\ll1$. In this
regime the above integral is dominated by small $\vec{Q}$ and we get
\begin{align}
  \label{eq:E20}
  P(\vec{R},\vec{R}\,') &= 
   \left(\mathcal{M}^{-1}\right)_{\vec{R},\vec{R}\,'}  
  \simeq \frac{a}{2} \int\frac{\D\vec Q}{(2\pi)^2}
  \frac{\EXP{\I \vec Q\cdot (\vec{R} - \vec{R}\,')}}
       {\gamma a^2 + \frac12\vec Q^2}
  \nonumber
   \\ 
  &=\frac{a}{2}\int_0^\infty\D t\,\EXP{-\gamma t}
  \frac{1}{2\pi t}\EXP{-\frac{a^2}{2t}(\vec{R} - \vec{R}\,')^2}
\end{align}
where integral over $\vec{Q}$ has been extended to $\RR^2$.
We expect $\mathcal{P}(x,x';t)$ to vary smoothly on the scale of the
lattice spacing, therefore we may write
\begin{equation}
  \label{eq:E21}
  \mathcal{P}(x,x';t)
  \underset{t\gg a^2}{\simeq}
  \frac{a}{2}\,
  \frac{1}{2\pi t}\EXP{-\frac{1}{2t}||x-x'||^2}
\end{equation}
where $||x-x'||$ designates the distance in  $\RR^2$ between the two
points $x$ and $x'$ of the square network.
This result call for two remarks. 
First, we see that the fact that the diffusion is constrained in the 1d
wires forming the square network leads to a continuum limit with a
renormalized diffusion constant $D^*=1/2$, if the diffusion constant
is $D=1$ in the wire (more generally $D^*=1/d$ where $d$ is the
effective dimensionality of the network).
Second, the probability presents a prefactor $1/2$ of similar origin
than the one in eq.~\eqref{eq:PdexxtChain}, ensuring normalization
condition. 
Each elementary plaquette, labelled by $\vec{R}\equiv(n,m)$, contains
2 wires. Therefore each wire of the network can be labelled with
$\vec{R}$ and an index $f$ taking two values (for horizontal or
vertical wires). 
We check that \eqref{eq:PdexxtSN} is correctly normalized~:
$\int_\mathrm{network}\D{}x\,\mathcal{P}(x,x';t)=\sum_{\vec{R}}\sum_f\int_{\mathrm{wire}\:(\vec{R},f)}\D{}x\,\mathcal{P}(x,x';t)\simeq\sum_{\vec{R}}2a\,\mathcal{P}(x,x';t)\simeq2a\sum_{\vec{R}}\frac{a}{2}\frac{1}{2\pi t}\EXP{-\frac{a^2}{2t}(\vec{R} - \vec{R}\,')^2}\simeq1$.

The generalization of (\ref{eq:E19},\ref{eq:E20},\ref{eq:E21})
to the other regular planar networks can be done.
We obtain~:
\begin{equation}
  \label{eq:PdexxtSN}
 \boxed{
  \mathcal{P}(x,x';t)
  \underset{t\gg a^2}{\simeq}
  \frac12\tan\left(\frac\pi{z}\right)\,
  \frac{a}{2\pi t}\EXP{-\frac{1}{2t}||x-x'||^2}
  }
\end{equation}
where the dimensionless factor 
$\frac12\tan\frac\pi{z}$, where $z$ is the coordination number of the
lattice,  has the interpretation of the area of the Wigner-Seitz
elementary cell $\mathcal{A}_z(1+\delta_{z,6})$, where $\mathcal{A}_z$
is defined in the next appendix,
divided by the number of bonds per cell.


\section{MC of planar regular networks}

\subsection{Planar regular networks}

We apply the formulae \eqref{eq:SpeDet} and \eqref{eq:PasMon} to the
case of infinite planar regular networks.
Let us denote by $z$ the coordination number of the network. The plane can be
covered by only three different tillings~: the triangular lattice
($z=6$), the square lattice ($z=4$) and the honeycomb lattice ($z=3$). 

Let us mention few properties of regular tillings of flat surfaces~: 
\begin{itemize}
\item the lattice of coordination number $z$ is a tilling by regular polygones
  with $p$ sides. $p$ is related to the coordination number by~\cite{Sti92}
  $(z-2)(p-2)=4$.
\item The area of a regular $p$-gone of side $a$ is 
  $\mathcal{A}_z=a^2\frac14p\cotg(\pi/p)=a^2\frac{z}{2(z-2)}\tan(\pi/z)$.
\item If boundary effects are neglected, 
  the number of bonds $B$ and the number of vertices $V$ of the planar
  network are
  related by $2B=zV$ ($z$ bonds issue from each vertex and each bond
  touch 2 vertices).
\end{itemize}
The matrix (\ref{MatM}) takes the form
$\mathcal{M}=\frac{\sqrt\gamma}{\sinh\sqrt\gamma\,a}N$ where the
matrix $N$ is given by 
\begin{equation}
  N_\ab = \delta_\ab\,z\,\cosh\sqrt\gamma\,a - a_\ab\,\EXP{-\I\theta_\ab}
\end{equation}
where reduced fluxes $\theta_\ab$ describe a uniform magnetic field
$\mathcal{B}$. 
We use eqs.~(\ref{eq:SpeDet},\ref{eq:PasMon}) to express the WL
correction 
$\Delta\tilde\sigma=-\frac2{\mathrm{Vol}}\derivp{}{\gamma}\ln{}S(\gamma)$,
where the volume is related to the number of bonds $\mathrm{Vol}=Ba$,
\begin{align}
  \label{eq:WLregnet}
  \Delta\tilde\sigma(\theta) &= 
  \left(\frac2z-1\right)\, L_\varphi\,
  \left(\coth\frac{a}{L_\varphi}-\frac{L_\varphi}{a}\right)
  \nonumber\\
  &- \frac{z}{B}\,L_\varphi\,\sinh\frac{a}{L_\varphi}\,
  \tr{\frac1{N(\gamma,\theta)}}
\end{align}
$\theta=4\pi\phi/\phi_0$ is the reduced flux per elementary plaquette.
The computation of this expression requires the knowledge of the
spectrum of the matrix $N$, {\it i.e.} of $a_\ab\,\EXP{-\I\theta_\ab}$
(Hofstadter problem~\cite{footnote18}).

\subsection{Continuum limit\label{app:ConLimPlanNet}}

We study the continuum limit~: for $L_\varphi\gg{}a$ the WL correction probes
large scales. Moreover when the flux per cell is much smaller that the
flux quantum, $\phi\ll\phi_0$, the result for the planar network
coincide with the result for a plane. Let us show that this is indeed
the case.
We note that the action of the adjacency matrix $a_\ab$ on a smooth
function can be replaced by the Laplacian~:
$a_\ab\longrightarrow\frac14za^2\,\Delta+z$.
Therefore in the presence of a weak magnetic field
$a_\ab\,\EXP{-\I\theta_\ab}\longrightarrow\frac14za^2\,(\nabla-2\I\,eA)^2+z$
involves the covariant derivative. The spectrum of this operator is
the Landau spectrum shifted by $-z$~:
$\mathrm{Spec}(-a_\ab\,\EXP{-\I\theta_\ab})\simeq\{za^2e\mathcal{B}(n+1/2)-z\,|\,n\in\NN\}$,
where each Landau level has a degeneracy
$d_\mathrm{LL}=\frac1\pi\,e\mathcal{B}\mathrm{Area}$,
``$\mathrm{Area}$'' being the total area occupied by the planar network.

In the limit $L_\varphi\gg{}a$ and $\theta\ll1$,
eq.~\eqref{eq:WLregnet} is dominated by 
the trace which is itself dominated by the bottom of the spectrum of
$-a_\ab\,\EXP{-\I\theta_\ab}$  and 
it rewrites 
\begin{align}
  \Delta\tilde\sigma
  &\simeq
  -a\frac{z}{B}\,\frac{e\mathcal{B}\mathrm{Area}}\pi\,
  \nonumber\\
  &\times
  \sum_{n=0}^{N_c}
  \frac{1}{z\frac{a^2}{2L_\varphi^2}+za^2e\mathcal{B}(n+1/2)}
  +\mathrm{cste}
\end{align}
where we have introduced a cutoff $N_c$ corresponding to
the number of the Landau level for which continuum limit fails~:
$za^2e\mathcal{B}(N_c+1/2)\sim{}z$, the width of the spectrum of the
adjacency matrix. This gives
$N_c\sim1/(a^2e\mathcal{B})\sim\phi_0/(\mathcal{B}a^2)$.  
Using that the total number of bonds is related to the number of elementary
plaquettes by 
$B=\frac{2z}{z-2}\frac12\times(\#$\:of\:plaquettes$)$ we find 
$\mathrm{Area}=\frac{a^2}{2}\tan(\pi/z)\:B$.
Finally the WL correction may be expressed with  the Digamma
function~\cite{gragra}~: 
\begin{align}
  \label{eq:ConLimPlanNet}
  \Delta\tilde\sigma(\theta\ll1)
  &\simeq \frac{a}{2\pi}  \tan\left(\frac\pi{z}\right)
  \\\nonumber
  &\times
  \left[
     \psi\!\left(\frac12+\frac{\phi_0}{4\pi\mathcal{B}L_\varphi^2}\right)
    -\ln\left(\frac{\phi_0}{\mathcal{B}a^2}\right)
  \right]
  +\mathrm{cste}
  \:.
\end{align}
This result indeed coincides with the result for a plane, 
eq.~\eqref{eq:Bergman}, apart for a factor $2$ in the argument of
Digamma functions. We think that it is
worth devoting a small paragraph to this point since it is related to
some interesting property of the continuum limit of the result for the
networks.

\vspace{0.25cm}

\noindent{\it The additional factor of $2$ and the continuum limit.--}
As mentioned in section~\ref{sec:sn} the WL correction in the network
can be written with a path integral as 
\begin{align}
  \label{eq:EncoreWL}
  \Delta\tilde\sigma(\theta)&=-2\int_0^\infty\D t\,\EXP{-t/L_\varphi^2}
 \\ \nonumber & \times
  \int_{x(0)=x}^{x(t)=x}\mathcal{D}x(\tau)\,
  \EXP{-\frac14\int_0^t\D\tau\,\dot{x}^2+\I\theta \mathcal{N}[x(\tau)]}
\end{align}
where $\mathcal{N}[x(\tau)]$ is the (algebraic) number of elementary
plaquettes surrounded the trajectories.
When one is interested in large time scale properties ($t\gg{}a^2$), 
eq.~\eqref{eq:PdexxtSN} shows that the path
integral over Brownian paths $x(\tau)$ in the square network can be
replaced by a path integral over planar Brownian curves
$\vec{r}(\tau)$ as  
\begin{align}
 & \int_{x(\tau)\in\mathrm{network}}\hspace{-1cm}\mathcal{D}x(\tau)\,
  \EXP{-\frac14\int\D\tau\,\dot{x}^2}  
  \nonumber\\
  &\hspace{0.5cm}\to
  a \frac12\tan\left(\frac\pi{z}\right)
  \int_{\vec{r}(\tau)\in\mathrm{plane}}\hspace{-0.75cm}\mathcal{D}\vec{r}(\tau)\,
  \EXP{-\frac1{2}\int\D\tau\,\dot{\vec{r}}\,^2}
  \:.
\end{align}

$\theta\mathcal{N}[x(\tau)]=2e\mathcal{B}\mathcal{A}[x(\tau)]$ coincides
with the magnetic flux enclosed by the trajectory multiplied by $2e$, where
$\mathcal{A}[x(\tau)]$ is the algebraic area enclosed by the curve
($\mathcal{A}[x(\tau)]=a^2\mathcal{N}[x(\tau)]$ for the square network).
Finally we can rewrite \eqref{eq:EncoreWL} as
\begin{align}
  \Delta\tilde\sigma(\theta\ll1)
  &\simeq-a\tan\left(\frac\pi{z}\right)\int_{a^2}^\infty\D t\,
  \EXP{-t/L_\varphi^2}
 \\ \nonumber & \times
  \int_{\vec{r}(0)=\vec{r}}^{\vec{r}(t)=\vec{r}}\mathcal{D}\vec{r}(\tau)\,
  \EXP{-\frac1{4D^*}\int_0^t\D\tau\,\dot{\vec{r}}\,^2
       +2\I e\mathcal{B}\mathcal{A}[\vec{r}(\tau)]}
\end{align}
with $D^*=1/2$.
Then we can use the well-known result~\cite{Ber84,AkkMon07}
\eqref{eq:Bergman} to get 
$
\Delta\tilde\sigma(\theta\ll1)\simeq\frac{a}{2\pi}
\tan(\frac\pi{z})
\big[
     \psi\big(\frac12+\frac{\phi_0}{8\pi\mathcal{B}D^*L_\varphi^2}\big)
    -\ln\big(\frac{\phi_0}{\mathcal{B}a^2}\big)
\big]
+$cste,
which precisely coincides with \eqref{eq:ConLimPlanNet}.

\subsection{Square network\label{app:DouRam}}

The WL correction can be expressed more explicitely for rational fluxes.
We recall briefly a derivation due to Dou\c{c}ot \&
Rammal~\cite{DouRam86} for the case of 
a square network of dimension $V=N_x\times{N_y}$. We start from
\begin{align}
  \Delta\tilde\sigma(\theta)
  &=-\frac{L_\varphi}{2}
  \bigg[
    \coth(\sqrt\gamma\,a) -\frac1{\sqrt\gamma\,a}
    \\\nonumber
  &+ 4\sinh(\sqrt\gamma\,a)
      \:\frac1{N_xN_y}\,\tr{ \frac1{N(\gamma,\theta)} }
  \bigg]
  \:.
\end{align}
We label the vertices $\alpha\equiv(n,m)$, $n$ for the position along
the horizontal direction and $m$ along the vertical direction. We can choose a
Landau gauge~: $\theta_{\alpha\beta}=m\theta$ for horizontal wires,
$\alpha\equiv(n,m)$ and $\beta\equiv(n+1,m)$, and $\theta_{\alpha\beta}=0$ for
vertical wires.

The computation of the MC for a square network submitted to a magnetic field
requires to consider the problem of determination of the spectrum of
$H_{\alpha\beta}=-a_{\alpha\beta}\EXP{-\I\theta_{\alpha\beta}}$
(Hofstadter problem).
In the Landau gauge where the flux is put along the $n$ axis 
the solution of $H\psi=\varepsilon\psi$ can be written as 
$\psi_{n,m}=\Phi_m\EXP{\I k_xn}$. We obtain the Harper equation
$\Phi_{m-1}+[\varepsilon+2\cos(k_x+m\theta)]\Phi_{m}+\Phi_{m+1}=0$.

If the flux is rational $\theta_{p,q}=2\pi{p}/q$ with $p,\,q\in\NN$, the Harper
equation is periodic with periodicity $q$. Therefore, writing $m=r+qs$ with
$s\in\ZZ$ and $r\in\{1,\cdots,q\}$ we can look for solutions of the form
$\Phi_m=\varphi_r\EXP{\I k_ys}$. The wave function $\varphi_r$ is solution of
a linear system $D_q\varphi=0$ where the hermitian $q\times{q}$-matrix $D_q$
is defined by
\begin{align}
  \left\{
    \begin{array}{llll}
     D_{r,r}   & = & \varepsilon+2\cos\left( k_x + 2\pi\frac{rp}{q} \right) 
               & \mbox{for } r=1,\cdots,q\\
     D_{r+1,r} & = & 1 & \mbox{for } r=1,\cdots,q-1\\
     D_{q,1}   & = & \EXP{\I k_y} &
    \end{array}
  \right.
  \:.
\end{align}
The secular equation $\det{D_q}=0$ gives the $q$ energy bands denoted 
$\varepsilon_r(\vec{k})$.
Interestingly the secular equation can be rewritten as
$\det{D_q}=(-1)^q[P_{p,q}(\varepsilon)-2(\cos qk_x+\cos k_y)]=0$, where 
$P_{p,q}(\varepsilon)$ is a polynomial of degree~\cite{ClaWan78} $q$.
For example 
$P_{1,1}(\varepsilon)=-\varepsilon$, 
$P_{1,2}(\varepsilon)=\varepsilon^2-4$, 
$P_{1,3}(\varepsilon)=-\varepsilon^3+6\varepsilon$, etc.

The weak localization correction involves
\begin{align}
  \label{fctGreenHof}
  &\frac1{N_xN_y}\,\tr{ \frac1{N(\gamma,\theta_{p,q})} }
  \nonumber\\
  &=\int_0^{2\pi}\frac{\D^2\vec{k}}{(2\pi)^2}\:\frac1q
   \sum_{r=1}^q\frac1{4\cosh(\sqrt\gamma\,a) + \varepsilon_r(\vec{k})}
  \\
  &= \frac1q\,\int_0^{2\pi}\frac{\D^2\vec{k}}{(2\pi)^2}\,
      \frac{P'_{p,q}(4\cosh(\sqrt\gamma\,a))}
             {P_{p,q}(4\cosh(\sqrt\gamma\,a))-2\cos k_x-2\cos k_y }
  \:,
\end{align}
where we have used the equality 
$
\sum_{r=1}^q\frac1{\varepsilon-\varepsilon_r}
=\frac{P'(\varepsilon)}{P(\varepsilon)}
$, 
valid for a polynomial of degree $q$~: 
$P(\varepsilon)=\prod_{r=1}^q(\varepsilon_r-\varepsilon)$.
Integration over $\vec{k}$ leads to the result (\ref{DouRam86}) 
of Dou{\c c}ot \& Rammal.

\end{appendix}



\begin{thebibliography}{10}
\bibitem{footnote1}
  Amplitude of Aharonov-Bohm (AB)
  oscillations can be studied by analysis of the correlation function
  of the magnetoconductance. Therefore AB oscillations of the
  conductance (period 
  $\phi_0$) are related to universal conductance fluctuations (UCF)
  while AAS oscillations of the averaged conductance (period
  $\phi_0/2$) are oscillations of the WL correction.
  See discussion in the appendix of Ref.~\cite{TexMon05b}.
\bibitem{footnote2} 
  It was shown in
  Ref.~\cite{TexMon05} on the case of the ring connected to arms
  that the crossover between the ``small
  coherence length regime'' $L_\varphi/L\ll1$ and the ``large coherence
  length regime'' $L_\varphi/L\gg1$ occurs in practice already for
  $L_\varphi\gtrsim{}L$. The study of this regime is far from academic
  since in metallic samples $L_\varphi$ can be made larger than
  $10\:\mu$m at $T\simeq20$mK, like in
  experiment~\cite{BauMalSchMaiEskSam05}.
\bibitem{footnote19}
 This question was first considered in a ring in Ref.~\cite{San89} who
 weighted properly the Cooperon in the wires. The rigorous proof and
 the generalization leading to a proper analysis of nonlocal effects
 was provided in Ref.~\cite{TexMon04}.
\bibitem{footnote3} 
  Factors 2 are missing in the $L_N$ given in AAK's
  original paper~\cite{AleAltGer99,AkkMon07}.
\bibitem{footnote4} 
  Let us estimate this length for the samples of 
  Ref.~\cite{SchMalMaiTexMonSamBau07} made of pure silver wires
  of width $w=60\:$nm and thickness $b=50\:$nm. Fermi wavelength in
  silver is $k_F^{-1}=0.083\:$nm 
  (i.e. DoS is $2\rho_0=18\:$eV$^{-1}$nm$^{-3}$).
  This corresponds to $N_c=33000$ channels.
  Diffusion constant was measured to be $D\simeq0.01\:$m$^2/$s corresponding 
  to elastic mean free path $\ell_e\simeq23\:$nm
  and leading to $L_T(T=1\:$K$)=0.27\:\mu$m.  We get
  $L_N(T=1\:$K$)=2.9\:\mu$m.
\bibitem{footnote5}
  Let us remind that $\ell_\mathrm{loc}\sim{}N_c\ell_e$ is the 
  localization length of the infinitely long
  wire~\cite{Dor82,MelPerKum88,Bee97}. The perturbation theory is 
  valid for~\cite{AltAro85} $L_N\ll\ell_\mathrm{loc}$
  (i.e. $L_T\ll{}L_N$), that is  above the temperature
  $T^*\sim{}D/\ell_\mathrm{loc}\sim1/(N_c^2\tau_ed)$.
\bibitem{footnote6}
  The experiment \cite{Mal06} has investigated the regime 
  $L_\varphi\gtrsim{}L$ where nonlocal effects are important, above 
  $1\:$K, when decoherence is dominated by
  electron-phonon scattering. The experimental MC was perfectly
  fitted by the MC obtained for a simple exponential relaxation.
\bibitem{footnote7}
  Note that the function $W(r,r')=P_d(r,r)-P_d(r,r')$ involved in this
  functional may present a divergent behaviour at short distance~:
  this is the case for the metallic film or the hollow cylinder studied
  in the last section for which $P_d(r,r')\sim\ln||r-r'||$.
  In this case one must come back to the assumption of classical
  electric fluctuations~: due to the Pauli principle, energy
  transfers between electrons
  are limited to $\omega\lesssim{}T$. This amounts to replace the Dirac
  of the correlator $\smean{V(r,t)V(r',t')}_V$ by a function of width
  $1/T$~: $\delta(t-t')\to\delta_T(t-t')$.
  The diverging term
  $\int\D\tau\,P_d(r(\tau),r(\tau))$ should be replaced by a term
  \cite{TreYevMarDelLer09}
  $\int\D\tau\D\tau'\,\delta_T(\tau-\tau')\,P_d(r(\tau),r(\tau'))$. 
  Therefore the $\delta_T$-function constrains the two trajectories
  to differ by at most the thermal length
  $||r(\tau)-r(\tau')||\lesssim{}L_T$.
  This is the reason why it is sufficient to cut off the divergence of
  $P_d(r,r)$ in $W(r,r')$ at scale $L_T$, according to
  Refs.~\cite{AltAroKhm82,AltAro85}.
\bibitem{footnote10}
  In Ref.~\cite{AltAro85} a
  distinction between inelastic and ``quasielastic'' processes is made. The
  former refer to high energy transfers (however limited by the Pauli
  principle) while the latter stands for low energy transfers. This
  separation is motivated by the low energy divergences in the usual
  perturbative treatment \cite{Eil84}. This has led AAK to
  propose the influence functional approach~\cite{AltAroKhm82}. 
  We however insist on the fact that the influence functional approach
  fully account for electron-electron
  interaction, what is clear on the more sophisticated
  version~\cite{MarDelSmiAmb07,DelMarSmiAmb07}, and not  
  only to quasielastic processes.
\bibitem{footnote11}
  A quantity frequently introduced is $g={(h/e^2)}/{R_\square}$,
  the ``dimensionless conductance'' (of a square connected at two
  opposite sides).
  In a plane (2DEG), the sheet resistance
  $R_\square^\mathrm{plane}=1/\sigma_0$ reads 
  $({h/e^2})/{R_\square^\mathrm{plane}} = \frac{2m^*D}{\hbar}=k_F\ell_e$,
  where $m^*$ is the effective mass of charge carriers. 
  In a thin film of thickness $b$, it  is given by 
  $\frac{h/e^2}{R_\square} = 4\pi\hbar\,\rho_0\,b\,D 
  = \frac2{3\pi}\,k_F^2b\,\ell_e$
  where $\rho_0$ is the DoS per volume and per spin channel.
  LEt us give order of magnitude~:
  for a film of silver of thickness $b=50\:$nm  with
  $\ell_e=23\:$nm we obtain~${(h/e^2)}/{R_\square}\simeq34000$.
\bibitem{footnote12}
  $\Delta\sigma=-\frac{2e^2}{\pi}\int_0^\infty\D{t}\,\mathcal{P}(t)\,(\EXP{-t/\tau_\varphi}-\EXP{-t/\tau_e})$.
  The``return probability'' is given by 
  $\mathcal{P}(t)=\frac1{\mathrm{Area}}\sum_n\EXP{-E_nt}$ where
  ``energies'' correspond to the spectrum of $-(\nabla-2\I\,eA)^2$~:
  $E_n=4e\mathcal{B}(n+1/2)$ for a degeneracy
  $\frac{e\mathcal{B}\mathrm{Area}}\pi$, 
  therefore
  $\mathcal{P}(t)=\frac{e\mathcal{B}}{\pi\sinh(2e\mathcal{B}t)}$. 
  Using  formula [3.541] of Ref.~\cite{gragra},
  $\int_0^\infty\D{t}\frac{\lambda}{\sinh\lambda{}t}(\EXP{-\gamma_1t}-\EXP{-\gamma_2t})=\psi\big(\frac12+\frac{\gamma_2}{2\lambda}\big)-\psi\big(\frac12+\frac{\gamma_1}{2\lambda}\big)$
  gives \eqref{eq:Bergman}.
\bibitem{footnote13}
  Note that this procedure can be avoided by computing the Cooperon
  beyond the  diffusion 
  approximation (by solving the Bethe-Salpether integral equation) as it has
  been discussed in Ref.~\cite{DmiKacGor97} (these authors have extended a
  previous work \cite{CasSha94} where some contributions were forgotten).
\bibitem{footnote15}
  for the silver samples of
  Refs.~\cite{SchMalMaiTexMonSamBau07,Mal06} the factor is
  $w/a\simeq0.09$ therefore $(h/e^2)/R_\square^\mathrm{net}\simeq3200$
  (compared to ${(h/e^2)}/{R_\square}\simeq34000$ for the film).
  From $D\simeq0.01\:$m$^2/$s we obtain 
  $L_T\simeq0.27\:\mu$m at $T=1\:$K. Therefore the various Nyquist
  lengths are $L_N(T=1\:$K$)\simeq2.9\:\mu$m for the infinite wire,
  $L_N^\mathrm{film}(T=1\:$K$)\simeq16\:\mu$m for the 2d film
  and 
  $L_N^\mathrm{net}(T=1\:$K$)\simeq6.4\:\mu$m for the square network.
\bibitem{footnote17}
  Note that if $\xi_n(\tau)\eqdef x_n(\tau) - x_n(t-\tau)$ we have
  $\xi_n(0)=-n$ and~$\xi_n(t)=n$.
\bibitem{footnote18}
  The Landau problem on a lattice was solved by Hofstadter \cite{Hof76} on the
  square lattice, by Claro \& Wannier \cite{ClaWan79} on the
  triangular lattice and on the honeycomb
  lattice by Rammal \cite{Ram85}.
  Note however that Hofstadter butterflies were already plotted in an older
  paper by Langbein~\cite{Lan69} (see also \cite{GerLan66}).
\end{thebibliography}

\end{document}